\documentclass[useAMS,usenatbib]{mnras} 

\def\aj{AJ}%
\def\araa{ARA\&A}%
\def\apj{ApJ}%
\def\apjl{ApJ}%
\def\apjs{ApJS}%
\def\ao{Appl.~Opt.}%
\def\aap{A\&A}%
\def\aaps{A\&AS}%
\def\mnras{MNRAS}%
\def\prd{Phys.~Rev.~D}%
\def\pasp{PASP}%
\def\ssr{Space~Sci.~Rev.}%
\def\jcap{J. Cosmology Astropart. Phys.}%

\newcommand{\LMU}{1}
\newcommand{\ECUniverse}{2}
\newcommand{\ANL}{3}
\newcommand{\KICPChicago}{4}
\newcommand{\AIfA}{5}
\newcommand{\Leiden}{6}
\newcommand{\MPE}{7}
\newcommand{\Stanford}{8}
\newcommand{\SLAC}{9}
\newcommand{\KIPAC}{10}
\newcommand{\MIT}{11}
\newcommand{\FNAL}{12}
\newcommand{\AAUChicago}{13}
\newcommand{\UMKC}{14}
\newcommand{\UFlorida}{15}
\newcommand{\Berkeley}{16}
\newcommand{\StonyBrook}{17}
\newcommand{\Arizona}{18}
\newcommand{\Melbourne}{19}
\newcommand{\CASA}{20}
\newcommand{\Ames}{21}
\newcommand{\INAFTrieste}{22}
\newcommand{\LSST}{23}
\newcommand{\CfA}{24}
\newcommand{\Harvard}{25}

\usepackage[export]{adjustbox}
\usepackage{graphicx,amsmath,amssymb,color,subfigure,multirow}
\usepackage[normalem]{ulem}

\makeatletter
\newlength{\abovecaptionskip}%
\makeatother
\usepackage{threeparttable}%
\usepackage{lastpage}
\usepackage{microtype}

\usepackage[svgnames]{xcolor}

\newcommand{\altaffilmark}[1]{$^{#1}$}

\bibpunct{(}{)}{;}{a}{}{,}

\graphicspath{{figures/}}

\renewcommand{\vec}[1]{\mbox{\boldmath$#1$}}

\newcommand{\dif}{\mathrm{d}}
\newcommand{\mgas}{\ensuremath{M_\mathrm{gas}}}
\newcommand{\hm}{h^{-1}_{70}}
\newcommand{\h}{h_{70}}
\newcommand{\gp}{\ensuremath{g^\prime}}
\newcommand{\rp}{\ensuremath{r^\prime}}
\newcommand{\ip}{\ensuremath{i^\prime}}
\newcommand{\et}{\ensuremath{e_\mathrm{t}}}
\newcommand{\ex}{\ensuremath{e_\times}}
\newcommand{\gt}{\ensuremath{{g_\mathrm{t}}}}

\newcommand{\mbeta}{\ensuremath{\langle\beta\rangle}}
\newcommand{\mbetasq}{\ensuremath{\langle\beta^2\rangle}}
\newcommand{\msun}{\ensuremath{\mathrm{M}_\odot}}
\newcommand{\mfive}{\ensuremath{M_{500}}}
\newcommand{\mtwo}{\ensuremath{M_{200}}}
\newcommand{\asz}{\ensuremath{{A_\mathrm{SZ}}}}
\newcommand{\bsz}{\ensuremath{{B_\mathrm{SZ}}}}
\newcommand{\csz}{\ensuremath{{C_\mathrm{SZ}}}}
\newcommand{\dsz}{\ensuremath{\sigma_{\ln \zeta}}}
\newcommand{\ay}{\ensuremath{{A_{Y_\mathrm{X}}}}}
\newcommand{\by}{\ensuremath{{B_{Y_\mathrm{X}}}}}
\newcommand{\cy}{\ensuremath{{C_{Y_\mathrm{X}}}}}
\newcommand{\dy}{\ensuremath{\sigma_{\ln Y_\mathrm{X}}}}
\newcommand{\yx}{\ensuremath{{Y_\mathrm{X}}}}
\newcommand{\am}{\ensuremath{{A_{M_\mathrm{g}}}}}
\newcommand{\bm}{\ensuremath{{B_{M_\mathrm{g}}}}}
\newcommand{\cm}{\ensuremath{{C_{M_\mathrm{g}}}}}
\newcommand{\dm}{\ensuremath{\sigma_{\ln M_\mathrm{g}}}}

\newcommand{\bwl}{\ensuremath{{b_\mathrm{WL}}}}
\newcommand{\wllocal}{\ensuremath{\sigma_\mathrm{WL, local}}}
\newcommand{\zl}{\ensuremath{{z_\mathrm{l}}}}

\definecolor{joerg}{rgb}{0.7, 0.4, 0.0}

\title[Weak Lensing Calibrated SZE and X-ray Scaling
Relations]{Sunyaev-Zel'dovich Effect and X-ray Scaling Relations from Weak-Lensing
  Mass Calibration of 32 SPT Selected Galaxy Clusters}

\author[J.\,P. Dietrich et
al.]{J.\,P. Dietrich\thanks{Email:dietrich@usm.lmu.de}\altaffilmark{\LMU,\ECUniverse},
  S. Bocquet\altaffilmark{\ANL,\KICPChicago}, 
  T. Schrabback\altaffilmark{\AIfA}, D. Applegate\altaffilmark{\KICPChicago},
  H. Hoekstra\altaffilmark{\Leiden}, 
  \newauthor
  S. Grandis\altaffilmark{\LMU,\ECUniverse},
  J.\,J. Mohr\altaffilmark{\LMU,\MPE,\ECUniverse}, 
  S.~W.~Allen\altaffilmark{\Stanford,\SLAC,\KIPAC},
  M.~B.~Bayliss\altaffilmark{\MIT},
  B.~A.~Benson\altaffilmark{\FNAL,\AAUChicago,\KICPChicago},
  \newauthor
  L.~E.~Bleem\altaffilmark{\ANL,\KICPChicago}
  M.~Brodwin\altaffilmark{\UMKC},
  E.~Bulbul\altaffilmark{\MIT},
  R.~Capasso\altaffilmark{\LMU,\ECUniverse},
  I. Chiu\altaffilmark{\LMU,\ECUniverse},
  \newauthor
  T.~M.~Crawford\altaffilmark{\KICPChicago,\AAUChicago},
  A.~H.~Gonzalez\altaffilmark{\UFlorida},
  T.~de~Haan\altaffilmark{\Berkeley},
  M.~Klein\altaffilmark{\LMU,\ECUniverse},
  \newauthor
  A.~von~der~Linden\altaffilmark{\StonyBrook},
  A.~B.~Mantz\altaffilmark{\KIPAC,\Stanford},
  D.~P.~Marrone\altaffilmark{\Arizona}
  M.~McDonald\altaffilmark{\MIT},
  \newauthor
  S.~Raghunathan\altaffilmark{\Melbourne},
  D.~Rapetti\altaffilmark{\CASA,\Ames},
  C.~L.~Reichardt\altaffilmark{\Melbourne},
  A.~Saro\altaffilmark{\LMU,\ECUniverse,\INAFTrieste},
  B.~Stalder\altaffilmark{\LSST},
  \newauthor
  A.~Stark\altaffilmark{\CfA},
  C.~Stern\altaffilmark{\LMU,\ECUniverse},
  C.~Stubbs\altaffilmark{\Harvard}
  \\
  \vspace{0.4cm}\\
  \parbox{\textwidth}{\large Affiliations are listed at the end of the paper}}
\begin{document}

\date{Accepted 2018 November 8. Received 2018 November 7; in original form 2017 November 14}
\pagerange{\pageref{firstpage}--\pageref{LastPage}}\pubyear{2018}
\maketitle
\label{firstpage}

\begin{abstract}
Uncertainty in the mass-observable scaling relations is currently the limiting
factor for galaxy cluster based cosmology. Weak gravitational lensing can
provide a direct mass calibration and reduce the mass uncertainty.
We present new ground-based weak lensing observations of 19 South Pole
Telescope (SPT) selected clusters at redshifts $0.29 \leq z \leq 0.61$ and
combine them with previously reported space-based observations of 13 galaxy
clusters at redshifts $0.576 \leq z \leq 1.132$ to constrain the cluster mass
scaling relations with the Sunyaev-Zel'dovich effect (SZE), the cluster gas
mass \mgas, and \yx, the product of \mgas\ and X-ray temperature.
We extend a previously used framework for the analysis of scaling relations
and cosmological constraints obtained from SPT-selected clusters to make use
of weak lensing information. We introduce a new approach to estimate the
effective average redshift distribution of background galaxies and quantify a
number of systematic errors affecting the weak lensing modelling. These errors
include a calibration of the bias incurred by fitting a Navarro-Frenk-White
profile to the reduced shear using $N$-body simulations. We blind the analysis
to avoid confirmation bias.
We are able to limit the systematic uncertainties to 5.6\% in cluster mass
(68\% confidence). Our constraints on the mass--X-ray observable scaling
relations parameters are consistent with those obtained by earlier studies,
and our constraints for the mass--SZE scaling relation are consistent with the
simulation-based prior used in the most recent SPT-SZ cosmology analysis.
We can now replace the \emph{external} mass calibration priors used in
previous SPT-SZ cosmology studies with a direct, \emph{internal} calibration
obtained on the same clusters.
\end{abstract}


\section{Introduction}
The cluster mass function, i.e. the abundance of clusters of galaxies as a
function of redshift and mass, is a sensitive cosmological probe
\citep[see][for a review]{allen11}. Because the cluster mass function is
sensitive to both the expansion history and the history of structure formation
in the Universe, cluster cosmology is in principle able to break degeneracies
between cosmological parameters arising in purely geometric probes such as the
primary Cosmic Microwave Background (CMB), baryonic acoustic oscillations, and
supernovae type Ia. Observable properties of galaxy clusters like X-ray
luminosity and temperature, optical richness, and the strength of the
Sunyaev-Zel'dovich Effect \citep[SZE,][]{sunyaev70,sunyaev72} have been shown
to scale with galaxy cluster mass following \emph{mass--observable scaling
  relations} (MOR). These scaling relations have intrinsic scatter around the
mean relationship between the observable, which is used as a proxy for cluster
mass, and the cluster mass, which has been used to parametrise the theoretical
cluster mass function. Cosmological constraints from cluster mass function
studies are currently limited by uncertainties in the mass--scaling relation
parameters.

Weak gravitational lensing offers the best opportunity to determine the
normalisation of the MOR as it can estimate projected cluster masses with near
zero bias on average \citep{corless09,becker11,bahe12}. The scatter between
lensing inferred cluster masses and true halo mass, however, is large and
typically exceeds the intrinsic scatter of the mass--observable relations
employed for cosmological purposes. Sources of this scatter include the shape
noise of the lensed background galaxies, correlated and uncorrelated
large-scale structure \citep[LSS,][]{hoekstra01, dodelson04, becker11} along
the line-of-sight, and halo triaxiality \citep{clowe04, corless07,
  meneghetti10}, the latter being the dominant source of scatter for massive
galaxy clusters. Therefore, large numbers of clusters are required to achieve
a competitive calibration of the normalisation of mass--observable scaling
relations. Several programs making use of gravitational lensing to this end
have published results \citep[e.g.][]{bardeau07, okabe10, hoekstra12,
  marrone12, applegate14, umetsu14, gruen14, hoekstra15, okabe16, battaglia16,
  applegate16, hilton18}, or are underway employing data from current
wide-field imaging surveys such as the Dark Energy Survey \citep{melchior17}
or the HyperSuprimeCam survey \citep{murata17}. Future surveys and
missions such as LSST\footnote{https://www.lsst.org/} \citep{lsst12},
\textit{Euclid}\footnote{https://www.euclid-ec.org/} \citep{laureijs11}, or
CMB-S4\footnote{https://www.cmb-s4.org/} \citep{abazajian16} will lead to much
tighter constraints while at the same time imposing much stricter requirements
for the control of systematic errors.

Here we describe the weak lensing analysis of 19 intermediate redshift
clusters selected from the 2500 square degree SPT-SZ survey \citep{bleem15},
five of which have already been presented in an earlier weak lensing study
\citep{high12}. After discussing these data in Section~\ref{sec:data} we
present our weak lensing methods in Sections~\ref{sec:weak-lensing-data} and
\ref{sec:weak-lensing-mass}, paying particular attention to controlling
systematic effects. In Section~\ref{sec:scaling-relations} we then combine our
19 clusters with 13 high redshift clusters from the SPT-SZ survey with
existing weak lensing data from the \textit{Hubble Space Telescope}
\citep[HST,][S18, hereinafter]{schrabback18}\defcitealias{schrabback18}{S18}
and X-ray data from the \textit{Chandra} X-ray satellite for 89 clusters to
perform a joint mass--observable scaling relations analysis using a newly
developed framework that self-consistently accounts for selection effects and
biases.

For quantities evaluated at a fixed cosmology we assume a flat $\Lambda$CDM
cosmology with $\Omega_\mathrm{m} = 0.3$, $\Omega_\Lambda=0.7$,
$H_0 = 70\, \h$\,km\,s$^{-1}$\,Mpc$^{-1}$, $\h=1$, throughout this paper. When
reporting cluster masses, denoted as $M_\Delta$, we follow the convention of
defining masses in terms of spherical overdensities that are a factor $\Delta$
above the critical density $\rho_\mathrm{c}(z)$ of the Universe at redshift
$z$. Likewise $r_\Delta$ corresponds to the radius of the sphere containing
the mass $M_\Delta = \frac{4}{3} \pi r_\Delta^3\rho_\mathrm{c}(z)$. We use
standard notation for statistical distributions, i.e. the normal distribution
with mean $\vec{\mu}$ and covariance matrix $\mathsf{\Sigma}$ is written as
$\mathcal{N}(\vec{\mu}, \mathsf{\Sigma})$ and $\mathcal{U}(a, b)$ denotes the
uniform distribution on the interval $[a, b]$.

\section{Data}
\label{sec:data}

\subsection{Cluster sample}
\label{sec:cluster-selection}

The South Pole Telescope \citep[SPT,][]{carlstrom11} is a 10-m telescope
located at the Amundsen-Scott South Pole Station. From 2007 to 2011 SPT
observed a contiguous 2500~sq.~deg. region in three bands (95, 150, and
220~GHz) to a fiducial depth of 18~$\mu$K-arcmin in the 150~GHz band. Details
of the survey strategy and data processing are provided elsewhere
\citep{staniszewski09,vanderlinde10,williamson11}. Galaxy clusters in the
survey were detected via their thermal SZE. The full cluster catalog of the
SPT-SZ survey was published in \citet{bleem15}. In the SPT-SZ survey 677
galaxy clusters were detected above signal to noise $\xi>4.5$ and 516 were
confirmed by optical and near-infraread imaging \citep{bleem15}. Of these, 415
were first identified by SPT, and 109 have been spectroscopically confirmed
\citep{ruel14, bayliss16}. The median mass of this sample is
$\mfive \approx 3 \times 10^{14}\,\msun$ with a median redshift of $0.55$ and
with the maximum above $1.4$ \citep{bleem15}. The selection function of the
survey is well understood and almost flat in mass at $z > 0.25$ with a
slightly higher sensitivity to lower mass systems at higher redshifts.

\begin{figure}
  \includegraphics[width=\columnwidth]{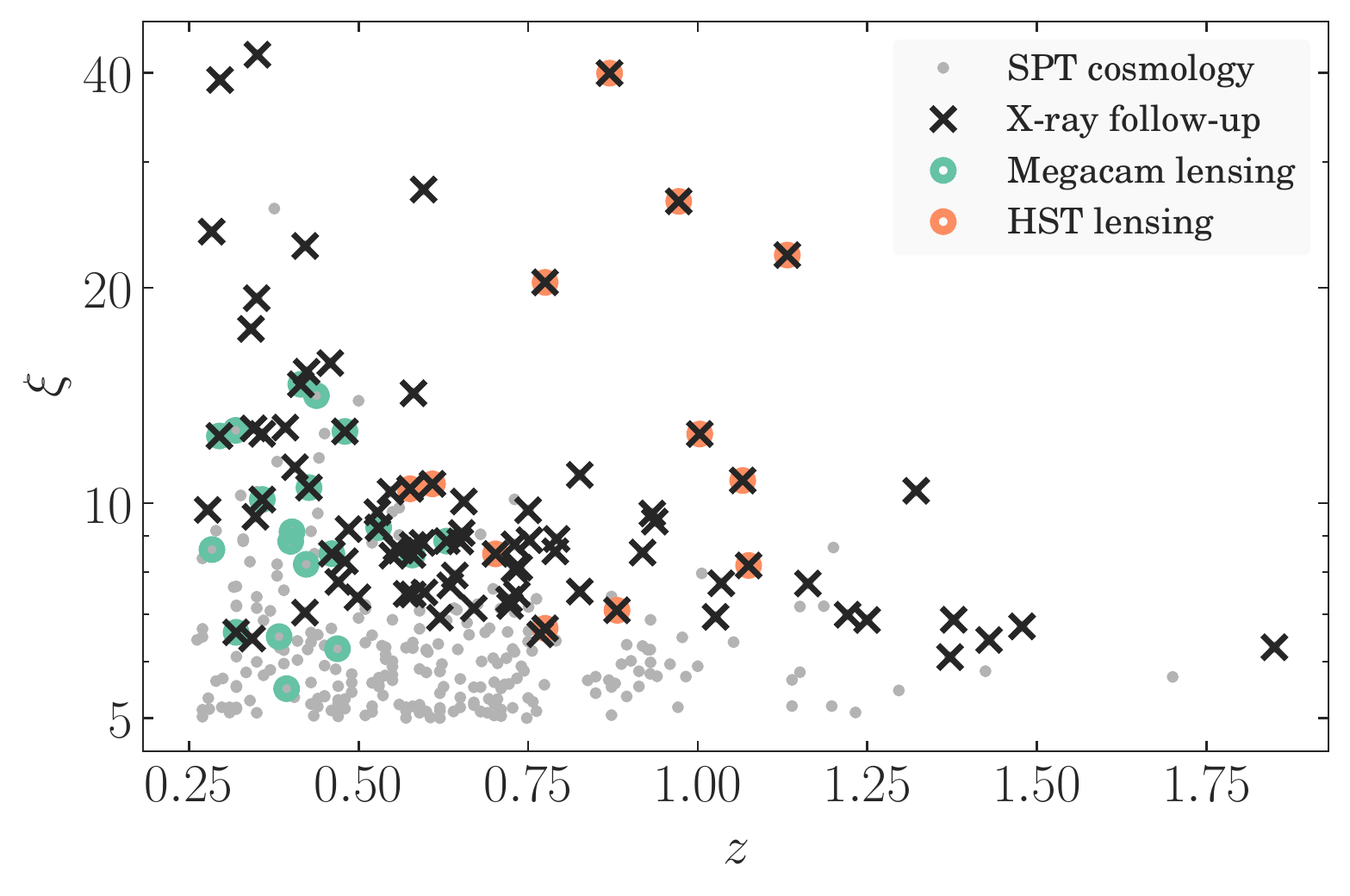}
  \caption{Overview of the SPT cosmology cluster sample, its coverage by X-ray
    data employed in this study, and the two weak lensing subsamples used in
    the scaling relation analysis in Sect.~\ref{sec:scaling-relations}. The
    axes show detection significance $\xi$ plotted against cluster redshift
    $z$.}
  \label{fig:sample_overview}
\end{figure}

\begin{table*}
  \begin{threeparttable}
  \caption{SPT clusters with lensing data used in this paper. Clusters
    observed with HST are imported from \citet{schrabback18} and are used in
    the scaling relations analysis (see Sects.~\ref{sec:scaling-relations} and
    \ref{sec:results}) only in combination with the Megacam cluster sample.}
  \begin{tabular}{lrrrrlc}\hline\hline
    Cluster & $\alpha$ (J2000.0) & $\delta$ (J2000.0) & $z$ & $\xi$ & Telescope
    & \textit{Chandra} data \\\hline
    SPT-CL\,J0000$-$5748 & 00:00:59.98 & $-57$:48:23.0 & 0.702 &  8.49 & HST &
                                                                               \checkmark
    \\
    SPT-CL\,J0102$-$4915 & 01:02:55.06 & $-49$:15:39.6 & 0.870 & 39.91 & HST &
                                                                               \checkmark \\
    SPT-CL\,J0234$-$5831 & 02:34:42.87 & $-58$:31:17.1 & 0.415 & 14.66 &
                                                                         Megacam & \checkmark \\  
    SPT-CL\,J0240$-$5946 & 02:40:38.54 & $-$59:46:10.9 & 0.400 & 8.84 &
                                                                        Megacam
    & \\
    SPT-CL\,J0254$-$5857 & 02:54:17.50 & $-$58:57:09.3 & 0.438 & 14.13 &
                                                                         Megacam 
    & \\
    SPT-CL\,J0307$-$6225 & 03:07:20.08 & $-$62:25:57.8 & 0.579 & 8.46 &
                                                                        Megacam
    & \checkmark \\
    SPT-CL\,J0317$-$5935 & 03:17:17.18 & $-$59:35:06.5 & 0.469 &  6.26 &
                                                                         Megacam
    & \\
    SPT-CL\,J0346$-$5439 & 03:46:53.93 & $-$54:39:01.9 & 0.530 & 9.25 &
                                                                        Megacam
    & \checkmark \\
    SPT-CL\,J0348$-$4515 & 03:48:17.70 & $-$45:15:03.5 & 0.358 & 10.12 &
                                                                         Megacam
    & \checkmark \\
    SPT-CL\,J0426$-$5455 & 04:26:04.78 & $-$54:55:10.8 & 0.642 & 8.85
                                                                    &
                                                                      Megacam
    & \checkmark \\
    SPT-CL\,J0509$-$5342 & 05:09:20.97 & $-$53:42:19.2 & 0.461 & 8.50 &
                                                                        Megacam
    & \checkmark \\
    SPT-CL\,J0516$-$5430 & 05:16:36.31 & $-$54:30:39.0 & 0.295 & 12.41 &
                                                                         Megacam
    & \checkmark \\
    SPT-CL\,J0533$-$5005 & 05:33:36.22 & $-50$:05:24.4 & 0.881 & 7.08 & HST &
                                                                              \checkmark
    \\
    SPT-CL\,J0546$-$5345 & 05:46:36.60 & $-53$:45:45.0 & 1.066 & 10.76 & HST &
                                                                               \checkmark
    \\ 
    SPT-CL\,J0551$-$5709 & 05:51:36.99 & $-$57:09:20.4 & 0.423 & 8.21 &
                                                                        Megacam
    & \checkmark\tnote{1} \\
    SPT-CL\,J0559$-$5249 & 05:59:42.02 & $-52$:49:33.6 & 0.609 & 10.64 & HST &
                                                                               \checkmark
    \\
    SPT-CL\,J0615$-$5746 & 06:15:51.60 & $-$57:46:34.7 & 0.972 & 26.42 & HST &
                                                                               \checkmark \\
    SPT-CL\,J2022$-$6323 & 20:22:06.25 & $-$63:23:56.1 & 0.383 & 6.51 &
                                                                        Megacam
    & \\
    SPT-CL\,J2030$-$5638 & 20:30:48.87 & $-$56:38:10.2 & 0.394 & 5.50 &
                                                                        Megacam
    & \\
    SPT-CL\,J2032$-$5627 & 20:32:19.37 & $-$56:27:28.9 & 0.284 & 8.61 &
                                                                        Megacam
    & \\
    SPT-CL\,J2040$-$5725 & 20:40:13.75 & $-$57:25:46.2 & 0.930 &  6.24 & HST &
                                                                               \checkmark
    \\ 
    SPT-CL\,J2106$-$5844 & 21:06:04.94 & $-$58:44:42.4 & 1.132 & 22.22 & HST &
                                                                               \checkmark
    \\ 
    SPT-CL\,J2135$-$5726 & 21:35:39.92 & $-$57:24:32.7 & 0.427 & 10.51 &
                                                                         Megacam
    & \checkmark \\
    SPT-CL\,J2138$-$6008 & 21:38:01.26 & $-$60:08:00.0 & 0.319 & 12.64 &
                                                                         Megacam
    & \\
    SPT-CL\,J2145$-$5644 & 21:45:52.37 & $-$56:44:51.2 & 0.480 & 12.60 &
                                                                         Megacam
    & \checkmark \\
    SPT-CL\,J2331$-$5051 & 23:31:50.59 & $-$50:51:50.0 & 0.576 & 10.47 & HST &
                                                                               \checkmark
    \\ 
    SPT-CL\,J2332$-$5358 & 23:32:25.37 & $-$53:58:03.1 & 0.402 & 9.12 &
                                                                        Megacam
    & \\
    SPT-CL\,J2337$-$5942 & 23:37:24.55 & $-$59:42:17.6 & 0.775 & 20.35 & HST &
                                                                               \checkmark \\
    SPT-CL\,J2341$-$5119 & 23:41:11.78 & $-$51:19:41.2 & 1.003 & 12.49 & HST &
                                                                               \checkmark \\
    SPT-CL\,J2342$-$5411 & 23:42:45:41 & $-$54:11:08.2 & 1.075 & 8.18 & HST &
                                                                              \checkmark \\
    SPT-CL\,J2355$-$5055 & 23:55:47.95 & $-$50:55:19.1 & 0.320 & 6.60 &
                                                                        Megacam
    & \checkmark \\
    SPT-CL\,J2359$-$5009 & 23:59:41.52 & $-$50:09:53.6 & 0.775 & 6.68 & HST &
                                                                              \checkmark \\
    \hline 
  \end{tabular}
  \begin{tablenotes}
    \item [1] \textit{Chandra} data excluded from the analysis. See Sect.~\ref{sec:x-ray-data}.
  \end{tablenotes}
  \label{tab:clusters}
\end{threeparttable}


\end{table*}

Cosmological constraints have been presented in \citet{dehaan16} based on the
``cosmology subset'' of the entire SPT-SZ cluster sample with redshift
$z > 0.25$ and detection significance $\xi > 5$. This $\xi > 5$ threshold
corresponds to a sample with 95\% purity from SZE selection alone. The mass
calibration employed in that analysis adopted information from the cluster
mass function together with information from X-ray observable
$\yx = \mgas T_\mathrm{X}$ available for 82 systems. The \yx--mass relation
calibration was informed from earlier weak lensing analyses of different
cluster samples \citep{vikhlinin09b,hoekstra15,applegate14}. We limit the
analysis in this paper to this cosmology subset.

We obtained pointed follow-up observations of 19 clusters in the redshift
range $0.29 \leq z \leq 0.61$ with the Megacam imager \citep{mcleod06} at the
Magellan Clay telescope. In the following we first describe these data and
their analysis before combining it with space-based HST weak-lensing follow-up
data of 13 SPT-SZ clusters in the redshift range $0.576 \leq z \leq
1.132$. \citep{schrabback18}.

\subsection{Optical data}
\label{sec:optical-data}

Our sample of 19 SPT clusters was observed with Megacam at the 6.5-m Magellan
Clay telescope. This sample includes 5 galaxy cluster observations previously
presented by \citet{high12}. This previous work also describes the observing
strategy, data reduction, and photometric and astrometric calibration in
detail. We briefly summarise the observing strategy for the remaining 14
clusters. These were observed in November 2011 through \gp, \rp, and \ip\
filters, for total exposure times of 1200~s, 1800~s, and 2400~s,
respectively. In \gp\ and \rp\ bands a three point diagonal dither pattern,
which covers the chip gaps, was used, while a five point linear dither pattern
was utilised for the \ip\ band exposures. As an exception from this strategy,
SPT-CL\,J0240$-$5946 was observed in 4 \rp\ band exposures.

Care was taken to observe the \rp\ band images, which are used to generate the
shear catalogues, in the most stable and best seeing conditions. Seeing values
for all \rp\ band images are given in Table~\ref{tab:seeing-distribution}. The
median seeing of our exposures is $0\farcs79$, the minimum and maximum values
are $0\farcs54$ and $1\farcs11$, respectively. The clusters observed with
Megacam were generally the most significant SPT cluster detections that were
known and visible at the time of the observing runs. An attempt was made to
observe higher redshift clusters during better seeing conditions.

As in \citet{high12}, a stellar locus regression code \citep{high09} and
cross-matching with the 2MASS catalogue \citep{skrutskie06} is employed in the
photometric calibration of our data. The resulting uncertainties on the
absolute photometric calibration and the colour measurements are
$\sim 0.05$\,mag and $0.03$\,mag, respectively.

\begin{table*}
  \begin{threeparttable}
  \centering
  \caption{PSF FWHM of individual cluster $r$-band exposures and $5\sigma$ limiting
    magnitude in a $2\arcsec$ diameter aperture of the $r$-band coadded
    image.}
  \begin{tabular}{lrrrrrr}\hline\hline
    Cluster & Exp. 1 & Exp. 2 & Exp. 3 & Exp. 4 & Avg. & lim. mag.\\\hline
    SPT-CL\,J0234$-$5831 & 0\farcs70 & 0\farcs81 & 0\farcs74 & --- & 0\farcs75 & 25.0 \\
    SPT-CL\,J0240$-$5946 & 0\farcs66 & 0\farcs74 & 
        0\farcs66 & 0\farcs73 & 0\farcs78 & 25.0 \\ 
    SPT-CL\,J0254$-$5857 & 0\farcs90 & 0\farcs87 & 0\farcs89 & --- & 0\farcs89 & 25.0 \\
    SPT-CL\,J0307$-$6225 & 0\farcs55 & 0\farcs59 & 0\farcs65 & --- & 0\farcs60 & 24.9 \\
    SPT-CL\,J0317$-$5935 & 0\farcs75 & 0\farcs73 & 0\farcs79 & --- & 0\farcs76 & 25.0 \\
    SPT-CL\,J0346$-$5439 & 0\farcs71 & 0\farcs73 & 0\farcs72 & --- & 0\farcs72 & 25.1 \\
    SPT-CL\,J0348$-$4515 & 0\farcs54 & 0\farcs54 & 0\farcs54 & --- & 0\farcs54 & 25.2 \\
    SPT-CL\,J0426$-$5455 & 0\farcs67 & 0\farcs59 & 0\farcs58 & --- & 0\farcs61 & 25.0 \\
    SPT-CL\,J0509$-$5342 & 0\farcs84 & 0\farcs79 & 0\farcs80 & --- & 0\farcs81 & 25.0 \\
    SPT-CL\,J0516$-$5430 & 0\farcs69 & 0\farcs69 & ---       & --- & 0\farcs69 & 24.8 \\
    SPT-CL\,J0551$-$5709 & 0\farcs79 & 0\farcs90 & 0\farcs85 & --- & 0\farcs85 & 25.0 \\
    SPT-CL\,J2022$-$6323 & 0\farcs88 & 0\farcs89 & 0\farcs97 & --- & 0\farcs91 & 25.1 \\
    SPT-CL\,J2030$-$5638 & 0\farcs87 & 0\farcs84 & 0\farcs80 & --- & 0\farcs83 & 25.1 \\
    SPT-CL\,J2032$-$5627 & 0\farcs92 & 0\farcs89 & 0\farcs79 & --- & 0\farcs86 & 24.8 \\
    SPT-CL\,J2135$-$5726 & 0\farcs88 & 0\farcs81 & 1\farcs00 & --- & 0\farcs90 & 24.7 \\
    SPT-CL\,J2138$-$6008 & 0\farcs90 & 1\farcs11 & 1\farcs02 & --- & 1\farcs01 & 24.5 \\
    SPT-CL\,J2145$-$5644 & 0\farcs80 & 0\farcs81 & 0\farcs82 & --- & 0\farcs82 & 25.0 \\
    SPT-CL\,J2332$-$5358 & 0\farcs80 & 0\farcs73 & 0\farcs73 & --- & 0\farcs75 & 25.1 \\
    SPT-CL\,J2355$-$5055 & 0\farcs66 & 0\farcs75 & 0\farcs76 & --- & 0\farcs73 & 25.0 \\\hline
  \end{tabular}
  \label{tab:seeing-distribution}
\end{threeparttable}

\end{table*}

\subsection{X-ray data}
\label{sec:x-ray-data}

The X-ray data in this work consist of 89 galaxy clusters observed with the
\textit{Chandra} satellite and is mostly identical to the data described in
\citet{dehaan16}. The reduction and analysis of these data is described in
detail in \citet{mcdonald13}. Changes in the data since this earlier SPT
publication include the addition of eight new clusters at $z > 1$
\citep{mcdonald17}, none of which currently have weak lensing data, and the
omission of SPT-CL\,J0551$-$5709. The latter cluster is part of our Megacam
sample. However, after the observations were obtained it was realized that
this cluster is indeed a projection of two clusters at different redshifts
\citep{andersson11}, the SPT selected cluster at $z=0.423$ and the local
cluster Abell~S0552 with a redshift of $z=0.09$ inferred from the cluster
red-sequence \citep{high12}. We thus exclude this cluster from the X-ray
analysis but not the weak lensing analysis, where the inclusion of such
projections is correctly accounted for (see
Sects.~\ref{sec:calibration-nfw-fits} and~\ref{sec:weak-lensing_scatter}).

Figure~\ref{fig:sample_overview} gives an overview of the different subsamples
in this study and their (partial) overlap. All 13 clusters with HST weak
lensing data \citepalias{schrabback18} have X-ray data, while this is the case
for only 10 out of the 19 clusters observed with Megacam after the exclusion of
SPT-CL\,J0551$-$5709. See also Table~\ref{tab:clusters} where all clusters
with lensing data are listed.

\section{Weak lensing analysis}
\label{sec:weak-lensing-data}

Weak gravitational lensing by massive foreground structures such as galaxy
clusters \citep[see][for a review of cluster lensing studies]{hoekstra13}
changes the observed ellipticities of background galaxies and imprints a
coherent shear pattern around the cluster centre. The azimuthally averaged
tangential shear at a distance $r$ from the cluster centre
\begin{equation}
  \label{eq:1}
  \gamma_\mathrm{t}(r) = \frac{\langle\Sigma(<r)\rangle -
    \Sigma(r)}{\Sigma_\mathrm{crit}} \;,
\end{equation}
depends on the mean surface mass density $\langle\Sigma(<r)\rangle$
inside and the surface mass density at this radius. This differential
surface mass density profile is scaled by the critical surface mass
density
\begin{equation}
  \label{eq:2}
  \Sigma_\mathrm{crit} = \frac{c^2}{4\upi G}\frac{1}{\beta D_\mathrm{l}}\;,
\end{equation}
where $c$ is the speed of light, $G$ is the gravitational constant,
$\beta = D_\mathrm{ls}/D_\mathrm{s}$ is the lensing efficiency and the $D_i$
are angular diameter distances, where `l' denotes the lens and `s' the source
galaxy.

The observable quantity is not the shear but the reduced shear
\begin{equation}
  \label{eq:3}
  g = \frac{\gamma}{1-\kappa}\;,
\end{equation}
where $\kappa = \Sigma/\Sigma_\mathrm{crit}$ is the dimensionless
surface mass density. A galaxy of intrinsic complex ellipticity
$\varepsilon^{(s)}$ is sheared by the reduced gravitational shear $g$
to have an observed (image) ellipticity \citep{seitz97}
\begin{equation}
  \label{eq:4}
  \varepsilon = \frac{\varepsilon^{(s)} + g}{1 + g^* \varepsilon^{(s)}}
  \approx
  \varepsilon^{(s)} + g \qquad g \ll 1 \;,
\end{equation}
so that, because $\langle \varepsilon^{s} \rangle = 0$, the expectation value
of $\varepsilon$ is $g$. 

We average the reduced shear over an ensemble of galaxies at different
redshifts. Strictly speaking redshifts for all background galaxies would be
required for the correct weighting with the geometric lensing efficiency
$\beta$. In the absence of such information, however, the average reduced
shear can be corrected to first order using \citep{seitz97}
\begin{equation}
  \label{eq:5}
  \frac{\langle g_\mathrm{cor} \rangle}{\langle g_\mathrm{true}
    \rangle} = 
  1 + \left(\frac{\mbetasq}{\mbeta^2} - 1\right) \kappa\,.
\end{equation}
The averages \mbeta\ and \mbetasq\ of the distribution of lensing efficiencies
can be computed from the redshift distribution of lensed galaxies (see
Sect.~\ref{sec:backgr-galaxy-select}).

\subsection{Shear catalogue creation}
\label{sec:shear-catal-creat}
Our shear analysis is based on the pipeline developed
for the Canadian Cluster Comparison Project \citep[CCCP;][]{hoekstra12}. In
this section we briefly review the main steps, but we refer the interested
reader to the more detailed discussion in \citet{hoekstra07} and in particular
the updates discussed in \citet{hoekstra15}, which used image simulations to
calibrate the bias in the algorithm to an accuracy of $1\text{--}2\%$. 

The observed galaxy shapes are biased, because of smearing by the point-spread
function (PSF): the seeing makes the galaxies appear rounder, whereas PSF
anisotropy will lead to coherent alignments in the observed shapes. Noise in
the images leads to additional biases \citep[e.g.][]{viola14}. To obtain
accurate cluster masses it is essential that the shape measurement algorithm
is able to correct for these sources of bias.

The shape measurement algorithm we use is based on the one introduced by
\citet{kaiser95b} and \citet{luppino97} with modifications described in
\citet{hoekstra98} and \citet{hoekstra00}. It uses the observed moments of the
surface brightness distribution to correct for the PSF. However, as shown in
\citet{hoekstra15} the measurements are still biased, predominantly because of
noise. These biases can be calibrated using image simulations. Because our
data cover a similar range in signal-to-noise ratio and seeing, we adopted the
correction parameters found by \citet{hoekstra15}.

Similar to what was done in \citet{hoekstra12}, we analyse each of the Megacam
exposures and combine the measurements in the catalogue stage, to avoid the
complex PSF pattern that would otherwise arise. We use \textsc{SExtractor}
\citep{bertin1996} to detect objects in the images and select objects with no
flags raised. We use the observed half-light radius to define the width of the
Gaussian weight function to measure the quadrupole (and higher) moments of the
surface brightness distribution of an object.

The next step is to find an adequate model to describe the spatial variation
of the PSF (both size and shape) as a function of the width of the weight
function used to analyse the galaxies \citep[see][for details]{hoekstra98}. To
quantify the properties of the PSF, we select a sample of bright, but
unsaturated, stars based on their half-light radius and shape. The number of
available stars varies from field to field and chip to chip with a median of
16 stars per chip and 519 stars per field. As shown in
Figure~\ref{fig:psf-diagnostics-2031_2} the PSF is anisotropic, and in many
cases it shows a coherent tangential pattern around the central parts of the
field-of-view. Such a pattern mimics the expected cluster lensing signal
(although that should decline with radius, rather than increase as is the case
for the PSF anisotropy). Therefore we have to take special care to model the
PSF \citep[also see][]{high12}.

To capture the dominant PSF pattern we fit a tangential pattern around the
centre of the focal plane, with a radial dependence that is a polynomial in
radius $r$ up to order 4, where the order was chosen based on a visual
inspection of the residuals. We also fit for a slope as a function of $x$ and
$y$. This model is fit to the full field-of-view. Inspection of the residuals
showed coherent variations on more or less the chip scale. We therefore also
fit a first order polynomial chip-by-chip in $x$ and $y$ to the residuals. The
resulting model is used to correct for the PSF. The tests of the performance
of the PSF model, described in more detail in \S\ref{sec:shear-catal-syst},
show that the model is adequate. We select galaxies to be objects with sizes
larger than the PSF. Specifically, we require that the half-light radius
$r_\mathrm{h}$ exceeds the half-light radius of the largest star selected from
the stellar locus in a magnitude-$r_\mathrm{h}$ plot by a factor
of $1.05$. Following \citet{hoekstra15} we also remove galaxies with
half-light radii larger than 1~arcsec, because many of these are blended
objects, biasing the shape measurements. 

This procedure is carried out for each exposure and bad regions are
masked. The resulting catalogues (typically three per cluster) are then
combined, with the shape measurements for objects that appear more than once
averaged accordingly. The averaging takes into account the measurement
uncertainties, thus naturally giving more weight to the better seeing
data. This results in a single catalogue of galaxy shapes that is used to
determine the cluster masses.

\subsubsection{Shear catalogue systematic tests}
\label{sec:shear-catal-syst}

\begin{figure*}
  \includegraphics[width=\textwidth]{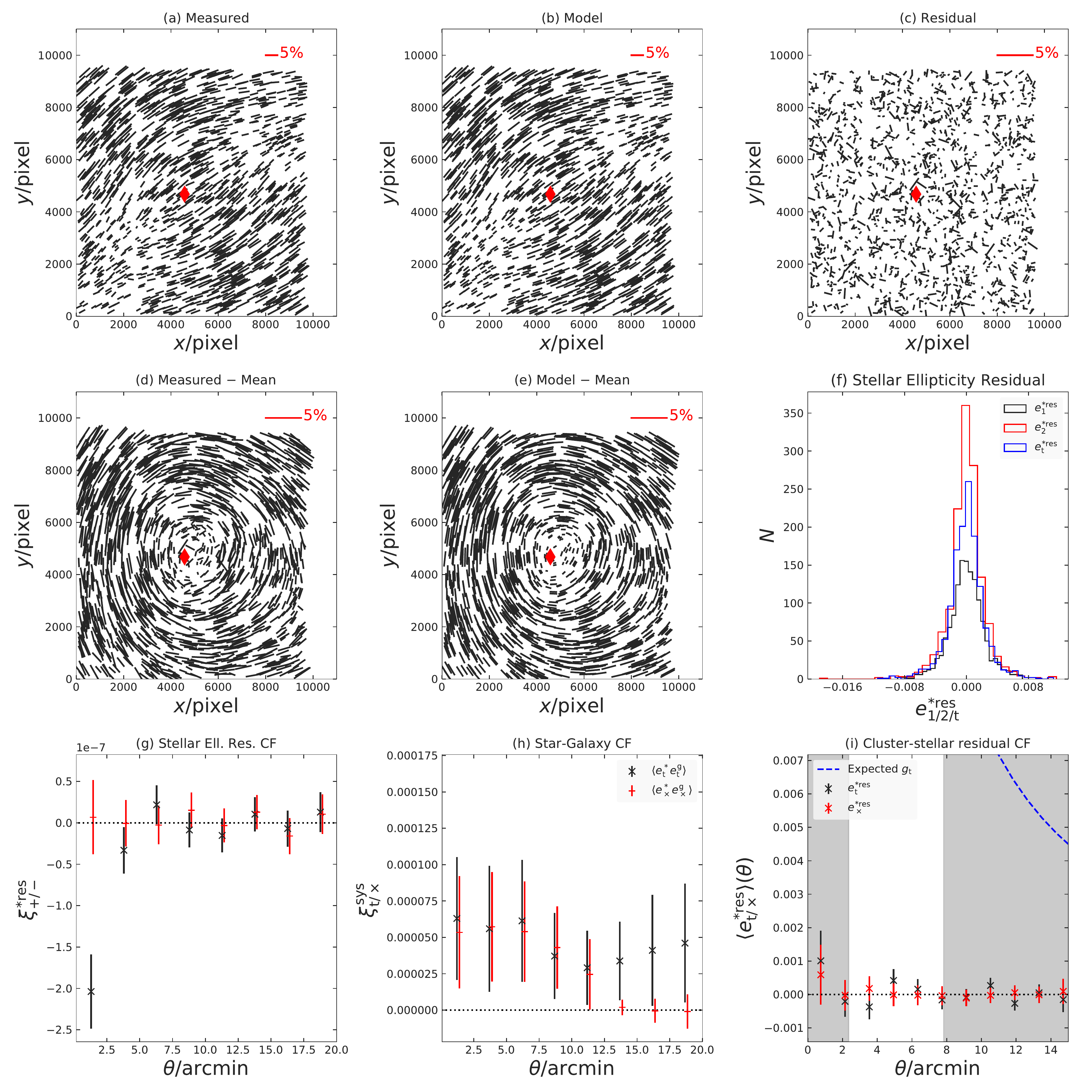}
  \caption{PSF correction diagnostic plots for the second exposure of
    SPT-CL\,J2030$-$5638. Red diamonds in panels (a)--(e) indicate the SZE
    derived cluster centre. \emph{Panel (a)}: Measured stellar ellipticity
    pattern; \emph{Panel (b)}: Model of the PSF pattern in panel
    (a); \emph{Panel (c)}: Residual between panels (a) and (b); \emph{Panel
      (d)}: Same as panel (a) with the mean ellipticity subtracted;
    \emph{Panel (e)}: Same as panel (b) with the mean ellipticity subtracted;
    \emph{Panel (f)}: Histogram of the stellar ellipticity residuals from
    panel (c) in the two Cartesian ellipticity components $e_{1/2}$ and the
    tangential ellipticity around the cluster centre $e_\mathrm{t}$;
    \emph{Panel (g)}: Ellipticity correlation functions $\xi_\pm$ of the
    stellar ellipticity residuals; \emph{Panel (h)}: Ellipticity correlation
    functions $\xi^\mathrm{sys}_{\mathrm{t}/\times}$ between measured stellar
    and corrected galaxy ellipticity; \emph{Panel (i)}: Ellipticity-position
    correlation function between stellar residual tangential and
    cross-component ellipticity, and the cluster centre. The blue dashed line
    shows a comparison to the expected tangential shear signal based on the
    SZE mass estimate of the cluster. The grey shaded regions are radii that
    are omitted in the NFW fitting procedure.}
  \label{fig:psf-diagnostics-2031_2}
\end{figure*}

We tested the PSF correction of the shear catalogues for a range of systematic
residuals to ascertain that these have negligible influence on our cluster
mass estimates. These are illustrated for the extreme case of the second
exposure of the cluster SPT-CL\,J2032$-$5627 in
Figure~\ref{fig:psf-diagnostics-2031_2}. This exposure shows a strong
tangential alignment of the stellar ellipticity pattern almost perfectly
centred on the cluster location. Left uncorrected, this PSF would lead to a
spurious cluster lensing signal and thus provides a good illustration of the
quality of our PSF correction. A randomly chosen, more representative example
of the Megacam PSF pattern and our diagnostic plots is shown in
Figure~\ref{fig:psf-diagnostics-0234_2} in
Appendix~\ref{sec:psf-residual-shear}. Because a constant shear will average
out in an azimuthal average around the cluster position we show the PSF
without its mean value across the field-of-view (FOV) in panel (d) and the
corresponding PSF model components in panel (e).

As a first diagnostic we examined the distribution of the residuals of
stellar ellipticity between the observed stellar ellipticity and the
smooth model describing the spatial variation of the PSF across the
focal plane. We verified that the mean of both Cartesian
ellipticity components as well as the tangential ellipticity residuals
with respect to the cluster centre are statistically compatible with
zero. A histogram showing the binned distribution of these three
ellipticity components is shown in panel (f) of
Figure~\ref{fig:psf-diagnostics-2031_2}. All exposures of all fields
pass this basic test.

Next we computed the correlation functions
\begin{equation}
  \label{eq:6}
  \xi_\pm = \langle \et \et \rangle \pm \langle \ex \ex \rangle
\end{equation}
for the residuals of stellar ellipticity, where the tangential and
cross-components are defined with respect to the line connecting the stars. We
used the tree code
\textsc{athena}\footnote{\url{http://www.cosmostat.org/software/athena/}} to
compute this and all other correlation functions. If the PSF model faithfully
represents the spatial variations of the actual PSF no correlation should
remain \citep{hoekstra04}. We find that this is generally the case for almost
all fields, except for the smallest bin of $\xi_+$, which is often slightly
negative, as seen in panel (g) of
Figure~\ref{fig:psf-diagnostics-2031_2}. This indicates some over-fitting of
the PSF on these scales, but the values of these bins are 2--3 orders of
magnitude below that of the cluster induced gravitational shears on the
angular scales of interest (cf. panel (i) in the same Figure). Moreover, this
overfitting happens on individual exposures and may not be coherent across all
three exposures, in which case it should approximately average out and its
real impact decreased even further.

A common diagnostic in cosmic shear analyses for the absence of
leakage from PSF ellipticity to the shear catalogue is the star-galaxy
correlation function \citep{bacon03}
\begin{equation}
  \label{eq:7}
  \xi^\mathrm{sys}_{\mathrm{t}/\times} = \frac{\langle e_i^* \gamma_i \rangle^2}{\langle
    e_i^*e_i^*\rangle} \;,
\end{equation}
which can be computed for the tangential ($i=\mathrm{t}$) and cross-components
($i=\times$) of the uncorrected stellar ellipticities $e^*$ and the observed
shear of the galaxies $\gamma$. For random fields, there should not be any
correlation between the stellar ellipticity and the measured shear. However,
observations centred on galaxy clusters are not random fields. The cluster
centre is a special location around which we expect a tangential alignment of
galaxies. The absence of a significant star-galaxy correlation thus indicates
that no PSF leakage into the shear catalogue occured; its presence, however,
would not be a cause for concern. Taking the covariance between spatial
correlation function bins into account, we find no significant deviations of
$\xi^\mathrm{sys}_{\mathrm{t}/\times}$ from zero.

Finally, in panel (i) of Figure~\ref{fig:psf-diagnostics-2031_2} we show the
tangential and cross-components of the residual stellar ellipticity around the
cluster centre in radial bins. A non-zero tangential component would
immediately bias our cluster mass measurements, while a non-zero
cross-component would render the diagnostic power of radially binned
cross-shear used later worthless. We find that these ellipticity profiles are
all consistent with zero mass for all exposures and fields. The occasional
outlier bin is more than one order of magnitude below the expected shear
signal.

The bias correction parameters derived in \citet{hoekstra15} and discussed in
the previous section are for a circular PSF and as shown in their appendix, in
the presence of PSF anisotropy, the smear polarisability is somewhat
biased. We therefore artificially boosted the smear polarisability by $4\%$
for each object to correct for this bias. We find that the cluster masses
estimated from the boosted catalogues, which are used in our analysis, are on
average $1.1\%$ higher than in the uncorrected catalogues, but not
significantly so because the mass scatter between boosted and unbooseted
catalogues is 2.5\%.

\subsubsection{Blind analysis}
\label{sec:blinded-analysis}

Attempting to measure cluster masses with gravitational lensing when other
estimates of the cluster mass -- such as SZE measurements -- are already known
presents the danger of the experimenter being influenced by confirmation
bias. A number of procedures described in the following sections required
careful checking of their behaviour with respect to varying input
parameters. Any experimenter is faced with the challenge of deciding when the
results of such tests are of sufficient quality. It is imperative that the
metric of this decision does not make use of the actual mass measurement. If
it did we would be more likely to stop testing our procedures when the cluster
masses seem to agree with our expectations from SZE measurements than when
there is a discrepancy. To avoid such experimenter bias, the practice of
``blind analyses'' has found wide-spread acceptance in particle and nuclear
physics \citep{2005ARNPS..55..141K} and is being adopted in cosmology as
well \citep[e.g.][]{vonderlinden14,hildebrandt17,desyr1cosmology}.

The analysis presented herein has been blinded so that no comparisons between
the weak lensing and SZE derived masses were made, which otherwise would have
allowed premature inferences of the weak lensing--X-ray observable scaling
relation parameters. At the same time we aimed to retain shear profiles that
resemble those of massive clusters to test our analysis pipeline with the
actual but blinded data. To ensure this we adopt the following procedure to
blind the normalisation and scatter of the scaling relation. First, a random
number $0.80 \leq x_\mathrm{l} < 0.95$ is drawn from a uniform probability
distribution. Then for each cluster $i$ a second random number $f_i$ is drawn
from the interval $[x_\mathrm{l}, 1)$. The shear values of each cluster are
multiplied by $f_i$. We enforce $f_i < 1$ to avoid unphysical shears; at the
same time $f_i$ cannot be very small to not wipe out the lensing signal. The
intrinsic ellipticity dispersion used in the calculation of the lensing
weights (see Sect.~\ref{sec:nfw-profile-fits}) is not rescaled, i.e., the
relative weighting of galaxies in any given cluster field is not changed by
the blinding procedure.

\subsubsection{Changes after unblinding}
\label{sec:chang-after-unbl}

Although great care was taken to unblind the shear profiles only after the
analysis was finalised, we realized that we inadvertently did not apply the
multiplicative shear bias correction. This biased our masses low by much more
than the average blinding factor turned out to be. The analysis we
present in this paper has the multiplicative shear bias correction applied. We
stress that these correction factors were already computed at the time of
unblinding and they remained unchanged by all further analysis changes.

We took the opportunity of this one very large shift in the analysis after
unblinding, corresponding to a $\sim 20\%$ shift in mass, to make two small
adjustements at the same time:
\begin{enumerate}
\item We transitioned from the unboosted PSF correction catalogues to the
  boosted smear polarisability (see Sect.~\ref{sec:shear-catal-syst}).
\item At the time of unblinding the 2500 sq. deg. SPT-SZ catalogue
  \citep{bleem15} was not finalized and we used centroids, redshifts, and
  estimated $\mfive^\mathrm{SZ}$ from the catalogue of \citet{andersson11}. We
  afterwards updated our analysis to use the quantities from the final SPT-SZ
  catalogue.
\end{enumerate}
Both of these changes lead to shifts at the $\sim 1$\% level in the absolute
mass scale. 

We also made changes to the scaling relation analysis scheme for our X-ray
data. Theoretical considerations as well as tests against mocks revealed that
the analysis scheme used in previous SPT cluster analyses led to a bias of the
X-ray slope toward steeper values. The updated analysis method is described in
Sect.~\ref{sec:scal-rel-pipeline} and was shown to produce unbiased
results. We note that our constraints on the slope are dominated by the
informative prior applied (see Sect.~\ref{sec:choices-priors}), and that we
choose a pivot point in the scaling relation that essentially decouples the
slope from the amplitude. Therefore, our final results are not much affected
by this change.

Finally, while this manuscript was edited for submission, one of us realized
that the blinding scheme described in the previous Section only has a
$\sim 2\%$ scatter on the mean blinding factor, while during the creation of
this work we assumed it to be in the $10\text{--}15\%$ range. The mean
blinding factor determines how well the true MOR normalisation is hidden from
us and is more important than the cluster-to-cluster blinding, which is indeed
large in our method. Our erroneous assumption kept us effectively blind during
the analysis. However, now that this flaw has been revealed, we strongly
advocate against using this scheme and advise to use a blinding scheme that
first determines the mean blinding factor from a random variable with a large
variance. 

\subsection{Background galaxy selection and critical surface mass density}
\label{sec:backgr-galaxy-select}
The reduced shear $g$ measured in weak-lensing data is a dimensionless
quantity. To connect it to the physical mass scales of our galaxy clusters we
need to determine the redshift distribution of the background galaxies, which
enters in the critical surface mass density (eq.~\ref{eq:2}). The three
Megacam passbands in which we have data are not sufficient to estimate
photometric redshifts for galaxies in our catalogues.

\begin{figure}
  \includegraphics[width=\columnwidth]{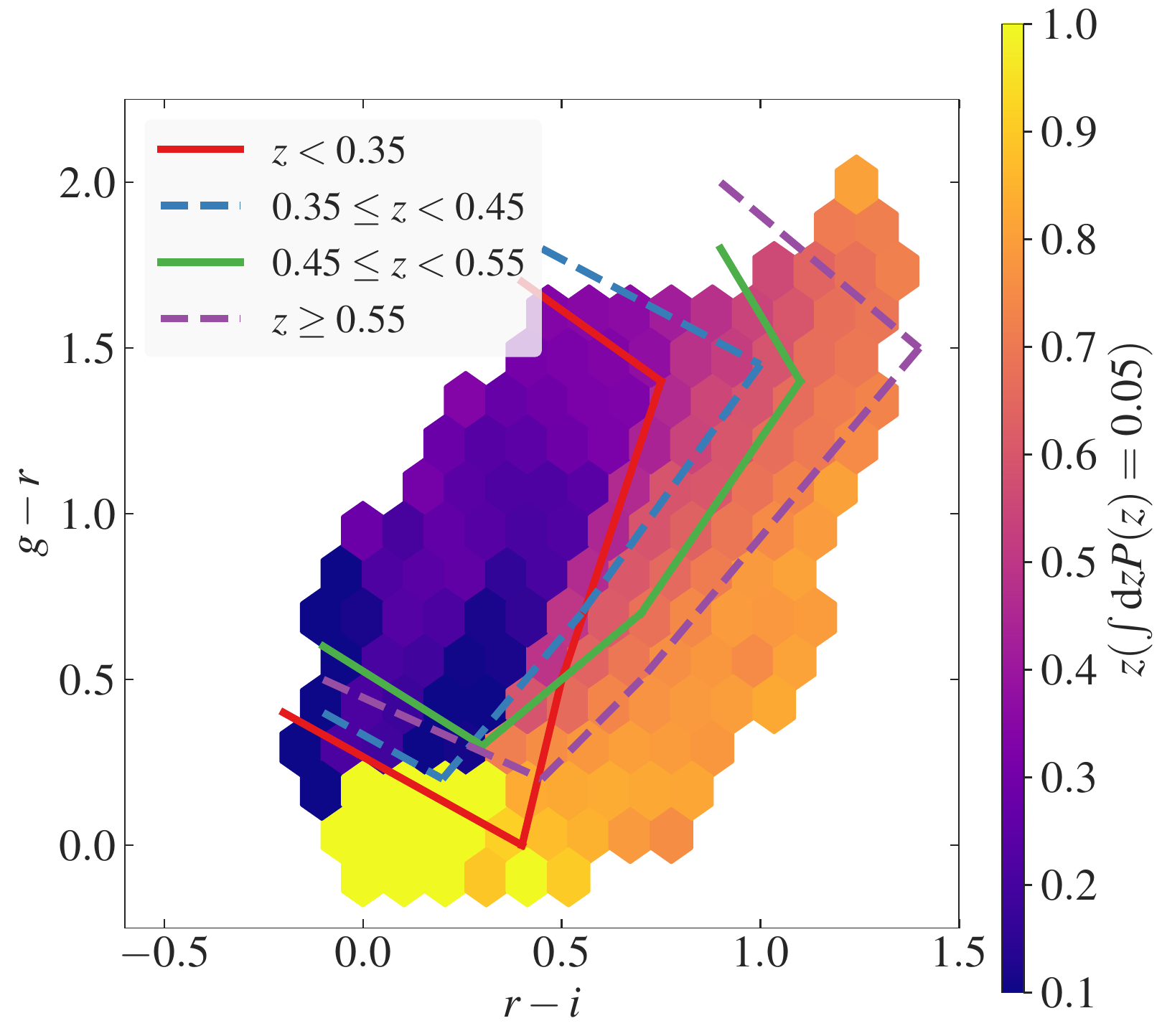}
  \caption{Colour cuts applied to the lensing catalogues to reject cluster and
    foreground galaxies, and redshift distribution properties of the COSMOS
    reference catalogue \citep{ilbert13}. Only galaxies to the top, right, and
    bottom of the indicated regions are kept for the lensing
    analysis. Different colours and line styles indicate the colour cuts
    applied to clusters at different redshifts. The colour coded map indicates
    the 5th per centile of the photometric redshift distribution of galaxies
    in the COSMOS catalogue at this particular point in colour-colour space. }
  \label{fig:color_cuts}
\end{figure}

We used redshift dependent colour cuts to reject likely foreground and cluster
galaxies. Rather than optimising these colour cuts for every cluster, we
divided the sample into four redshift slices. The polygons that define our
color cuts are illustrated in Fig.~\ref{fig:color_cuts}. These are based on
the color cuts defined in an earlier SPT weak-lensing study \citep{high12} and
were constructed in the same way. The density distribution of galaxies in the
CFHTLS Deep Field 3 \citep{coupon09} with $i<25$\,mag was plotted in
$(g-r, r-i)$ colour-colour space for (1) galaxies with photometric redshifts
$|z_\mathrm{phot} - z_\mathrm{l}| < 0.05$ (``non-sources'') and (2) for all
other galaxies (``sources''). Polygons were drawn by hands to reject the
majority of non-source galaxies. More sophisticated approaches to select only
background galaxies have been proposed, e.g. by \citet{okabe16} and
\citet{medezinski18a}, but the present scheme is sufficient for our purposes
and its efficacy is demonstrated by the background map in
Fig.~\ref{fig:color_cuts}. Additionally, we rejected all galaxies with
$i < 20.5$\,mag from the lensing catalogue because such bright galaxies are
very unlikely to be background galaxies.

We use an external catalogue with well calibrated photometric redshifts to
estimate the redshift distributions of the lensing catalogues. By applying the
same cuts we use for the shear catalogues to the reference catalogue and by
matching galaxy properties such as magnitude and size, we can draw photometric
samples from the reference catalogue. Their photometric redshifts can then be
used to determine the effective redshift distribution of our lensing
catalogues.

We used a version of the COSMOS30 photometric redshifts catalogue
\citep{ilbert13}, which makes use of additional near-infrared photometry
provided by the UltraVISTA survey \citep{mccracken2012}. We transformed the
magnitudes in the catalogue to the CFHT system, to which our Megacam data was
calibrated by using the colour terms from \citet{capak07}
\begin{equation}\label{eq:8}
\begin{split}
  g& = g^+ -0.084(g^+ - r^+) - 0.007\\
  r& = r^+ +0.019(g^+ - r^+) - 0.001\\
  i& = i^+ +0.018(g^+ - r^+) - 0.005 \;.
\end{split}
\end{equation}
This catalogue is complete to $i \gtrsim
24.5$\,mag. Consequently, this is the limit we must adopt when doing a
faint magnitude cut on the shear catalogues. We further impose the
following constraints on galaxies in the reference catalogue:
\begin{enumerate}
\item No flags set in $i$-band;
\item Full width at half maximum $\text{(FWHM)} > 2$\,px, to reject the
  stellar locus;
\item Unsaturated in $g$-, $r$-, and $i$-band;
\item Above $5\sigma$ detection in $g$-, $r$-, and $i$-band, to reject
  spurious objects;
\item Same colour cuts as for the lensing catalogue;
\item $z < 5$, to reject objects with unrealistic photo-$z$ estimates. 
\end{enumerate}
We emphasise that cuts (iii) and (iv) remove only objects from the COSMOS
catalogue that cannot be present in our lensing catalogues because they are
either rejected by the bright magnitude cut on the lensing catalogue or are
too faint to be detected in our Megacam data where we require detection in all
three passbands.

Galaxies in the shear catalogue have weights assigned to them (see
Section~\ref{sec:nfw-profile-fits}). These are taken into account in all
lensing derived quantities. Simply sampling from the reference catalogue such
that the samples reproduce the photometric properties of the shear catalogue
without taking the lensing weights into account could bias the computation of
$\langle\beta\rangle$ and $\langle\beta^2\rangle$. The lensing weight chiefly
depends on signal-to-noise ratio (SNR) and to a lesser degree on the object
size. We thus have to map lensing weights to the $\beta$ distribution of
COSMOS galaxies with the same magnitude-size distribution as in the shear
catalogue.

Our version of the COSMOS catalogue (P. Capak, priv. comm.) has a column with
the object FWHM on the $i$-band detection image, which has not been convolved
to homogenise the PSF across passbands. Assuming that atmospheric seeing
causes a simple Gaussian convolution, we added the size of convolution kernels
in quadrature to achieve the same seeing in the reference catalogue as the
field seeings in Table~\ref{tab:seeing-distribution}. This is almost always
possible because the average seeing in the COSMOS field is $0\farcs57$ and
thus less than the seeing in our fields, with the one exception of
SPT-CL\,J0348$-$4514, which has a seeing of $0\farcs54$. In this case the
COSMOS detection FWHM column was left unaltered.\footnote{We ignore the
  wavelength mismatch between our seeing values measured in the $r$-band and the
  FWHM of the COSMOS objects detected in the $i$-band. In standard seeing
  theory the difference in FWHM is only $\sim 4\%$.}

We developed an algorithm to infer (from the COSMOS catalogue) the expected
$\beta$ distribution for galaxies with the magnitude and size distribution of
objects in the cluster field shear catalogues, correctly applying the lensing
weights. This algorithm first constructs a joint probability distribution in
\ip--size--lensing weight space from the observed shear catalogue for each
cluster field. Then a random deviate from this distribution is drawn and the
closest match in \ip\ magnitude and size in the COSMOS catalogue is found and
the redshift of the matched COSMOS object is assigned to the random
deviate. In this respect the algorithm is similar to photo-$z$ methods based
on nearest-neighbour identification in multi-color space
\citep[e.g.,][]{lima08, cunha09b}, except that we require that galaxies in the
reference catalogue follow the same magnitude and size distribution whereas
those other methods only used color information. With a redshift (from COSMOS)
and a lensing weight (from the random deviate), we can now compute weighted
\mbeta\ and \mbetasq. In detail the algorithm works as follows:

\begin{description}
\item We construct a Gaussian kernel density estimator (KDE) of the density
  distribution in $i-\text{FWHM}-\text{weight}$ space from the shear
  catalogue. The number of lensing galaxies with weights below a
  characteristic value drops sharply. This discrete feature of the density
  distribution, as well as the sharp magnitude cut at $i=24.5$\,mag are not
  well represented by a smooth KDE. To avoid biases at the edges of the
  probability density distribution, we mirror the size and magnitude
  distributions at their extreme values. This ensures that we have smooth
  distributions, which can be well described by a Gaussian KDE.

  \item We then draw random samples from this KDE. Samples in the mirrored
  quadrants are flipped back into the original quadrant. For each
  random sample we identify the COSMOS galaxy that minimises the
  quantity
  \begin{align}
    d &= \left[
        \left(\frac{i_\mathrm{sample} - 
        i_\mathrm{COSMOS}}{\sigma_i}\right)^2 \right. \nonumber \\
      & \qquad \left. + \left(\frac{\text{FWHM}_\mathrm{sample} 
        - \text{FWHM}_\mathrm{COSMOS}}{\sigma_\mathrm{FWHM}}\right)^2
        \right]^{1/2}\;,
        \label{eq:9}
  \end{align}
  where the $\sigma_x$ with $x \in \{i, \mathrm{FWHM}\}$ are the
  standard deviations of the $i$-band and FWHM distributions in the
  shear catalogue. This sample galaxy is assigned the weight drawn from
  the KDE and $\beta$ and $\beta^2$ for this galaxy are computed. We
  verify that the samples drawn in this way from the reference catalogue
  are distributed consistently with the lensing catalogues by computing
  the Kolmogorov-Smirnov test for the marginal distributions in size
  and $i$-band magnitude.

\item The first two moments of the $\beta$ distribution are then
  computed as weighted average of $\beta$ and $\beta^2$ using the
  lensing weights. These values are reported in
  Table~\ref{tab:redshift-betas}. 
\end{description}

We tested the ability of this procedure to correctly reproduce input
distributions that are very different from the intrinsic COSMOS30
galaxy properties. We divided the COSMOS30 reference into two halves
and created mock catalogues from one of the halves. To create the mock
catalogues we subsampled from the first half such that the magnitude
distribution $P(i)$ follows the linear distribution
\begin{equation}
  \label{eq:10}
  P(i) = \frac{2(i - 20.5)}{(25.0 - 20.5)^2} \;,\qquad 20.5 \leq i \leq 25.0\;,
\end{equation}
the size distribution is log-normal
$\ln P(\text{FWHM})\sim \mathcal{N}(1, 0.0625)$, and the lensing
weights are distributed according to $P(w) \sim 20 - \exp(-w)$. Just
like the actual shear catalogues, these magnitude and weight
distributions have sharp cut-offs to test the unbiasedness of our
mirroring approach.

Following the construction of the KDE as described above, we resampled from
the second half of the reference catalogues and compared the estimated values
of \mbeta\ and \mbetasq\ to the known values of the mock catalogues. We find
that our resampling slightly underestimates the true values of \mbeta\ between
$0.3\%$ and $0.9\%$ as a function of redshift. At the median redshift of the
cluster sample the bias is $0.5\%$. The values of \mbetasq\ are biased low
between $0.3\%$ and $0.6\%$ with a bias of $0.5\%$ at $z=0.42$. This level of
bias is negligible for our analysis. It is caused by a slight oversampling of
bright galaxies with redshifts $z < \zl$.

We consider a number of effects contributing to uncertainties in our estimates of
\mbeta\ and \mbetasq. First, COSMOS is a small field and variations between the LSS in
COSMOS and the lines-of-sight to our galaxy clusters (``cosmic variance'') can lead to
biases. We computed \mbeta\ separately from all four CFHTLS fields and took the variance
between these fields as our estimate for the impact of cosmic
variance.\footnote{Although we include the effect of \mbetasq\ in our mass calibration,
  it is generally completely negligible for the radial ranges employed in this work. We
  therefore do not separately quote the small uncertainties on \mbetasq.} We find
$\sigma_{\mbeta, \mathrm{CV}} = 0.0082$. The CFHTLS photometric catalogues do not come
with object size information. Computing the variance among CFHTLS fields only rather
than also with the COSMOS fields allows us to isolate the impact of cosmic variance from
the influence of object size.

Second, even with the high quality of the photometric redshifts of the COSMOS30
catalogue, some biases may exist, particularly at the faint, high-redshift end. To
evaluate this, we matched our COSMOS catalogue with the 3D-HST catalogue
\citep{momcheva16}. This catalogue contains redshifts based on spectroscopic, grism, and
photometric redshift estimates. We limited our comparison to objects for which the
3D-HST catalogue lists either spectroscopic or grism redshifts to which the COSMOS
redshifts may be reliably compared. We have $1980$ objects of this type in common with
their catalogue. We first computed the additional uncertainty stemming from only
randomly sampling $1980$ objects from the COSMOS photo-$z$ catalogue. This is
$\sigma_{\mbeta, \text{sample}} = 0.0013$. We then recomputed \mbeta\ for all clusters
using only the $1980$ 3D-HST redshifts and find
$\langle(\mbeta_\mathrm{COSMOS} - \mbeta_\text{3D-HST}) / \mbeta_\mathrm{COSMOS}\rangle
= 0.6\%$, which is fully consistent with no redshift bias, up to the sampling
uncertainty, where the outer average runs over all clusters. This test is reliable as
long as any possible redshift bias in the COSMOS catalogue is not different for objects
with or withour spectroscopic or grism redshifts. At the present we have no indication
for such a type dependent bias but also cannot confidently rule out that hitherto
undiscovered biases in the COSMOS catalogue exist for faint objects.

Third, we also investigate the impact of the uncertainties of the photometric
calibration on our estimation of the lensing efficiency by shifting the
relative and absolute photometry within the systematic calibration
uncertainties. We find an additional uncertainty of $\sigma_{\mbeta,
  \mathrm{pc}} = 0.0018$.

Finally, a more recent version of the COSMOS photo-$z$ catalogue
\citep{laigle16} was published after we finalized our data vectors. We
verified that this catalogue gives consistent results for \mbeta\ and
\mbetasq\ with $\Delta\mbeta = 0.2\%$ and $\Delta\mbetasq = -0.2\%$ and treat
the difference between these catalogues as an additional source of
uncertainty, $\sigma_{\mbeta, \mathrm{NC}} = 0.002$. 

We add all four $\sigma_{\mbeta, i}$ in quadrature and arrive at a final
uncertainty of $\sigma_\mbeta = 0.0087$. Cluster mass scales with
$M \propto \gamma^{1 / \Gamma}$, where the exponent $\Gamma$ depends on the
cluster centric radius. For a wide range of radial fitting ranges
$\Gamma = 0.75$ \citep{melchior17} and hence the systematic uncertainty on
cluster mass due to uncertainty on the redshift distribution of background
galaxies is $1.2\%$. We confirmed this value by rescaling the tangential shear
by a factor of $1.0087$, fitting NFW profiles (see
Section~\ref{sec:nfw-profile-fits}) to the unscaled and scaled shear profiles,
and comparing the mass estimates.

\begin{figure}
  \includegraphics[width=\columnwidth]{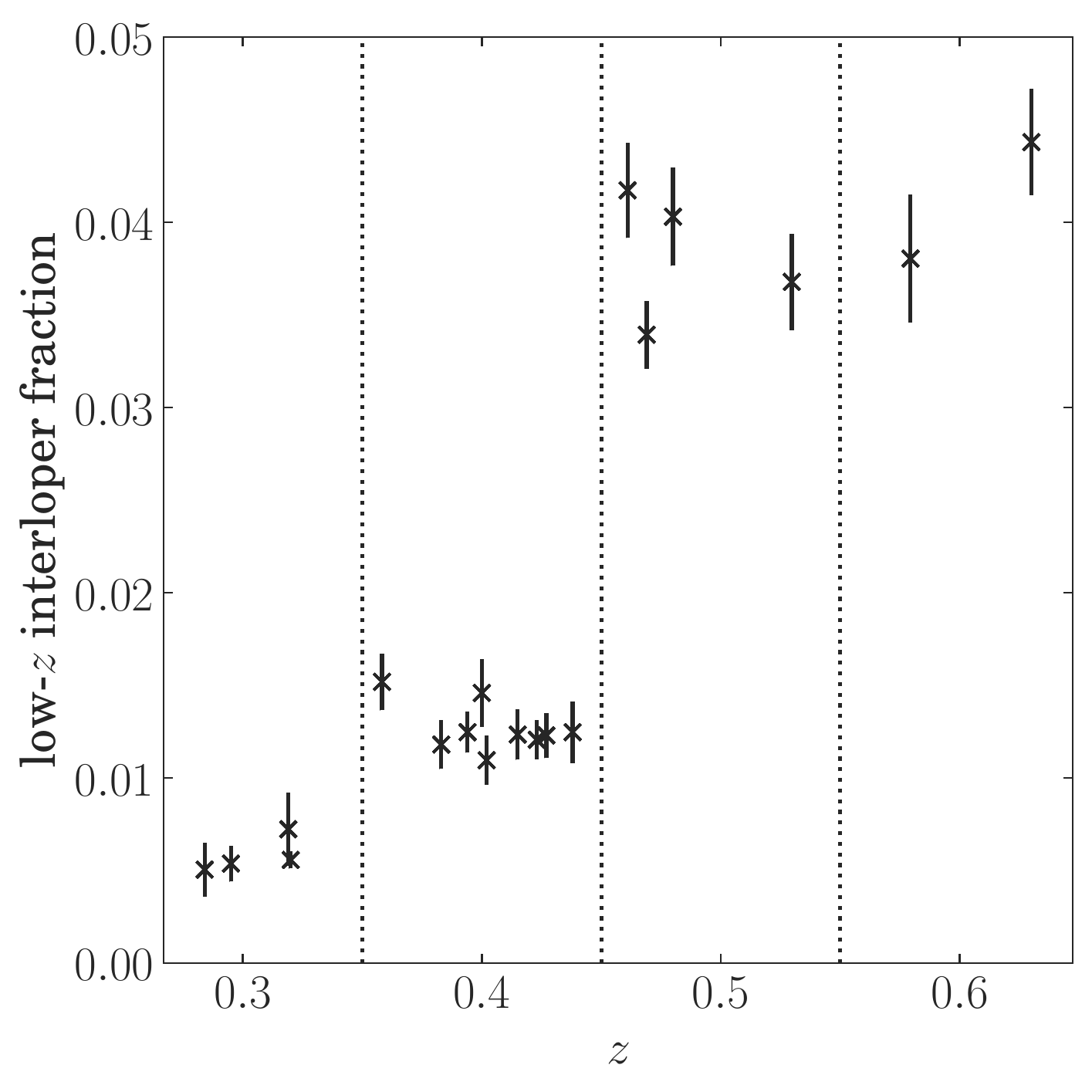}
  \caption{Fraction of low-redshift galaxies surviving our colour cuts as a
    function of redshift estimated by sampling from the reference
    catalogue. The error bars are the standard deviation of the mean number of
    low-$z$ interloper galaxies. The vertical dotted lines indicate the
    transitions from one colour cut to another as illustrated in
    Fig.~\ref{fig:color_cuts}.}
  \label{fig:contamination_fraction}
\end{figure}
Our sampling from the reference catalogue also enables us to estimate
the fraction of foreground galaxies surviving our colour cuts and
diluting the shear signal without biasing it. This is shown in
Fig.~\ref{fig:contamination_fraction}. The fraction of low-redshift
interlopers is below 2\% for clusters at redshift $z< 0.45$. At higher
redshifts it jumps to $\sim 5\%$. It is possible to optimise the colour
cuts to keep the low-$z$ contamination at $\sim 2\%$ also for the
$0.45 < z < 0.55$ redshift bin, but this optimisation happens at the
cost of an increased contamination of the shear catalogue by cluster
galaxies, as we discuss in detail in the next Section. 

\begin{table}\centering
  \caption{Cluster redshift, source galaxy lensing efficiency, and density after
    color cuts.}  
    \begin{tabular}{lrrrr}
\hline\hline
Cluster Name & $z_\mathrm{l}$  & $\langle \beta \rangle$ & $\langle \beta^2 \rangle$ & $n_\mathrm{gal}$ \\
 & & & & (arcmin$^{-2}$) \\
\hline
SPT-CL\,J0234$-$5831 & 0.41 & 0.48 & 0.25 & 12.1 \\
SPT-CL\,J0240$-$5946 & 0.40 & 0.50 & 0.27 & 12.3 \\
SPT-CL\,J0254$-$5857 & 0.44 & 0.46 & 0.23 & 11.1 \\
SPT-CL\,J0307$-$6225 & 0.58 & 0.40 & 0.18 & 7.9 \\
SPT-CL\,J0317$-$5935 & 0.47 & 0.46 & 0.23 & 9.2 \\
SPT-CL\,J0346$-$5439 & 0.53 & 0.40 & 0.18 & 13.1 \\
SPT-CL\,J0348$-$4515 & 0.36 & 0.56 & 0.32 & 12.1 \\
SPT-CL\,J0426$-$5455 & 0.63 & 0.35 & 0.14 & 8.9 \\
SPT-CL\,J0509$-$5342 & 0.46 & 0.46 & 0.23 & 11.7 \\
SPT-CL\,J0516$-$5430 & 0.29 & 0.60 & 0.37 & 9.3 \\
SPT-CL\,J0551$-$5709 & 0.42 & 0.48 & 0.24 & 8.4 \\
SPT-CL\,J2022$-$6323 & 0.38 & 0.51 & 0.28 & 7.4 \\
SPT-CL\,J2030$-$5638 & 0.39 & 0.50 & 0.27 & 9.0 \\
SPT-CL\,J2032$-$5627 & 0.28 & 0.60 & 0.37 & 8.4 \\
SPT-CL\,J2135$-$5726 & 0.43 & 0.47 & 0.24 & 9.7 \\
SPT-CL\,J2138$-$6008 & 0.32 & 0.54 & 0.31 & 4.0 \\
SPT-CL\,J2145$-$5644 & 0.48 & 0.44 & 0.21 & 9.9 \\
SPT-CL\,J2332$-$5358 & 0.40 & 0.51 & 0.27 & 11.5 \\
SPT-CL\,J2355$-$5055 & 0.32 & 0.57 & 0.34 & 10.1 \\
\hline
\end{tabular}

    \label{tab:redshift-betas}
\end{table}

\subsection{Cluster contamination correction}
\label{sec:clust-cont-corr}

Sampling from the reference catalogue -- as in the preceding section -- allows
us to estimate the background properties and foreground contamination of the
shear catalogues. However, it does not allow us to estimate the contamination
by cluster galaxies remaining after the colour cuts in the shear catalogues,
because cluster galaxies are a very significant overdensity in redshift space
not present in the reference catalogue. Contamination of the shear catalogues
by cluster galaxies dilutes the shear signal as these galaxies are not lensed
and show no specific alignment \citep[e.g.][]{sifon15}. Thus, they should be
counted as contributing $\beta = 0$ in the estimation of the lensing
efficiency.

We implement and test two different methods to estimate the contamination
fraction in our cluster sample. All methods looking at radial variations of a
population must carefully keep track of areas not available for observations
of that population \citep{simet15, hoekstra15}. We therefore use the mask
files created for the magnification study of \citet{chiu16}, where details on
their generation are provided. Briefly, regions covered by extended bright
objects are automatically masked by \textsc{SExtractor} while satellite trails
and diffraction spikes are manually masked. We determine the cluster
contamination fraction in radial bins and correct the bin area for masked
pixels in both methods, keeping track of the area covered by bright galaxies
not already included in the pixel masks. An increased incidence of blending
could in principle also lead to a depletion of object detections in higher
density environments. The simulations of \citet{chiu16b} show that this is not
a problem for the choice of radial range ($0.75\text{--}2.5$\,Mpc) considered
in the present study.

\subsubsection{Number density profile}
\label{sec:numb-dens-prof}

As in \citet{applegate14} -- based on an approach by \citet{hoekstra07} -- the
radial profile of contaminating galaxies is modeled as
\begin{equation} \label{eq:11}
  f_\mathrm{cl}(r) = 
  \frac{n_\mathrm{cl}(r)}{n_\mathrm{cl}(r)
    +n_\mathrm{gal}} 
   = f_{500} \exp\left(1 - \frac{r}{r_{\mathrm{SZ}, 500}}\right) \;,
\end{equation}
where $f_{500}$ is the contamination fraction at $r_\mathrm{SZ, 500}$, the SZE
determined radius $r_\mathrm{500}$, and $n_\mathrm{gal}$ is the number density
of background galaxies. An important consideration in our case is that this
approach does not rely on measuring the background number density of galaxies
far away from the cluster centres but treats it as a free parameter. The
virial radius of most clusters in our sample is only slightly smaller than the
FOV of Megacam affording us no area completely free from cluster galaxies.

As \citet{applegate14} we see an upturn in the number density in most cluster
fields towards the centre. Per field measurements of cluster contamination
fractions are nevertheless too noisy to be meaningful and we adopt their
strategy of fitting all clusters simultaneously with a single contamination
fraction $f_{500}$ and per field $n_\mathrm{gal}$ values. We radially bin
the shear catalogue in angular bins of width $1\arcmin$ from the cluster
centre out to a maximum radius of $12\arcmin$. We assume Poisson errors on the
binned number counts. After binning we rescale the bin locations to units of
the SZE derived $r_{500}$ of each cluster. We emphasise that this is the only
step in our analysis that depends on an SZE derived cluster mass. Its only
purpose is to correct for the range in cluster mass and any systematic
covariance between the weak lensing derived cluster masses and their SZE based
estimates introduced by this scaling are subdominant to the relatively large
statistical errors on the contamination fraction.

\begin{table}
  \centering
  \caption{Cluster contamination fractions at $r_{500}$ show no dependence on inner
    radial cuts indicating that the fits are not affected by decreasing
    catalogue completeness towards the cluster centre.}
  \begin{tabular}{llr}
    \hline
    Method & Bin Rejection & \multicolumn{1}{c}{$f_{500}$} \\\hline
           & $<2\arcmin$   & $(4.8 \pm 2.5)\%$\\
    \citet{applegate14} & $<3\arcmin$ & $(5.5 \pm 3.1)\%$\\
                        & $<r_\mathrm{SZ, 500}$ & $(4.9 \pm 3.2)\%$ \\\hline
    \citet{gruen14}     & ---                 & $(2.3 \pm 1.7)\%$ \\\hline
  \end{tabular}
  \label{tab:contam_frac}
\end{table}
We reject some inner bins in the fitting procedure because we do not fit the
shear profiles all the way to the centre. Among other reasons, we try to
minimise the impact of cluster miscentring, which would also affect the number
density profiles. Another effect that could potentially be important in the
inner bins but was verified to be of negligible influence in our analysis is
the impact of magnification \citep{chiu16b}.

Table~\ref{tab:contam_frac} shows the result of performing the fit in this way
and removing a varying number of bins close to the cluster
centre.\footnote{For most clusters in our sample
  $r_\mathrm{SZ, 500} >3\arcmin$.} We find no dependence of $f_{500}$ on
the inner fit radius, indicating that over the radial ranges considered here
the catalogue completeness does not change strongly. The error bars are
estimated by bootstrap resampling from the cluster sample. The estimated
background galaxy number densities are reported in
Table~\ref{tab:redshift-betas}.

To test a possible redshift dependence of the cluster contamination we
split the sample at $z = 0.45$ where the foreground contamination
shows a strong jump when we transition to a different colour cut
regime. We find $f^{z>0.45}_{500} = (4.1 \pm 4.82)\%$ and
$f^{z<0.45}_{500} = (5.0 \pm 2.9)\%$ when the fit is restricted to
$r > 2\arcmin$. Both numbers are consistent with each other and the
value reported in Table~\ref{tab:contam_frac} if $r > 2\arcmin$ is
imposed.

Additionally, we test whether we can reduce the foreground contamination by
adjusting the colour cuts without adversely affecting the cluster
contamination fraction. We remove the colour cut transition at $z=0.45$ and
apply the colour cuts used for objects in the redshift range
$0.35 \leq z < 0.45$ over the range $0.35 \leq z < 0.55$ instead. Indeed this
reduces the foreground contamination for the four clusters in this bin to
$\lesssim 2\%\text{--}3\%$. At the same time we notice a significant
steepening of the number density profiles of these four clusters, 
indicating an increased contamination by cluster galaxies. On the one hand the
dilution of the shear signal by foreground galaxies is reliably taken care of
by setting their $\beta = 0$ in the estimation of $\langle \beta \rangle$ and
$\langle \beta^2 \rangle$. On the other hand we know that the reference field
cannot be a faithful representation of the galaxy density in redshift space in
the presence of a massive cluster. Given the low SNR of our $f_{500}$
measurement and the relative straightforwardness of the redshift sampling in
Section~\ref{sec:backgr-galaxy-select} we prefer to optimise our colour cuts
for rejection of cluster galaxies.

\subsubsection{Redshift space decomposition}
\label{sec:redsh-space-decomp}

An alternative method to fitting an analytical number density profile was
proposed by \citet{gruen14}. Briefly, they looked at the probability
distribution of the lensing efficiency $\beta$ and decomposed the observed
probability distribution at a given cluster centric radius $r$ into the
cluster and the field galaxy probability distribution
\begin{equation}
  \label{eq:12}
  p(\beta, r) = f_\mathrm{cl}(r) p_\mathrm{cl}(\beta, r) + \left(1 -
  f_\mathrm{cl}(r)\right) p_\mathrm{f}(\beta)\;,
\end{equation}
where $f_\mathrm{cl}(r)$ is the radially dependent cluster contamination
fraction. Once $p(\beta, r)$, $p_\mathrm{cl}(\beta, r)$, and
$p_\mathrm{f}(\beta)$ are known, the contamination fraction can be
found by simple $\chi^2$-minimisation. We additionally imposed the
constraint that $f_\mathrm{cl} \in [0, 1]$. This method works if its two
underlying assumptions are fulfilled: 
\begin{enumerate}
  \item the redshift distribution of galaxies is constant over the image;
  \item the cluster and field probability distributions $p_\mathrm{cl}$ and
    $p_\mathrm{f}$ are sufficiently independent such that the full
    distribution function $p(\beta)$ can be written as a linear combination of
    the two.
\end{enumerate}

It is reasonable to assume that the first condition is met in our case,
because our images have a homogeneous depth per field and cover only a small
solid angle. We experimentally verified that the second condition is also
fulfilled by plotting $p_\mathrm{cl}(\beta)$ and $p_\mathrm{f}$. We estimated
these distributions from the reference catalogue in the manner described by
\citet{gruen14}, which we summarise here.

The distributions $p(\beta, r)$ and $p_\mathrm{cl}(\beta, r)$ are
estimated in annuli around the cluster centre. We chose 9 bins of
width $1\arcmin$ starting at the cluster centre. In each bin, for
every object in the shear catalogue with magnitudes $\{g, r, i\}$ we
take galaxies with
$\sqrt{(\Delta g)^2 + (\Delta r)^2 + (\Delta i)^2} < 0.1$\,mag from
the reference catalogue. For each such sample we compute the probability
$P_\mathrm{cl}$ that the respective object is a cluster galaxy by
assigning it the fraction of sample galaxies that have
$|z - z_\mathrm{l}| \leq 0.06 (1 + z_\mathrm{l})$. Also, for every
sampled galaxy, we compute $\langle\beta\rangle$ from the COSMOS
sample. The unweighted histogram of these $\langle\beta\rangle$ values
is $p(\beta)$. The histogram weighted by the $P_\mathrm{cl}$ values is
$p_\mathrm{cl}(\beta)$.

The probability distribution of $\beta$ for field galaxies is estimated
in a similar fashion. For each object in the shear catalogue at a large
distance from the cluster -- we choose $r > 10\arcmin$ -- samples are drawn in
the same way. The probability $P_\mathrm{f}$ that a galaxy is a field
galaxy is assigned the fraction of sample objects with
$z < z_\mathrm{l} - 0.06 (1 + z_\mathrm{l})$. Again, the value of
$\langle\beta\rangle$ of the samples for each shear catalogue object is
computed. A histogram weighted by the probabilities $P_\mathrm{f}$ is the
distribution $p_\mathrm{f}(\beta)$. Following \citet{gruen14}, the choice of
$0.06 (1+\zl)$ as separation here and for computing the probability that a
galaxy is a cluster galaxy is based on the $2\sigma$ uncertainty of the
photometric redshifts in our reference catalogue. Varying this parameter does
not systematically influence our estimates of the contamination fraction.

\begin{figure}
  \includegraphics[width=\columnwidth]{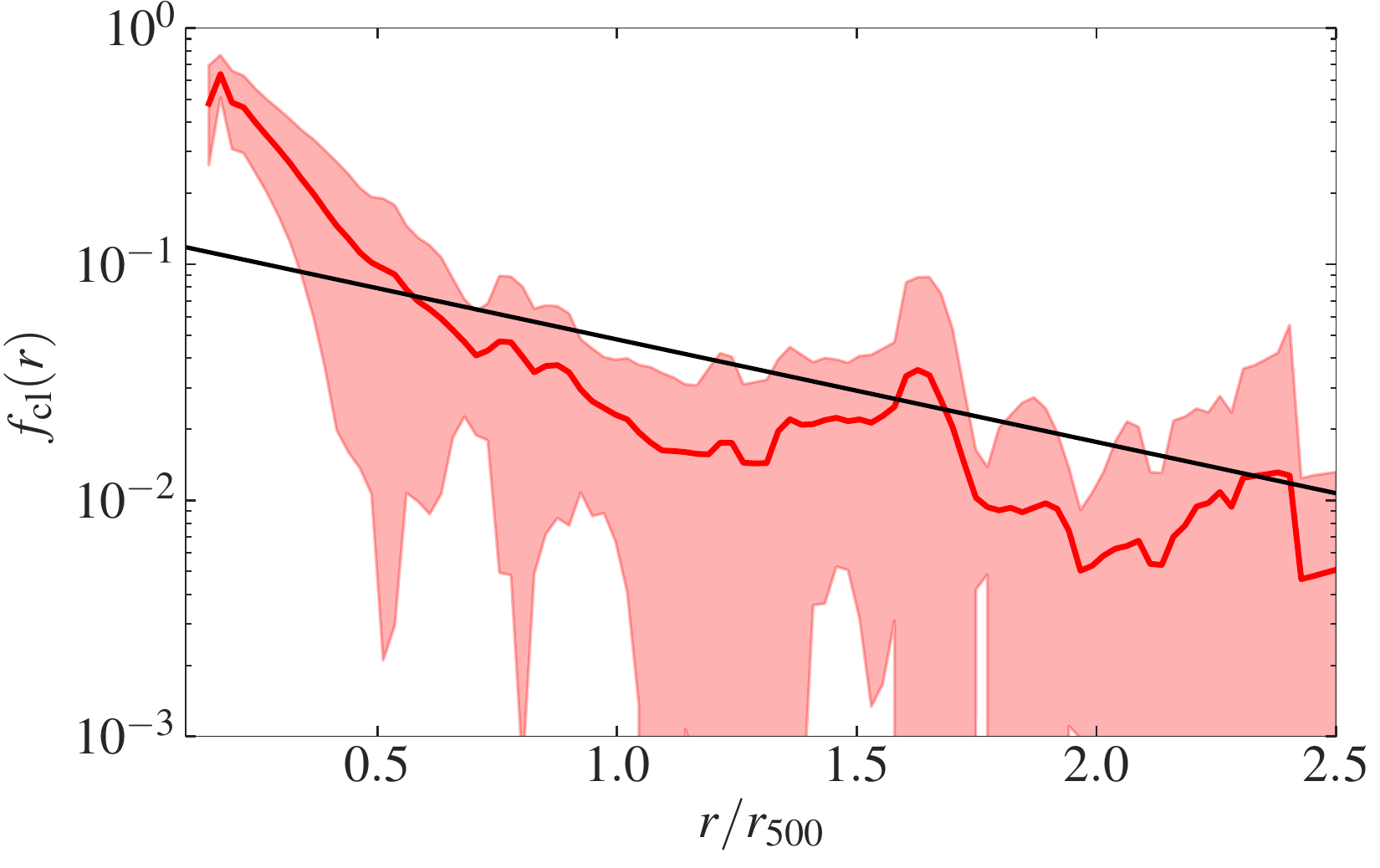}  
  \caption{Contamination correction derived by the method of
    \citet{gruen14}. The thick red line is a robust estimation of the mean of
    all clusters with its error range indicated by the shaded region. The
    solid black line is the exponential model derived in
    Sect.~\ref{sec:numb-dens-prof} for comparison.}
  \label{fig:fcl_gruen}
\end{figure}

Figure~\ref{fig:fcl_gruen} shows the radial contamination profile fraction
derived in this way for the ensemble of all clusters. Like in the case of the
exponential contamination model we found that individual cluster estimates are
very noisy and that there is no obvious redshift trend. Instead of correcting
each cluster profile with its own noisy contamination profile
$f_\mathrm{cl}(r)$, we estimate an average contamination profile and its error
using the robust location and scale estimator of \citet{beers90}.

We compare the impact both methods have when they are applied over different radial
ranges in the process of fitting NFW profiles \citep[see
Sect.~\ref{sec:nfw-profile-fits}]{navarro97} to the tangential shear. We measure the
relative change of mass compared to a profile fit ignoring the contamination
correction. In all cases the outer radial range considered is $2.5$\,Mpc and the inner
radius takes the values listed in Table~\ref{tab:fcl_mass}. We conclude that both
methods agree to better than 2\% outside $0.65\,r_{500}$. As one would expect larger
corrections are necessary if one decreases the inner radius of the shear profile
analysis. Nevertheless, we find that the purely empirical decomposition method is
significantly steeper than the exponential model at smaller radii indicating that the
latter is not a good model and the actual contamination profile is more similar to the
cored $1/r$ profile employed in \citet{hoekstra15}. We take the difference of $0.9\%$ in
mass (see the last line of Table~\ref{tab:fcl_mass}, which uses the inner radius later
employed in this work) between both methods considered here as an upper limit on the
impact of the systematic uncertainty of the contamination correction.

We also tested for the existence of mass dependent trend in the mean $f_{500}$ by
splitting the cluster sample into two equal sized bins along the detection significance
$\xi$. The contamination fractions measured in both bins are statistically
indistinguishable and fully consistent with the one determined for the whole cluster
sample, excluding any significant mass trend at the current level of uncertainty.

\begin{table}
  \centering
  \caption{Impact of the cluster contamination correction on the mean cluster mass
    relative to no correction for various choices of inner radial fitting
    range. $r_{500}$ is derived from the SZE based mass 
    estimate in \citet{bleem15}.}
  \begin{tabular}{rrr}\hline
    inner radius   & \citet{applegate14}& \citet{gruen14}\\\hline
    $0.5\,r_{500}$ & 3.8\% & 3.6\% \\
    $0.7\,r_{500}$ & 2.0\% & 0.1\% \\
    $500$\,kpc    & 3.7\% & 3.1\% \\
    $750$\,kpc    & 2.5\% & 1.6\% \\\hline
  \end{tabular}
  \label{tab:fcl_mass}
\end{table}

\section{Weak-lensing mass measurements}
\label{sec:weak-lensing-mass}
We present reconstructions of the projected mass density in
Section~\ref{sec:mass-reconstr-maps} and constrain the mass of our
galaxy clusters with fits to analytical shear profiles in
Section~\ref{sec:nfw-profile-fits}. As we will discuss, these fits
represent biased mass estimators, which can be calibrated with
simulations. First, however, the uncorrected fits can be compared to
mass estimates obtained for a subset of our clusters from the same
data by \citet{high12}.

\subsection{Mass reconstruction maps}
\label{sec:mass-reconstr-maps}

Cluster mass maps are often instructive to assess the weak-lensing detection
of a galaxy cluster and to compare light and mass distributions. We used the
finite field inversion method of \citet{SeitzSchneider01} to obtain
reconstructed $\kappa$-maps from the observed shear fields with a smoothing of
$2\arcmin$, which was selected based on the visual impression of the
reconstructed maps. To compute the noise levels of the surface mass density
reconstruction we create 800 realisations of the shear catalogues with
randomly rotated galaxy ellipticities while keeping the absolute value of the
ellipticity and the galaxy positions fixed. The variance of these random maps
is used as a noise estimator for each pixel, although pixels within the
smoothing scale are of course highly correlated. Dividing the $\kappa$-map by
noise maps created in this way gives SNR maps whose contours we show in the
left panels of Figures~\ref{fig:kappa-shear-0234}--~\ref{fig:kappa-shear-2355}
in Appendix~\ref{sec:mass-reconstr-prof}.

In these figures we compare the weak lensing significance contours with
significance contours of filtered SPT-SZ maps and significance contours of the
density of colour selected red sequence cluster galaxies. Although the SNR of
the WL reconstruction is low, in most cases we find good agreement between the
SPT and the WL centroid. Sizeable offsets between those are expected due to
shape noise and smoothing of an asymmetric mass distribution with a symmetric
kernel \citep{dietrich12} even in the absence of collisional processes
separating the dark matter and gas components of a galaxy cluster
\citep[e.g.,][]{clowe06}. The only noteworthy case in this gallery is
SPT-CL\,J2355$-$5055 (Fig.~\ref{fig:kappa-shear-2355}) whose field shows
another cluster west of the SPT detection in the galaxy density contour with
almost identical colours and an elongated structure extending NE from this
second cluster. These are not detected by SPT but seem to be broadly traced,
albeit at very low significance, by the mass reconstruction.

\subsection{NFW profile fits}
\label{sec:nfw-profile-fits}

Average density profiles of galaxy clusters in cosmological simulations are
known to follow a universal density profile
\begin{equation}
  \label{eq:13}
  \rho(r) = \frac{\delta_c \rho_\mathrm{crit}}{
    (c\, r / r_{200})(1 + c r / r_{200})^2}\;,
\end{equation}
first described by \citet{navarro97} to a very good
approximation. Here $r$ is the three-dimensional radius from the
cluster centre, $\rho_\mathrm{crit}$ is the critical density of the
Universe at the cluster redshift, $r_{200}$ is the radius at which the
enclosed mean density is $200 \rho_\mathrm{crit}$, $c$ is the
concentration parameter, which determines how fast the density profile
turns over from $\propto r^{-1}$ to $\propto r^{-3}$, and $\delta_c$
is a characteristic overdensity
\begin{equation}
  \label{eq:14}
  \delta_c =\frac{200}{3} \frac{c^2}{\ln(1 + c) - c / (1 + c)}\;.
\end{equation}

Although the NFW profile is a very good approximation of the average density
profile of galaxy clusters \citep[e.g.][]{johnston07}, better fitting
descriptions exist. The \citet{einasto65} profile is a better description of
the density profile close to the centre. At large radii ($>r_{200})$
correlated large-scale structure leads to systematic deviations from the NFW
profile \citep{johnston07}. For the radial ranges of interest in this work,
however, the original NFW profile with its well known lensing properties
\citep{bartelmann96,brainerd00} is a sufficiently good description of isolated
haloes. We will calibrate the impact of deviations from spherical NFW profiles
using simulations (cf. Section~\ref{sec:calibration-nfw-fits}).

We fit spherical NFW profiles to the binned tangential shear over the range of
$750$~kpc to $2.5$~Mpc. Going further inwards would increase our sensitivity
to miscentring \citep[e.g.,][]{johnston07,mandelbaum10}, the cluster
contamination correction (see Table~\ref{tab:fcl_mass}), and the
mass--concentration relation, which is difficult to measure using weak lensing
data alone. Going further outwards, deviations from an NFW profile become more
pronounced \citep{becker11} due to correlated \citep{johnston07} and
uncorrelated \citep{hoekstra03,dodelson04} LSS. We choose the SZE peak
position as cluster centre for the Megacam cluster sample and the X-ray
centroid as cluster centre for the HST sample. We use 8 linearly spaced bins
over this radial range and compute weighted averages of the reduced shear in
each bin
\begin{equation}
  \label{eq:15}
  \langle g_i \rangle = \frac{\sum_n w_n g_{i,n}}{\sum_n w_n}\;,\qquad i \in
  \{\mathrm{t}, \times\}\;,
\end{equation}
using the lensing weights
\begin{equation}
  \label{eq:16}
  w = \frac{{P^\gamma}^2}{\sigma_\varepsilon^2 {P^\gamma}^2  + (\Delta e)^2}\;,
\end{equation}
where $P^\gamma$ is the shear polarisability \citep{hoekstra98},
$\sigma_\varepsilon$ is the intrinsic ellipticity dispersion, which we
fixed to $0.25$, and $\Delta e$ is the error estimate for the
polarisation \citep{hoekstra00}. The errors of the mean shear in each
bin are computed as
\begin{equation}
  \label{eq:17}
  \frac{1}{\sigma_{\langle g_i \rangle}^2} = \sum_n w_n\;.
\end{equation}
We use the weighted average of the radial galaxy positions in a bin as
the effective bin location. We verified that the number of radial bins
and their location has no systematic influence on the recovered
cluster masses for a wide range of binning schemes, if we restrict the
fitting procedure to the chosen radial range of
$0.75~\mathrm{Mpc} < r < 2.5~\mathrm{Mpc}$.

We correct the binned tangential shear for the remaining contamination with
cluster galaxies via
\begin{equation}
  \label{eq:18}
  \langle g_\mathrm{t,cor} \rangle (r) = 
  \frac{\langle g_\mathrm{t}\rangle(r)}{(1 - f_\mathrm{cl}(r))}\;,
\end{equation}
where we use the mean contamination fraction of all clusters derived
from the method of \citet{gruen14} in
Sect.~\ref{sec:redsh-space-decomp}. We propagate the uncertainties of
this $f_\mathrm{cl}(r)$ profile to the reduced shear error estimates,
eq.~(\ref{eq:17}).

When fitting the model to the observed reduced shear profile, we treat the NFW
model as a one-parameter family with $\mtwo$ being the only free parameter and
fix the concentration parameter $c$ to exactly follow a mass--concentration
scaling relation with no intrinsic scatter. Specifically we adopt the
$M\text{--}c$ relation of \citet{diemer15}. This choice is justified by recent
observational constraints on the $M\text{--}c$ relation for the mass and
redshift range of the Megacam cluster sample \citep{merten15, cibirka17} and
by measurements of the concentration in the HST sample itself
\citepalias{schrabback18}.

Observed shear profiles with best fit NFW models are presented in
Appendix~\ref{sec:mass-reconstr-prof}. For all clusters, the
cross-shear is consistent with zero, as expected for shear catalogues
that are not significantly affected by systematics.

\subsection{Comparison with earlier mass measurements}
\label{sec:comp-with-earl}

Our weak lensing analysis of Megacam data is a significant expansion of an
earlier analysis of a subset of five clusters \citep{high12}. A comparison of
the mass estimates obtained is a natural part of our analysis. Although we use
the same data as \citet{high12} our analysis differs in a few key features as
described in the previous sections. Most importantly these are: (i) new shear
catalogues with new PSF model and multiplicative shear bias correction; (ii)
updated estimates for \mbeta\ and \mbetasq; (iii) improved estimation of the
cluster contamination correction; (iv) different mass--concentration scaling
relation.

\begin{figure}
  \includegraphics[width=\columnwidth]{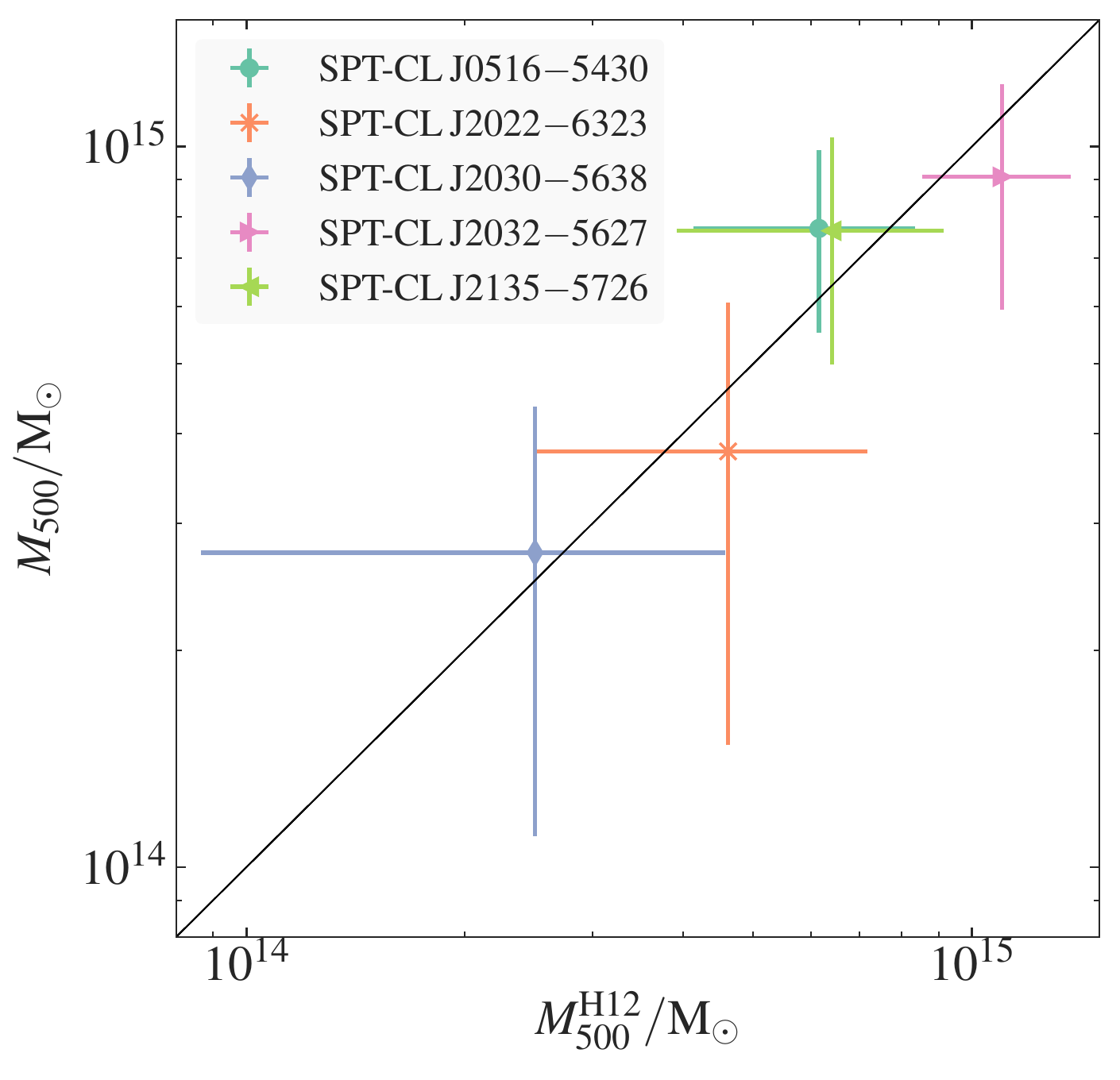}
  \caption{A comparison of the spherical NFW masses evaluated at
    $R_{500, \mathrm{WL}}$ between \citet{high12} and this work. The
    black horizontal line show the one-to-one relation and is not a
    fit to the data points.} 
  \label{fig:high12-comparison}
\end{figure}

Nevertheless, the mass estimates from this previous work and our analysis are
in agreement. Figure~\ref{fig:high12-comparison} shows a comparison of the
$M_{500}$ masses obtained from NFW fits of \citet{high12} and our masses
estimates. The weighted difference is
$\langle M_{500} - M^\mathrm{H12}_{500}\rangle = (0.0 \pm 1.3) \times
10^{14}\,\msun$.
Given the changes in the analysis mentioned above, we consider this agreement
to be a coincidental. We also emphasize that these changes were made to make
our mass estimates more robust and obtain better limits on the systematic
uncertainties of our analysis procedures.

One example where our new methods lead to significantly different results from
the one described in \citet{high12} is the \mbeta\ estimation for shallow
fields whose completeness drops sharply before our limiting magnitude of
$i=24.5$\,mag. SPT-CL\,J2138$-$6008 is one such field not present in
\citet{high12} in which the cluster mass would have been overestimated by
$\sim 14\%$ in the original analysis, leaving everything else the analysis
pipeline unchanged.

\subsection{Calibration of the NFW fits with simulations}
\label{sec:calibration-nfw-fits}

As mentioned in the previous section, systematic deviations from the NFW
profile and miscentring lead to biased mass estimates when fitting an NFW
profile to the tangential shear. Furthermore, halo triaxiality
\citep{clowe04,corless07} and projected LSS lead to additional scatter. We
characterise the relation between measured weak lensing mass and true mass
with a bias parameter \bwl,
\begin{equation}
  \label{eq:19}
  M_\mathrm{WL} = \bwl M_{500}\;,
\end{equation}
and scatter $\sigma_\mathrm{WL}$. This scatter consists of two
components: (i) a local component caused by the aforementioned deviations
from a spherical NFW profile and correlated LSS, \wllocal, assumed to be
log-normal in weak-lensing mass at fixed true mass; (ii) scatter caused the
projection of uncorrelated LSS, $\sigma_\mathrm{WL, LSS}$.

Our approach to calibrate \bwl\ and \wllocal\ is to create an ensemble of
simulated observations that match the observational properties of a random
subset of cluster fields and then apply the same measurement technique as we
do to the real data. In general, we are aiming to reconstruct the probability
distribution $P(M_{\mathrm{WL}}|M_{\mathrm{true}})$, which can then be
included in forward probabilistic modelling of the cluster sample. However, we
simplify the relation as stated above to one log-normal distribution that is
the same for all observed cluster fields. Any residuals from such an
oversimplification are still insignificant compared to the statistical
precision of our dataset.

To build our simulated observations for one observed cluster field, we start
with the $N$-body simulations from \citet{becker11}. These are $1$\,Gpc boxes
with $1024^3$ dark matter particles with a mass of
$6.98 \times 10^{10}\,\msun$ each. We cut out $400\,h^{-1}$\,Mpc long boxes
centred on the most massive 788 haloes with
$M_{500,c} > 1.5\times10^{14}\,h^{-1}\msun$ from the $z=0.5$ snapshot.
Particles are projected to form 2D mass maps that are then used to create
shear maps via Fast Fourier transform. The observed $\mbeta$ from a cluster
observation is used to scale the shear and $\kappa$ maps appropriately. Random
Gaussian noise is added to the shear map to match the observed shape noise in
the observations. Because in our real observations we fit a 1-D profile, we
select an ``observed'' cluster centre for each simulation map. We assume that
the displacement between the true projected centre of the simulated cluster
and the ``observed'' centre is randomly oriented with respect to the
underlying structure, a reasonable assumption given the noise sources of SPT
observations and the statistical power of this sample. Centre offsets are
randomly chosen following the form specified by \citet{song12}, a Gaussian
distribution with a width dependent on the SPT beam size and the core radius
of the matched filter used to detect the observed cluster. The simulated 1-D
profiles are then fit with an NFW model as in the data analysis.

We assume that $P(M_{\mathrm{WL}}|M_{\mathrm{true}})$ follows a log-normal
distribution with location and scale parameters $\mu =\ln b_{\mathrm{WL}}$ and
$\sigma = \sigma_{\mathrm{WL, local}}$, respectively. For the set of simulated
fields, we find the maximum \textit{a posteriori} location for the probability
distribution

\begin{equation}
\label{eq:20}
\begin{split}
  & P(\bwl, \sigma_{\mathrm{WL, local}} | \mathrm{mocks}) \\
  & \propto \prod_i
  \int P(\bwl, \sigma_\mathrm{WL, local} |M_{\mathrm{WL}}) P(M_\mathrm{WL} |
  \mathrm{mock}_i) 
  \dif M_{\mathrm{WL}} \;.
\end{split}
\end{equation}

Uninformative priors were used for the parameters of interest. Simulated
observations were also created and analysed using the $z=0.25$ snapshot from
\citet{becker11} as well as the Millennium-XXL simulations
\citep{angulo12}. No significant trends were seen between snapshots or
simulations. We also did not see any significant trend with the observational
properties of each observed field, including the amount of shape noise or
different filter core size. Our final bias ($\bwl = 0.938 \pm 0.028$) and
scatter ($\sigma_\mathrm{WL, local} = 0.214 \pm 0.040$) are then the average
across the random subset of cluster fields targeted for mock up when the
mass--concentration relation of \citet{diemer15} is used.

\subsection{Impact of the mass--concentration relation}
\label{sec:impact-mass-conc}

Weak lensing data often provide poor constraints on the concentration
parameter of the NFW profile. The shear signal is determined by the enclosed
mass at each radius, and the NFW scale radius is typically interior to the
innermost radius at which the shear is reliably measured. Without observing
the mass profile shape around the scale radius, via the shear profile, our
analysis can only provide very weak lower bounds on the
concentrations. Because most mass--concentration relations in the literature
seem to agree that there is a lower bound on concentrations at
$c \sim 2\text{--}3$, we only fit for the NFW mass of our clusters and enforce
that they follow a mass--concentration scaling relation. Any mismatch between
this relation and true galaxy clusters then introduces another source of
systematic error that we need to take into account.

We can estimate the sensitivity of our analysis to uncertainty of published
mass--concentration relation by carrying out the NFW fit bias analysis of the
previous section for different fixed concentrations. We find that the average
mass bias at concentrations $c = 5$ and $c = 3$ is $\bwl = 0.978$ and
$\bwl = 0.907$, respectively, implying $\dif \bwl / \dif c |_{c=4} =
-0.0355$. Using Gaussian error propagation on eq.~(\ref{eq:19}) we obtain
\begin{equation}
  \label{eq:21}
  \left(\frac{\sigma_M}{M_\mathrm{true}}\right)^2 = \frac{1}{\bwl^2}
  \left(\frac{\dif \bwl}{\dif c}\right)^2 \sigma_c^2\;.
\end{equation}
Because we calibrated the bias resulting from NFW fits in
Sect.~\ref{sec:calibration-nfw-fits} using our chosen $M\text{--}c$ relation,
namely the one of \citet{diemer15}, the systematic uncertainty is not given by
how well this relation describes the actual cluster sample, but by how
faithfully the simulated clusters represent true clusters in the Universe. The
simulations used in the previous section are Dark Matter only and thus the
question is how much would the concentrations for clusters of the mass and
redshift in our sample and redshift be impacted by baryonic
effects. \citet{duffy10} constrain this to an upper limit of $10\%$.
Evaluating eq.~(\ref{eq:21}) we set $\sigma_c|_{c=4}=0.4$ and obtain a mass
uncertainty due to the mass--concentation relation of $1.5\%$.

\begin{table}
  \begin{threeparttable}
    \centering
    \caption{Overview of all known systematic error sources and their
      contributions to the overall systematic error budget. The different
      error sources are added in quadrature to obtain the total systematic
      error estimate (68\% confidence).}
    \begin{tabular}{lrl}\hline
      Error Source & Impact on Mass&Reference\\\hline
      Multiplicative shear bias & 2\% & \S~\ref{sec:shear-catal-creat}\\
      PSF boost correction & 2.5\% & \S~\ref{sec:shear-catal-syst}\\
      $\mbeta$ and $\mbetasq$ estimation & 1.2\% & \S~\ref{sec:backgr-galaxy-select}\\
      Contamination correction & 0.9\% & \S~\ref{sec:clust-cont-corr}\\
      NFW mass bias & $2.8\%$ & \S~\ref{sec:calibration-nfw-fits}\\
      $M$--$c$ relation & $1.5\%$ & \S~\ref{sec:impact-mass-conc}\\
      Miscentring distribution & $3\%$ & \S~\ref{sec:impact-misc-model}\\\hline
      Total & 5.6\%\\\hline
    \end{tabular}
    \label{tab:systematic-errors}
  \end{threeparttable}
\end{table}

\subsection{Impact of the miscentring model}
\label{sec:impact-misc-model}

Our baseline model for the distribution of offsets between the SZE peak
position, which we use as the cluster centre in our analysis of the Megacam
data, and the true cluster centre is the analytical form of \citet{song12}
described in Sect.~\ref{sec:calibration-nfw-fits}. We estimate limits on the
impact on the mass calibration of this miscentring uncertainty by running the
NFW bias analysis of Sect.~\ref{sec:calibration-nfw-fits} with a different
miscentring model. We use a miscentring distribution adopted from the analysis
of \citet{saro14} but based on cosmological hydrodynamical simulations with
both large volume and high resolution \citep[see, e.g.][]{bocquet16,
  gupta17}. This includes a mock SZE signal and a simulation of the SPT
cluster detection procedure, which uses the multi-frequency adaptive filter
method \citep{melin06}. Briefly, a $\beta$-profile with $\beta = 1$ is used as
cluster template, with 12 different core radii $\theta_\text{core}$, the same
as used by SPT \citep{bleem15}. The highest signal-to-noise peaks within in
the larger of $\theta_\text{core}$ or $1\arcmin$ are picked as individual
cluster candidates with the peak position as the centre. The SZE peaks
identified in this way are matched to the projected halo centre, which is the
most bound particle.

For this miscentring distribution and the \citet{diemer15} $M\text{--}c$
relation, we find a weak lensing bias $\bwl = 0.960 \pm 0.027$. From the
difference to the \bwl\ value in our baseline analysis, we conservatively
assume an uncertainty of $3\%$ in the weak lensing bias parameter.

\subsection{Summary of systematic uncertainties}
\label{sec:summ-syst-uncert}

We now briefly summarise all contributions to our systematic uncertainty
budget. An overview is presented in
Table~\ref{tab:systematic-errors}. Broadly, these fall into two categories,
observational uncertainties and modelling uncertainties. We considered
observational biases in Sect.~\ref{sec:weak-lensing-data}. The first two of
these four pertain to how well we can measure shear. Based on
\citet{hoekstra15} the impact on mass of the multiplicative shear bias was
estimated to be $<2\%$. Additionally, the shear calibration of
\citet{hoekstra15} was derived for a circular PSF. For the strongly
anisotropic PSF in our data, an additional boost to the smear polarisability
was suggested by \citet{hoekstra15} to avoid biases. Applying this correction
led to an additional scatter of $2.5\%$ in mass.

The second set of observational systematics is due to uncertainties in the
redshift estimates of galaxies. First, uncertainties in \mbeta\ and \mbetasq\
come from cosmic variance of the reference field, uncertainties in the
photometric redshifts of the reference field itself, and uncertainties in our
photometric calibration. This contributes $1.2\%$ to our systematic
errors. Second, cluster galaxies evading our colour-colour cuts dilute the
shear signal. We model this small signal in Sect.~\ref{sec:clust-cont-corr}
using two approaches. We propagate the uncertainties of the model we judged to
be more reliable to the \emph{statistical} error budget and treat the
difference between the two models as a source of systematic uncertainty. This
difference amounts to $0.9\%$ in mass.

We considered the second category of modelling errors in
Sects.~\ref{sec:calibration-nfw-fits}--~\ref{sec:impact-misc-model}. We
discussed three sources of modelling errors. First, a bias incurred by
fitting an NFW profile following a fixed mass--concentration relation to shear
profiles that could deviate from an NFW profile, e.g. from correlated LSS,
miscentring, and obeying a different $M\text{--}c$ relation. We calibrate this
bias factor $b_\mathrm{WL}$, eq.~(\ref{eq:19}), on $N$-body simulations and
use its uncertainty of $2.8\%$ as the systematic error number in
Table~\ref{tab:systematic-errors}. Second, we used previous estimates
\citep{duffy10} by how much the concentration of simulated dark-matter only
haloes may depart from the true cluster concentration to estimate the impact
of baryonic effects on mass. This amount to $1.5\%$ in our error
budget. Finally, we studied how much uncertainties in our miscentring model
affect the mass estimates. The $2.8\%$ error on $\bwl$ quoted above is only
the uncertainty of the NFW mass bias calibration for our chosen miscentring
baseline model. Replacing this model with another leads to a different
estimate of \bwl. We take this difference of $3\%$ as uncertainty caused by
the choice of miscentring model.

Because the various sources of systematic uncertainties are not expected to be
correlated, we sum them in quadrature to obtain final systematic error budget
of $5.6\%$.

\section{Mass--observable scaling relations and likelihood function}
\label{sec:scaling-relations}

We use our cluster data set, containing SZE, X-ray, and weak-lensing
measurements, to constrain the mass--observable relations for all
observables. We consider two different observables for the X-ray scaling
relations, the gas mass \mgas\, and \yx. Because both observables share the
same gas mass measurements, they are not independent, and we do not run any
fits for both X-ray relations simultaneously; rather, we either fit for one or
the other relation. In the following, we discuss all mass--observable
relations, the likelihood function, and our choice of priors.

\subsection{SZE and X-ray scaling relations}
\label{sec:sze-x-ray}

Galaxy clusters in the SPT-SZ survey were detected via their thermal SZE in
the 95 and 150~GHz maps via a multi-scale matched filter technique
\citep{melin06}. The observable used to quantify the cluster SZE signal is
$\xi$, the detection significance maximised over all filter scales. These
filter scales are a set of 12 linearly spaced values from $0\farcm5$ to
$2\farcm5$ and the filter scale that maximises the detection significance is
associated with the cluster core radius $\theta_\mathrm{c}$. Due to noise
bias, $\xi$ is a biased estimator of SNR. Therefore, an unbiased SZE
significance $\zeta$ is introduced, corresponding to the signal-to-noise at
the true cluster position and filter scale \citep{vanderlinde10}. For
$\xi> 2$,
\begin{equation}
  \label{eq:22}
  \zeta = \sqrt{\langle\xi\rangle^2 -3}
\end{equation}
describes the relation between $\xi$ and $\zeta$, with scatter described by a
Gaussian of unit width, where the average is taken over many noise
realizations.

The unbiased SNR $\zeta$ can be related to cluster mass by the
mass--observable scaling relation
\begin{equation}
  \label{eq:23}
  \zeta = \asz \left(\frac{0.7 \mfive}{3\times10^{14}\,\msun
      \hm}\right)^\bsz \left(\frac{E(z)}{E(0.6)}\right)^\csz\;,
\end{equation}
where $A_\mathrm{SZ}$ is the normalisation, $B_\mathrm{SZ}$ the mass slope,
$C_\mathrm{SZ}$ the redshift evolution and $E(z) = H(z)/H_0$. An additional
parameter $\dsz$ describes the intrinsic scatter in $\zeta$, which is assumed
to be log-normal and constant as a function of mass and redshift.

We also relate the X-ray observables to cluster mass via mass--observable scaling
relations
\begin{equation}
\label{eq:24}
\begin{split}
  \frac{\yx}{10^{14}\,\msun \, \mathrm{keV}} =& \ay
  \left(\frac{\mfive}{5\times10^{14}\,\msun \sqrt{0.7\h}}\right)^\by\\
  &\times \left(\frac{E(z)}{E(0.6)}\right)^\cy
\end{split}
\end{equation}
and
\begin{equation}
\label{eq:25}
\frac{\mgas }{5\times10^{14}\,\msun } = \am \left(\frac{\mfive}{5\times10^{14}\,\msun
      }\right)^\bm \left(\frac{E(z)}{E(0.6)}\right)^\cm \;,
\end{equation}
and assume a corresponding log-normal scatter \dy\ (\dm) in \yx\ (\mgas) at
fixed mass. Note that we use the same redshift pivots as for the SZE scaling
relation, but apply a slightly larger pivot point in mass, approximately
corresponding to the median mass of the subsample with available X-ray
observations. Also note that the parametrisation of the \yx-mass relation we
use here departs from the one used in previous work by the SPT collaboration
\citep[e.g.][]{dehaan16}. We write \yx\ as a function of mass so that all
mass--observable-relations~(\ref{eq:23})--~(\ref{eq:25}) have the observable
on the left-hand side.

\subsection{Weak-lensing modelling systematics}
As discussed in Section~\ref{sec:calibration-nfw-fits} we assume a relation
between the \emph{weak lensing mass} that is obtained from fitting an NFW
profile to our shear data and the unobservable, \emph{true mass}
$M_\mathrm{WL} = \bwl \mfive$. The normalisation \bwl\ and the scatter about
this mean relation are calibrated taking modelling and measurement
uncertainties into account; we use numerical simulations for the modelling
part. As our weak-lensing data set consists of two subsamples -- Megacam and
HST -- with slightly different measurement and analysis schemes, we expect
some systematics to be shared among the entire sample, while we expect others
to affect each subsample independently.

All simulation calibrated quantities $x$ come with an estimate $\hat{x}$ and
at least one source of uncertainty $\Delta x$ on this estimate. Instead of
applying a prior $\mathcal{N}(\hat{x}, (\Delta x)^2)$ on $x$ we write
$x = \hat{x} + \delta \Delta x$ and leave $\delta$ as a free parameter in our
MCMC chain with a prior $\sim \mathcal{N}(0, 1)$. We describe this in detail
below for the weak lensing bias and local sources of scatter.

\subsubsection{Weak-lensing bias}
We model the weak-lensing bias as two independent components: mass model and
measurement systematics. We calibrate the amplitude of the bias due to mass
modelling against numerical simulations, and model the measurement systematics
such that we expect zero bias. For our likelihood analysis, we parametrise the
weak-lensing bias as
\begin{equation}
  \label{eq:26}
  \begin{split}
    b_{\mathrm{WL}, i} &= b_{\mathrm{sim}_i}
    + \delta_\mathrm{WL, bias}\, \Delta b_{\mathrm{mass\,model}_i}
    + \delta_i\, \Delta b_{\mathrm{shear cal,}\,N(z)_i},\\
    &i \in \{\text{Megacam, HST}\}\;,
  \end{split}
\end{equation}
where $b_\mathrm{sim}$ is the mean expected bias due to the mass modelling,
$\Delta b_{\mathrm{mass\,model}}$ is the uncertainty in our calibration of
$b_\mathrm{sim}$, and $\Delta b_{\mathrm{shear cal,}\,N(z)}$ is the quadrature
sum of the uncertainties in shear calibration and in the determination of the
distribution of background galaxies; $\delta_\mathrm{WL, bias}$,
$\delta_\mathrm{Megacam}$, and $\delta_\mathrm{HST}$ are free parameters in our
likelihood. With this parametrisation, we put Gaussian priors of unit width
centred at zero $\mathcal N(0,\,1)$ on the three parameters
$\delta_\mathrm{WL, bias}$, $\delta_\mathrm{Megacam}$, and
$\delta_\mathrm{HST}$. We investigate a possible redshift dependence of
$b_{\mathrm{sim}_i}$ and $\Delta b_{\text{mass model}}$ and find no
indications for it, so we treat these terms as redshift independent.

Due to the different observing strategies for the Megacam and HST samples, the
mean expected biases $b_{\mathrm{sim}_i}$ are determined for each sample
separately. The uncertainty on the mass model
$\Delta b_{\mathrm{mass\,model}_i}$ is modeled as the quadrature sum of the
uncertainty obtained from the numerical simulations, the uncertainty in the
$M-c$ relation, and the uncertainty due to miscentring. These uncertainties
are determined in identical ways for both subsamples (although the numbers
differ), and so we adopt a common fit parameter $\delta_\mathrm{WL,
  bias}$.
This effectively correlates the uncertainties due to mass modelling between
both samples. The shear calibration and determination of the distribution of
background galaxies, however, is independent for each sample, and we therefore
adopt a fit parameter $\delta_\text{Megacam/HST}$ for each sample.

\subsubsection{Weak-lensing scatter}
\label{sec:weak-lensing_scatter}

We decompose the weak-lensing scatter into two components: uncorrelated LSS
modeled by a normal distribution and scatter intrinsic to the NFW modelling of
the lensing halo. The latter term includes scatter due to the miscentring
distribution, halo triaxiality, and correlated LSS. Our motivation for this
approach is twofold: First, the simulations used to calibrate the bias and
scatter in Sect.~\ref{sec:calibration-nfw-fits} are not full light cones and
do not capture the entirety of projected large-scale structure. Second, these
simulations indicate that this local scatter is well described, at least for
our purposes, by a log-normal distribution, while uncorrelated LSS leads to an
additional Gaussian scatter contribution to the tangential shear. We model the
latter term as Gaussian scatter on the cluster mass, although this is not
entirely correct as the relation between cluster mass and shear is non-linear
\citep[see also][]{hoekstra03}. The combination of log-normal local scatter
and normal non-local scatter gives us enough flexibility to model the true
mass scatter, which is also neither exactly normal nor log-normal.

We calibrate the local, log-normal scatter against simulations. The Megacam
and HST samples have different scatter properties, but these numbers are
calibrated against the same simulations, and therefore share the same
systematics. We use
\begin{equation}
\label{eq:27}
\begin{split}
\sigma_{\mathrm{local}_i} = \sigma_{\mathrm{sim}_i}+ \delta_\mathrm{WL, scatter}\, \Delta
\sigma_{\mathrm{sim}_i}\;,\quad i \in \{\text{Megacam, HST}\}
\end{split}
\end{equation}
where $\Delta \sigma_{\mathrm{sim}_i}$ is the uncertainty of the simulation
calibrated scatter $\sigma_{\mathrm{sim}_i}$ and $\delta_\mathrm{WL, scatter}$
is a free parameter in our likelihood, on which we apply a Gaussian prior
$\mathcal N(0,\,1)$.

We estimate the uncorrelated LSS contribution to the weak-lensing scatter in
our NFW fits of the Megacam data by calculating the variance of the surface
mass density inside our fit aperture following the prescription presented in
\citet{hoekstra01}. A key difference between our work and that of
\citet{hoekstra01} is that they compute the variance inside an aperture for
the aperture mass statistics while we perform NFW fits to the shear
profile. The aperture mass is radially weighted average of the mass inside a
cylinder where the weight is given by a fixed filter function chosen by the
user. To adapt the prescription of \citet{hoekstra01} to our case we weigh the
surface mass density power spectrum by an NFW profile representing the average
mass and redshift of the Megacam cluster sample. For a cluster with
$M_{200} = 8 \times 10^{14}\,\msun$ at $z=0.4$ we obtain
$\sigma_{\mathrm{WL, LSS}_\mathrm{Megacam}} = 9 \times 10^{13}\,\msun$.

This value is close to and slightly larger than the average value reported for
the HST clusters,
$\sigma_{\mathrm{WL, LSS}_\mathrm{HST}} = 8 \times 10^{13}\,\msun$. This may seem
suprising at first because the lensing catalog of the HST is much deeper than
the Megacam data and consequently integrates over more large-scale
structure. The apertures employed in the lower redshift Megacam sample are,
however, larger than in the HST sample. This more than compensates our
shallower redshift distribution. 

In our analysis we only use the mean value
$\sigma_{\mathrm{WL, LSS}_\mathrm{HST}} = 8 \times 10^{13}\,\msun$ of the LSS scatter
values reported in \citetalias{schrabback18} as the mean of a Gaussian prior rather than
an individual prior for each cluster. This reduces computational complexity and the
impact on our analysis is negligible because the various sources of scatter are (almost)
fully degenerate so that tiny deviations from reality in one scatter term are easily
absorbed by another. The Gaussian prior for $\sigma_{\mathrm{Megacam}_i}$ is centred on
the value computed above. Both priors have a standard deviation of
$\Delta\sigma_{\mathrm{WL, LSS}_i} = 10^{13}\,\msun$, based on the estimated scatter of
$\sigma_{\mathrm{WL, LSS}_\mathrm{HST}}$ \citepalias{schrabback18}.

As mentioned in Sect.~\ref{sec:calibration-nfw-fits}, the bias $\bwl$ and
scatter $\sigma_\mathrm{WL, local}$ depend on the miscentering model one
adopts. In general the centroid of the X-ray emission of the intra-cluster
medium is expected to be a more reliable indicator of the true cluster centre
than the SZE peak position based on observations with a relatively broad
beam. The HST sample has X-ray data for all clusters and thus we choose the
X-ray positions and their corresponding bias and scatter values from
\citetalias{schrabback18} as input to our analysis. The Megacam sample is not
fully covered by \textit{Chandra} data. For these data we take the SZE peak
position as the cluster centre.

\subsection{Likelihood function and analysis pipeline}
\label{sec:scal-rel-pipeline}

We simultaneously constrain the SZE and X-ray scaling relations (four
parameters each) and the weak lensing model (six parameters) using an
extension of the framework described in \citet{bocquet15a}. We summarize the
main points of their likelihood function and discuss our extensions.
All fit parameters are also listed in Table~\ref{tab:priors}.

The translation of the weak lensing observable, i.e. the reduced shear \gt\
into a physical mass scale depends on the cosmological parameters in a number
of ways. First, the critical density of the Universe at the cluster redshift
enters the NFW profile. Second, the translation of the angular shear profile
into a radial shear profile measured in physical distances depends on the
distance-redshift relation. Similarly, the distance-redshift relation enters
the computation of the critical surface mass density,
eq.~(\ref{eq:2}). Finally, while many mass--concentration relations are
strictly speaking valid only for the cosmological parameters for which they
were derived, the $M\text{--}c$ relation of \citet{diemer15} we employ, has an
explicit cosmological dependence.

We use the observed reduced shear with cluster contamination correction
applied, $g_\mathrm{t, cor}$, and redshift distribution $N(z)$ as input to the
weak lensing portion of the likelihood code, which then computes \mbeta,
\mbetasq, and fits an NFW profile as described in
Section~\ref{sec:nfw-profile-fits} at every sample point of the MCMC chain. In
this way the cosmology dependence of the NFW shear profile due to the
evolution of the critical density with redshift and the redshift-distance
relation are taken into account.

Our cluster sample is SZE-selected. To properly take selection effects into
account, for each cluster $i$ in our sample, we evaluate the likelihood
\begin{equation}
\label{eq:28}
\begin{split}
&P(\mathrm{X}_i,M_{\mathrm{WL}_i}|\xi_i, z_i, \vec p) =\\
&\left[\int\int \dif M \dif\zeta P(\xi|\zeta)
  P(\mathrm{X}_i,M_{\mathrm{WL}_i},\zeta | M, z_i, \vec p) P(M|z_i,\vec
  p)\right]\Big|_{\xi_i}  
\end{split}
\end{equation}
where, for simplicity, we denote the X-ray observable as X, $P(M|z,\vec p)$ is
the halo mass function at redshift $z$, and $\vec p$ is the vector of
cosmological and scaling relation parameters. The multiplication with the
halo mass function is a necessary step to account for the Eddington bias.

The term $P(\mathrm{X},M_\mathrm{WL},\zeta | M, z, \vec p)$ contains the
mass--observable relations defined in Sect.~\ref{sec:sze-x-ray} as well as the
intrinsic scatter about each relation. Extending the original analysis
framework \citep{bocquet15a}, we allow for correlated scatter between all
observables. Namely \dsz, \dy, and $\sigma_\mathrm{WL}$ are linked by
correlation coefficients $\rho_\mathrm{SZ-Y}$, $\rho_\mathrm{SZ-WL}$, and
$\rho_\mathrm{WL-Y}$, so that the intrinsic covariance matrix is
\begin{equation}
  \label{eq:29}
  \begin{split}
    \mathsf{\Sigma}_Y & = 
    \begin{pmatrix}
      \dsz^2 & \dsz\dy & \dsz\wllocal \\
      \dsz\dy & \dy^2 & \dy\wllocal \\
      \dsz\wllocal & \dy\wllocal & \wllocal^2 \\
    \end{pmatrix} \\
    & \circ
    \begin{pmatrix}
      1 & \rho_\mathrm{SZ-X} & \rho_\mathrm{SZ-WL} \\
      \rho_\mathrm{SZ-X} & 1 & \rho_\mathrm{WL-Y} \\
      \rho_\mathrm{SZ-WL} & \rho_\mathrm{WL-Y} & 1 \\
    \end{pmatrix}
    \;,
  \end{split}
\end{equation}
and equivalently for \dm. We put flat priors allowing the full range from $-1$
to $1$ on all three correlation coefficients with the additional restriction
that the combination of all three coefficients must be physically allowed. In
practice, we compute eq.~(\ref{eq:28}) on a three-dimensional
grid in $\mathrm{X}$, $M_\mathrm{WL}$, and $\zeta$. To make this step
computationally efficient, we i) choose an optimal range in mass for each
cluster, informed by its measured SZE, X-ray, and weak-lensing signals and the
current set of scaling parameters $\vec p$, to avoid parts of the observable
space with effectively zero probability, ii) only perform this
three-dimensional computation for clusters that actually have all three
measurements, otherwise computing the (cheaper) two-dimensional version, and
iii) employ a Fast Fourier Transform convolution.

For each cluster in our sample, we compare the predicted
$P(\mathrm{X},M_{\mathrm{WL}}|\xi, z, \vec p)$ with the actual measurement and
extract the probability of consistency.

The X-ray measurements consist of a radial profile $\mgas(r)$ and a
global temperature measurement $T_\mathrm{X}$ from which a
$\yx(r) = \mgas(r) T_\mathrm{X}$ profile is computed. To account for the
radial dependence of the measurement and the variation of the modeled
$r_{500}$ throughout the parameter space, we define a fiducial radius
$r_{500}^\mathrm{fid}$ for each cluster, and evaluate the likelihood at this
radius. Note that the X-ray scaling relations eqs.~(\ref{eq:24}) and
(\ref{eq:25}) predict the X-ray quantity at $r_{500}$. To translate this model
prediction to $r_{500}^\mathrm{fid}$, we use the fact that the X-ray profile
can be well approximated by a power-law in radius \citep[see
also][]{mantz16}. With this, the prediction at $r_{500}^\mathrm{fid}$ becomes
\begin{equation}
\label{eq:30}
\yx(r_{500}^\mathrm{fid}) \equiv 
\left(\frac{r_{500}^\mathrm{fid}}{r_{500}}\right)^\text{slope}\, \yx(\mfive,z,\vec p),
\end{equation}
where $r_{500}$ is derived from \mfive. The measurement uncertainty in
$\yx(r_{500}^\mathrm{fid})$ is captured by a log-normal distribution.

For the weak-lensing data, we forward-model from $M_\mathrm{WL}$ to the
observed reduced shear \gt: we convolve $P(M_\mathrm{WL})$ with the Gaussian
LSS noise (Section~\ref{sec:weak-lensing_scatter}), and then compute the
reduced shear $\gt(M_\mathrm{WL}, r_j)$ for each radial bin $r_j$ following
eqs.~(\ref{eq:1})--(\ref{eq:3}). Finally, for each radial bin, we compute the
likelihood of the measurement given $\gt(r_j)$, and multiply the likelihoods
of all bins.

Ultimately, we sum the log-likelihoods for all clusters. The total likelihood
function (up to an additive constant) then is
\begin{equation}
  \label{eq:31}
  \ln \mathcal L = \sum_{i=1}^{N_\mathrm{cl}}
  \ln P(\mathrm{X}_i, \gt_i | \xi_i, z_i,\vec{p}) \;,
\end{equation}
where $i$ runs over all clusters. It is important to note that the measured
cluster abundance does not enter the likelihood function or analysis in this work.

\begin{table*}
  \centering
  \caption{Parameters and priors used in the scaling relations analysis. The
  weak-lensing parametrisation is such that the fit parameters rescale the expected
  central values and uncertainties.}
  \begin{tabular}{llll}\hline
    Parameter & Prior & Parameter & Prior \\\hline
    \multicolumn{4}{l}{\emph{SZE and \yx}}\\
    \asz & $1 / \asz$                & \ay & $1 / \ay$ \\
    \bsz & $\mathcal{N}(1.63, 0.1^2)$ & \by & const. \\
    \csz & const.                  & \cy & $\mathcal{N}(0.702, 0.351^2)$ \\
    \dsz & $\mathcal{N}(0.13, 0.13^2)$ & \dy & $\mathcal{N}(0.12, 0.08^2)$ \\\hline
    \multicolumn{4}{l}{\emph{SZE and \mgas}}\\
    \multicolumn{2}{c}{\multirow{4}{*}{SZE as above}} & \am & $1 / \am$\\
         & & \bm & const. \\
         & & \cm & $\mathcal{N}(0.0, 0.2^2)$ \\
         & & \dm & $\mathcal{N}(0.12, 0.08^2)$ \\\hline
    \multicolumn{4}{l}{\emph{Weak-lensing systematics}}\\
    $\delta_\mathrm{WL, bias}$ & $\mathcal{N}(0,1)$ \\
    $\delta_\mathrm{Megacam}$ & $\mathcal{N}(0,1)$ \\
    $\delta_\mathrm{HST}$ & $\mathcal{N}(0,1)$ \\
    $\delta_\mathrm{WL, scatter}$ & $\mathcal{N}(0,1)$ \\
    $\sigma_{\mathrm{WL, LSS}_\mathrm{Megacam}} / \msun$ &
                                                           $\mathcal{N}(9\times10^{13},10^{26})$ \\ 
    $\sigma_{\mathrm{WL, LSS}_\mathrm{HST}} / \msun$ &
                                                       $\mathcal{N}(8\times10^{13},10^{26})$ \\\hline 
    \multicolumn{4}{l}{\emph{Correlated scatter}}\\
    $\rho_\mathrm{SZ-X}$ & $\mathcal{U}(-1, 1)$ \\
    $\rho_\mathrm{SZ-WL}$ & $\mathcal{U}(-1, 1)$ \\
    $\rho_\mathrm{WL-X}$ & $\mathcal{U}(-1, 1)$ \\
    Eq. (\ref{eq:29}) & $\mathrm{det}(\mathsf{\Sigma}) > 0$ \\\hline
    \multicolumn{4}{l}{\emph{Cosmology}}\\
        \multicolumn{1}{l}{$[\Omega_\mathrm{m}, \sigma_8]$} & 
              \multicolumn{3}{l}{$\mathcal{N}\left([0.291, 0.783], 
                          \bigl(\begin{smallmatrix}
                            0.0016 & -0.0010 \\ -0.0010 & 0.0013 \\
                          \end{smallmatrix}
                          \bigr)\right)$}\\
    \multicolumn{1}{l}{$H_0 / (\mathrm{km\,s^{-1}\,Mpc^{-1}})$} & 
                                                                \multicolumn{3}{l}{$\mathcal{N}(73.8, 2.4^2)$} \\
    \hline
  \end{tabular}
  \label{tab:priors}
\end{table*}

We emphasize that all sample selection effects are accounted for in our
likelihood framework. Equation~(\ref{eq:31}) is only evaluated for clusters
that pass the SZE and redshift selection functions ($\xi>5$ and $z>0.25$), and
the target selection of the follow-up observations (X-ray and weak lensing) is
not based on these observables themselves (e.g., X-ray properties or weak
lensing strength). Obviously, one must not reject follow-up observations that
did not lead to a detection of the cluster. This frequently happens in the
weak lensing observable due to its large scatter. Our forward modelling
approach naturally deals with clusters whose radial shear profile is
consistent with zero or less.

We use the \textsc{emcee} \citep{2013PASP..125..306F} implementation of the
affine-invariant ensemble sampler algorithm \citep{2010CAMCS...5...65G} to
evaluate the likelihood function of eq.~(\ref{eq:31}). We use an ensemble of
192 walkers and discard the first five autocorrelation lengths of the chain as
burn-in period. We consider chains to be converged if no evolution of the mean
and standard deviation are visible in trace plots and if the
\citet{1992StaSc...7..457G} criterion is $\hat{R}<1.1$ for all parameters.

\subsection{Test on mock catalogues}
\label{sec:test-mock-catalogues}

We test that our implementation of the calibration framework described above
recovers unbiased parameter estimates using mock galaxy cluster
catalogs. These are created by Poisson sampling the halo mass function over
the redshift range of the SPT-SZ cluster sample. The SZE detection
significance $\ln \zeta$ and the follow-up quantities $\yx$ and weak-lensing
mass are drawn together from a multivariate normal distribution according to
the fiducial scaling relation parameters including the full intrinsic
covariance matrix of eq.~(\ref{eq:29}). NFW shear profiles are generated from
the weak-lensing mass set in this way. In the mock catalogues we select the 80
most significant clusters to have \yx\ as follow-up observable. For the
weak-lensing follow-up we either select the 19 most significant clusters or
randomly sample from all significances. In this way we also verify the
independence of the recovered scaling relation parameters on the follow-up
strategy.

We generate mock catalogues for an SPT-SZ-like 2500 sq. deg. survey and for a
survey 10 fold the size of the actual SPT-SZ survey. For all cases we recover
the input scaling relations within $1\sigma$ uncertainty. Additionally, these
mock catalogues allow for predictions about which parameters our dataset will
be able to constrain and choose appropriate priors for those parameters where
the information content is too low to give meaningful constraints.

\subsection{Choices of priors}
\label{sec:choices-priors}
In analyzing the scaling relations described above we aim to put informative
priors only on parameters that our data cannot constrain. In addition to
testing the constraining power of our data by running Monte Carlo chains with
different prior choices we also create mock realisations of the SPT + X-ray +
weak lensing catalogs to ascertain that the real data behave as expected from
these simulations.

The weak lensing bias \bwl\ and the overall scaling of the per cluster bias
factors of the \citetalias{schrabback18} samples are obviously fully
degenerate with the normalisations of the scaling relations we aim to
constrain. We therefore put Gaussian priors with widths corresponding to the
uncertainties obtained from the calibration with simulations on them. Also the
various sources of intrinsic scatter cannot be disentangled by our analysis,
and we fix them using Gaussian priors.

Putting noninformative priors on the mass slopes \bsz, \by, and \bm\ and the
redshift evolution coefficients \csz, \cy, and \cm\, we learn that our data
are not able to obtain meaningful constraints for these parameters. Our mock
catalogues confirm that -- given the current data set -- we should not expect
to be able to constrain these parameters. We therefore choose a Gaussian prior
$\bsz \sim \mathcal{N}(1.63, 0.1^2)$. The mean and central values are
determined by running a full cosmological analysis of the SPT cosmology sample
plus the weak-lensing data sets, similarly to what was done in the recent
SPT-SZ cosmology analysis \citep{dehaan16}. Using the cluster number count
data we constrain the mass slope \bsz, obtaining those values for its mean and
uncertainty. We choose to put flat priors on \by/\bm\ because these are
constrained through their degeneracy with \bsz\ once \bsz\ is fixed.

We use a prior $\cy \sim \mathcal{N}(0.70, 0.35^2)$ to encode our belief that
the X-ray gas in clusters evolves (approximately) self-similarly. These values
correspond to the self-similar exponent $-2/5$ in the form of the \yx--mass
relation chosen by \citet{vikhlinin09b} and allow for 50\% scatter around
self-similarity. We put a flat prior on \csz\ because it is degenerate with
\cy. Likewise for the \mgas\ scaling relation we assume no redshift evolution
with the same uncertainty as for \cy, i.e. we set
$\cm \sim \mathcal{N}(0, 0.2^2)$.

This leaves the normalisations \asz, \ay, and \am\ to be determined. Because
these are the parameters we are chiefly interested in, we put non-informative
priors on them. Specifically, because the scaling relations are linear in
log-space and the non-informative prior on the intercept of a line is flat,
the non-informative prior for the normalisation of a power law is proportional
to $1 / A_i, i \in \{\mathrm{SZ}, \mathrm{Y}, \mathrm{M}\}$.

Finally, we note that the scaling relation parameters are mildly cosmology
dependent. This is due to the distance-redshift relation as well as the
critical density at a given redshift being dependent on cosmology. In our
analysis we marginalize over the uncertainty of the parameters most affecting
these two quantities, $\Omega_\mathrm{m}$, $\sigma_8$, and $H_0$. For the
first two, our prior is a bivariate Gaussian describing the degeneracy between
these parameters, based on the posterior probability distribution of the
cosmology chain of \citet{dehaan16}. For the Hubble constant we choose the
\citet{riess11} value of $H_0 = (73.8 \pm 2.4)\,\mathrm{km\,s^{-1}\,Mpc^{-1}}$
as our prior, which was also utilized in \citet{dehaan16}. We list all priors
in Table~\ref{tab:priors}.

\section{Results and discussion}
\label{sec:results}

\begin{figure*}
  \centering
  \includegraphics[width=\textwidth]{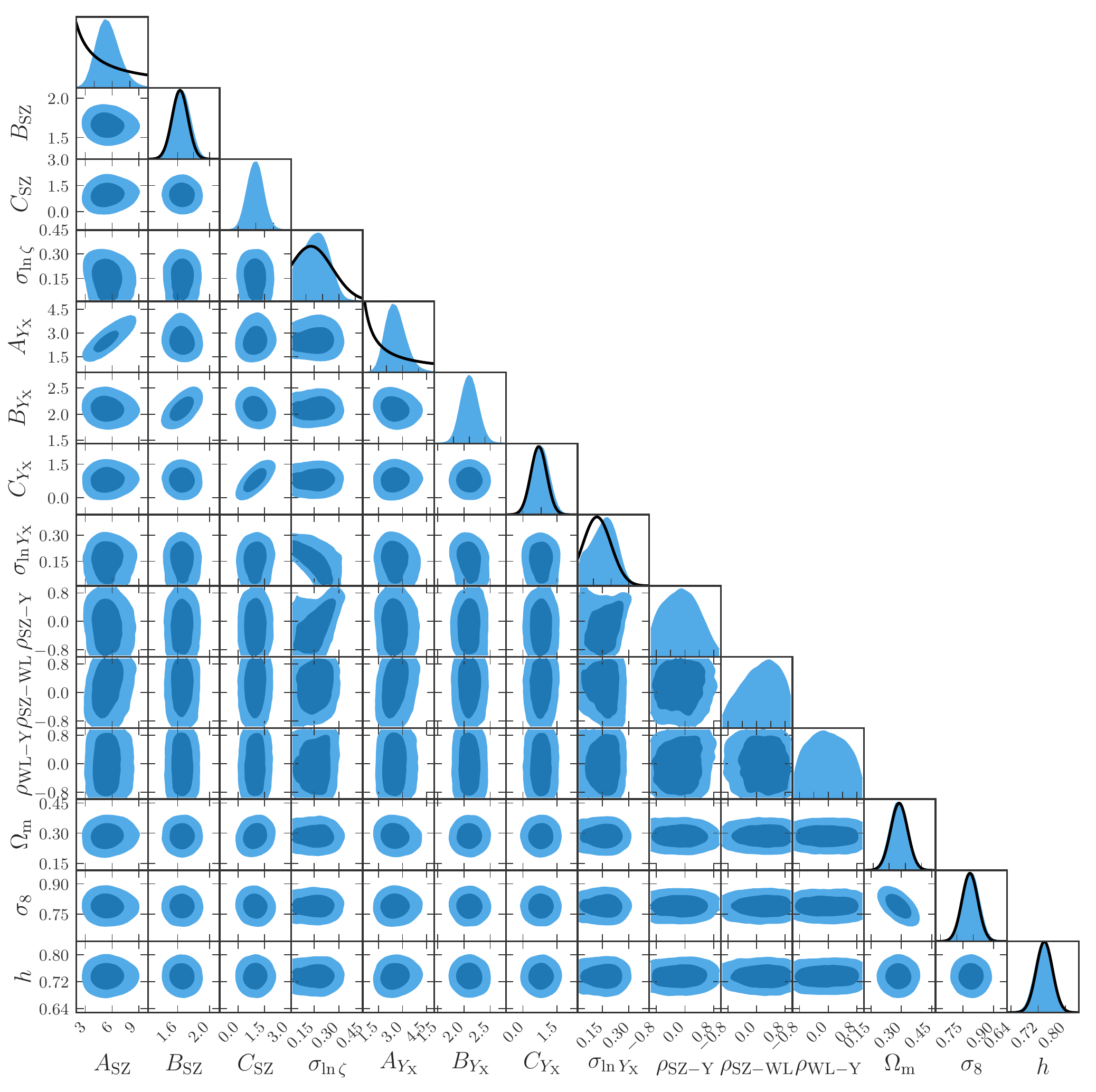}
  \caption{Parameter constraints for the SZE and \yx\ scaling relation
    parameters. Solid black lines are the priors imposed on parameters (see
    Sect.~\ref{sec:choices-priors}). We show here the correlated scatter
    coefficients and the cosmological parameters varied within the prior
    ranges (see Table~\ref{tab:priors}) and omit the lensing nuisance
    parameters due to space constraints. They are shown for the \mgas\ scaling
    relations analysis in Fig.~\ref{fig:contours_Mgas} and are virtually
    identical to the ones omitted here.}
  \label{fig:contours_Yx} 
\end{figure*}

\begin{figure*}
  \centering
  \includegraphics[width=\textwidth]{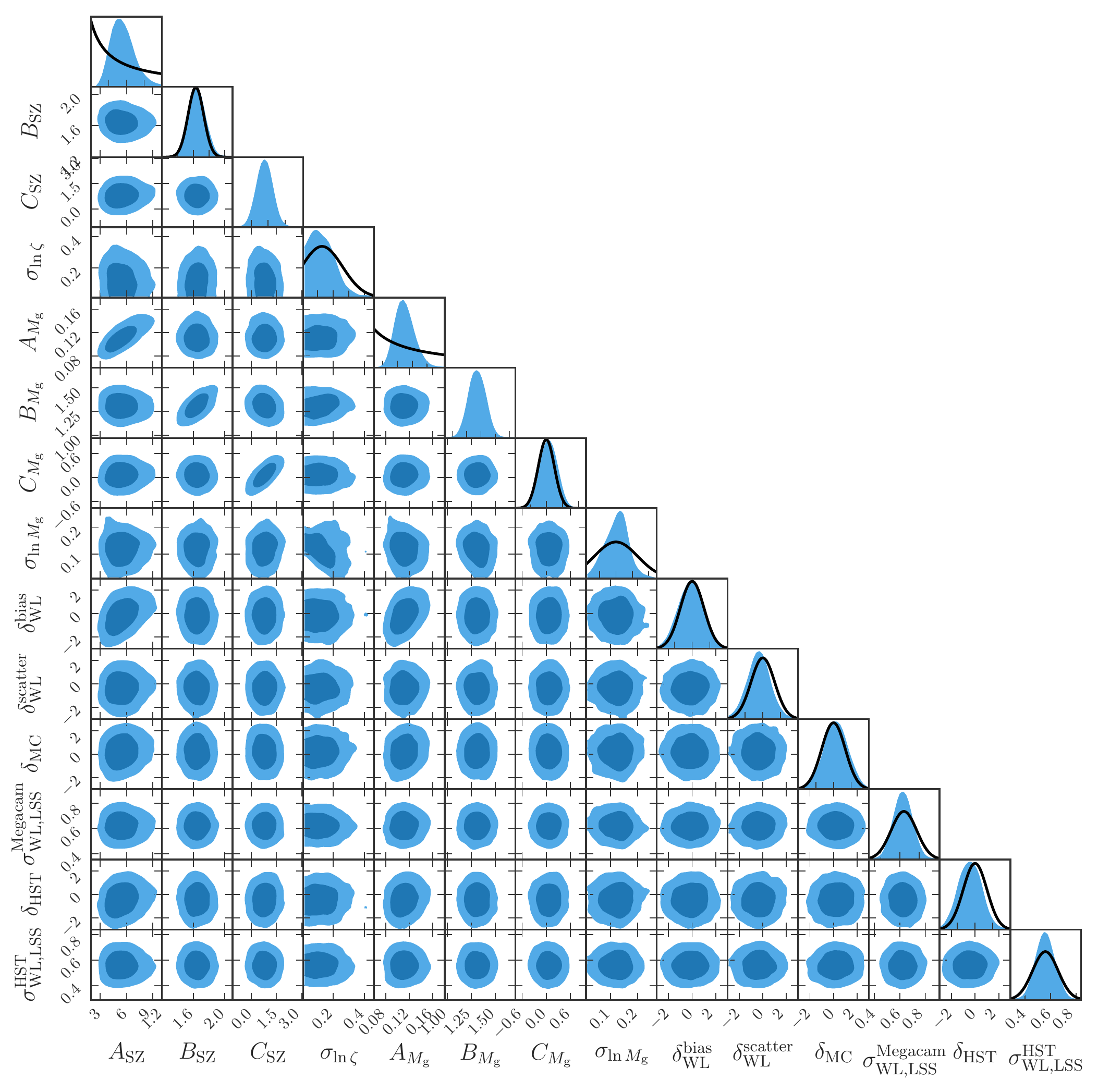}
  \caption{Same as Fig.~\ref{fig:contours_Yx} for the SZE and \mgas\ scaling
    relations. Here we show the lensing nuisance parameters omitted from
    Fig.~\ref{fig:contours_Yx} and omit the correlation coefficients of the
    scatter and the cosmological parameters instead.}
  \label{fig:contours_Mgas}
\end{figure*}

We show parameter constraints for the \yx\ and \mgas\ analyses in
Figs.~\ref{fig:contours_Yx} and \ref{fig:contours_Mgas}, respectively, and
summarize the best fit scaling relations parameters and their 68\% credible
intervals in Table~\ref{tab:scaling-rel-params}. Our key results are the
normalisations of the mass--SZE and mass--X-ray scaling relations, which
directly affect the systematic uncertainty limits of the SPT cluster cosmology
results \citep[Bocquet et al. in prep]{dehaan16}. The best-fit \asz\ values of
the \yx\ and \mgas\ chain are almost identical to each other at
$\asz = 5.56^{+0.96}_{-1.35}$ and $\asz = 5.57^{+0.90}_{-1.41}$, as one would
expect because these numbers are essentially set by the weak-lensing
calibration. We will discuss these results in detail below.

\begin{figure*}
  \centering
  \includegraphics[width=\textwidth]{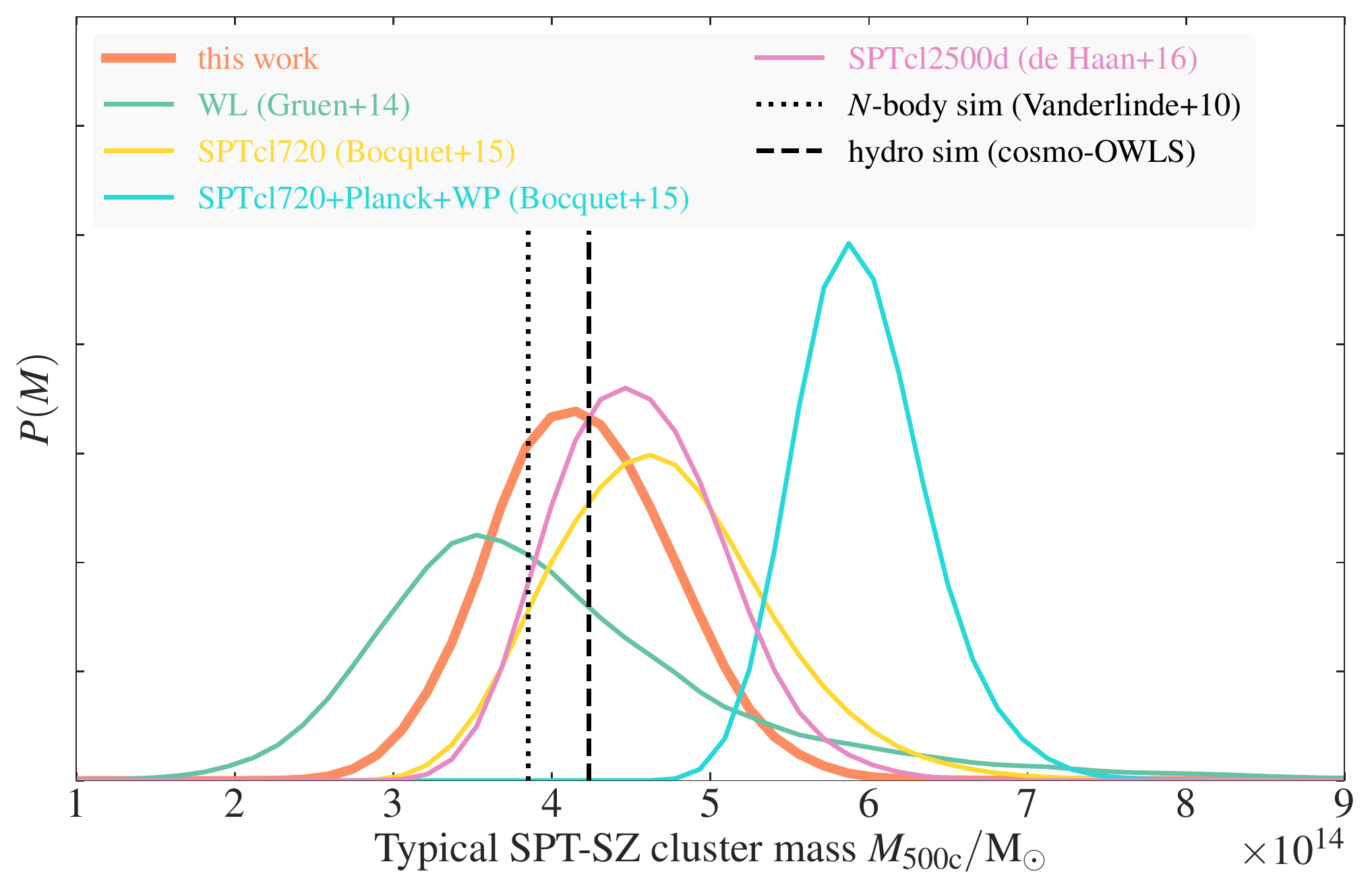}
  \caption{The probability distribution of the mass $M_{500}$ of a typical
    (median) SPT cluster with $\xi=6.5$ at $z = 0.5$ according to different
    mass calibration efforts. The vertical lines correspond to the predictions
    from simulations in \citet[dotted line]{vanderlinde10} and the cosmo-OWLS
    simulation \citep[dashed line]{lebrun14}. The mass scale in this work
    agrees equally well with both simulation predictions.}
  \label{fig:mass-scale}
\end{figure*}

For the mass--SZE scaling relation a comparison to earlier works is best
illustrated by looking at the probability distribution of the mass of a
typical SPT-SZ selected cluster (Fig.~\ref{fig:mass-scale}). Our measurement
of $\asz = 5.56_{-1.35}^{+0.96}$ is in agreement both with the
simulation-based prior of $\asz = 6.01 \pm 1.80$ used in early SPT-SZ work
(\cite{vanderlinde10} who used $N$-body simulations and a gas model from
\citet{shaw09}) and the updated prior $\asz = 5.38 \pm 1.61$ based on the
cosmo-OWLS hydrodynamic simulations \citep{lebrun14} and used in the latest
SPT-SZ cluster cosmology analysis \citep{dehaan16}.

We also compare our value of \asz\ to normalisations obtained from data in other
works. Outside the SPT collaboration, \citet{gruen14} measured weak-lensing
masses of SPT and \textit{Planck} selected galaxy clusters using the
Canada-France-Hawaii Telescope Legacy Survey and pointed follow-up
observations using WFI at the 2.2\,m ESO/MPG telescope. Their
$\asz=6.0_{-1.8}^{+1.9}$ is in excellent agreement with ours. \citet{gruen14}
find a slightly shallower mass slope ($\bsz = 1.25_{-0.28}^{+0.36}$) than we
adopt from the 2500 sq. deg. SPT-SZ cosmology analysis \citep{dehaan16} and
more in line with the expectation from simulations. Our pivot points are,
however, identical so that we can directly compare normalisations, except for
a slight mismatch in \csz, which was also held fixed in their analysis but at
a value of $\csz = 0.83$, which is about $1\sigma$ below our value.

\begin{table*}
  \caption{Marginalized scaling relations parameter constraints for the
    $\zeta$-$\mfive$ scaling relation 
    and the \yx-\mfive\ scaling relation (top half) and the \mgas-\mfive\ 
    scaling relation (bottom half). The values reported are the mean of the
    posterior and the shortest 68\% credible interval.}
  \begin{tabular}{lrlr}\hline
    Parameter & Value & Parameter & Value \\\hline
    \asz & $5.56^{+0.96}_{-1.35}$ & \ay & $2.57^{+0.44}_{-0.67}$\\
    \bsz & $1.656^{+0.092}_{-0.101}$ & \by & $2.11^{+0.14}_{-0.16}$  \\
    \csz & $0.96^{+0.41}_{-0.43}$ & \cy & $0.80^{+0.33}_{-0.35}$ \\
    \dsz & $0.155^{+0.084}_{-0.079}$ & \dy & $0.154^{+0.083}_{-0.065}$ \\\hline
    \asz & $5.57^{+0.90}_{-1.41}$ & \am & $0.112^{+0.012}_{-0.017}$ \\
    \bsz & $1.648^{+0.094}_{-0.103}$ & \bm & $1.310^{+0.080}_{-0.084}$ \\
    \csz & $0.79\pm0.43$ & \cm & $0.06^{+0.19}_{-0.20}$ \\
    \dsz & $0.131^{+0.053}_{-0.100}$ & \dm & $0.120^{+0.044}_{-0.039}$ \\\hline
  \end{tabular}
  \label{tab:scaling-rel-params}
\end{table*}

Our normalisation of the mass--SZE scaling relation is also in good agreement
with earlier SPT work \citep{bocquet15a,dehaan16}. Visually, the largest
disagreement is with the SPT cluster cosmology analysis of \citet{bocquet15a}
when it is combined with the first release of the \textit{Planck} primary CMB
cosmology results \citep{planck_cosmology_14}. The combination of the velocity
dispersion based MOR normalisation constraints of \citet{bocquet15a} with the
CMB data leads to a shift in $\Omega_\mathrm{m}\text{--}\sigma_8$ orthogonal
to the cluster SPT cosmology constraints on these parameters. As a result the
normalisation of the mass-SZE MOR shifts accordingly to account for the
implied different cluster mass scale leading to the difference seen in
Fig.~\ref{fig:mass-scale}. 

For a quantitative comparison we follow \citet{bocquet15a} to compute the
significance of the difference of two distributions. We randomly draw points
from two distributions and compute the difference $\delta$ between pairs of
points. We use this to estimate the probability distribution $P_\delta$ of
these differences and compute the likelihood that zero is within this
distribution. Assuming a normal distribution this likelihood is then converted
to a significance. The lower normalisation parameter \asz, corresponding to
higher cluster masses, inferred from a joint cosmological analysis of the SPT
cluster sample and \textit{Planck} CMB data sets \citep{bocquet15a} disagrees
with our result at the $2.6\sigma$ level.

We emphasise that the change in normalisation in \citet[yellow to cyan line in
Fig.~\ref{fig:mass-scale}]{bocquet15a} when the underlying cosmology shifts is
caused by the self-calibration of the scaling relations from cluster number
counts. In this work we adopt a cosmology with $\Omega_\mathrm{m}, \sigma_8$,
and $H_0$ close to the results of \citet{bocquet15a} \emph{without} the
\textit{Planck} data added. Because we use the cluster mass function only for
the Eddington bias correction and not for self-calibration of the MOR, small
changes in the cosmological parameters do not have any big impact on our
recovered normalization \asz. In particular, changing the cosmological
parameters to the ones obtained from SPT clusters with \textit{Planck} data
\citep{bocquet15a}, changes our normalisation by less than 1\% and does not
bring it into better agreement with their lower \asz\ value.

\begin{figure}
  \includegraphics[width=\columnwidth]{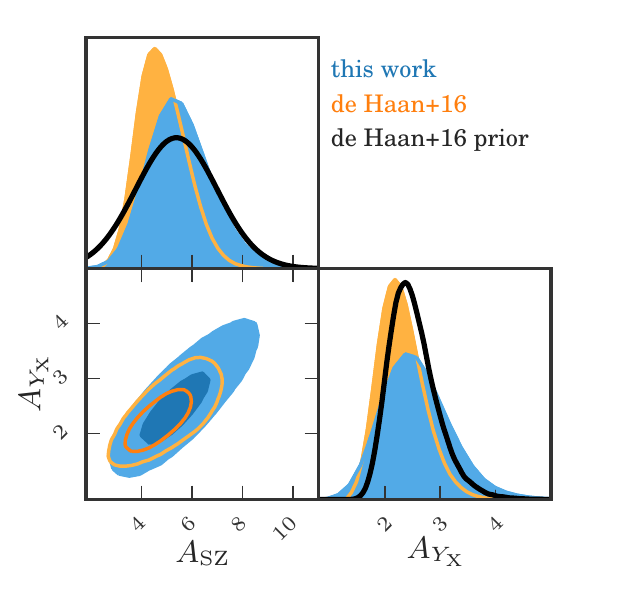}
  \caption{Comparison of our posterior probability distributions of the
    normalisations of the SZE and \yx\ scaling relations with the posterior
    and priors of \citet{dehaan16} converted to our parametrisation of the
    mass--\yx\ scaling relation (\ref{eq:24}).}
  \label{fig:normalisation_comp}
\end{figure}

Also used in the SPT-SZ cosmology analysis is the mass--\yx\ scaling
relation. As for the mass--$\zeta$ relation our marginalized posterior for the
normalisation is in very good agreement with the prior utilised in the
cosmology analysis \citep{dehaan16}. This is an important result as the prior
was based on an \emph{external calibration} of the normalisation of the
mass--\yx\ scaling relation, namely the \citet{vikhlinin09b} scaling relation
updated with the weak lensing mass calibration of the Weighing the Giants
(WtG) and CCCP projects \citep{vonderlinden14, applegate14, hoekstra15}. We
are now able to confirm that these priors were appropriate for the cosmology
analysis based on an \emph{internal
  calibration}. Figure~\ref{fig:normalisation_comp} shows a comparison of the
marginalized and joint posterior probability distributions for the
normalisations \asz\ and \ay\ in comparison with the results obtained by
\cite{dehaan16} and the priors used in this previous SPT work. Our posterior
distributions are a little broader than theirs and consequently we do not yet
expect that our mass calibration efforts will lead to tighter cosmological
constraints with the current data set (Bocquet et al., in prep.). We note,
however, that the width of the \asz\ posterior distributions of
\citet{dehaan16} is narrower than their prior range. This indicates that their
constraint on the MOR normalisation benefits from self-calibration. We do not
use this self-calibration from the number counts of galaxy clusters (see
Sect.~\ref{sec:scal-rel-pipeline}) and thus obtain broader posterior
distributions given the still relatively small sample of SPT clusters with
weak-lensing information.

\begin{figure}
  \includegraphics[width=\columnwidth]{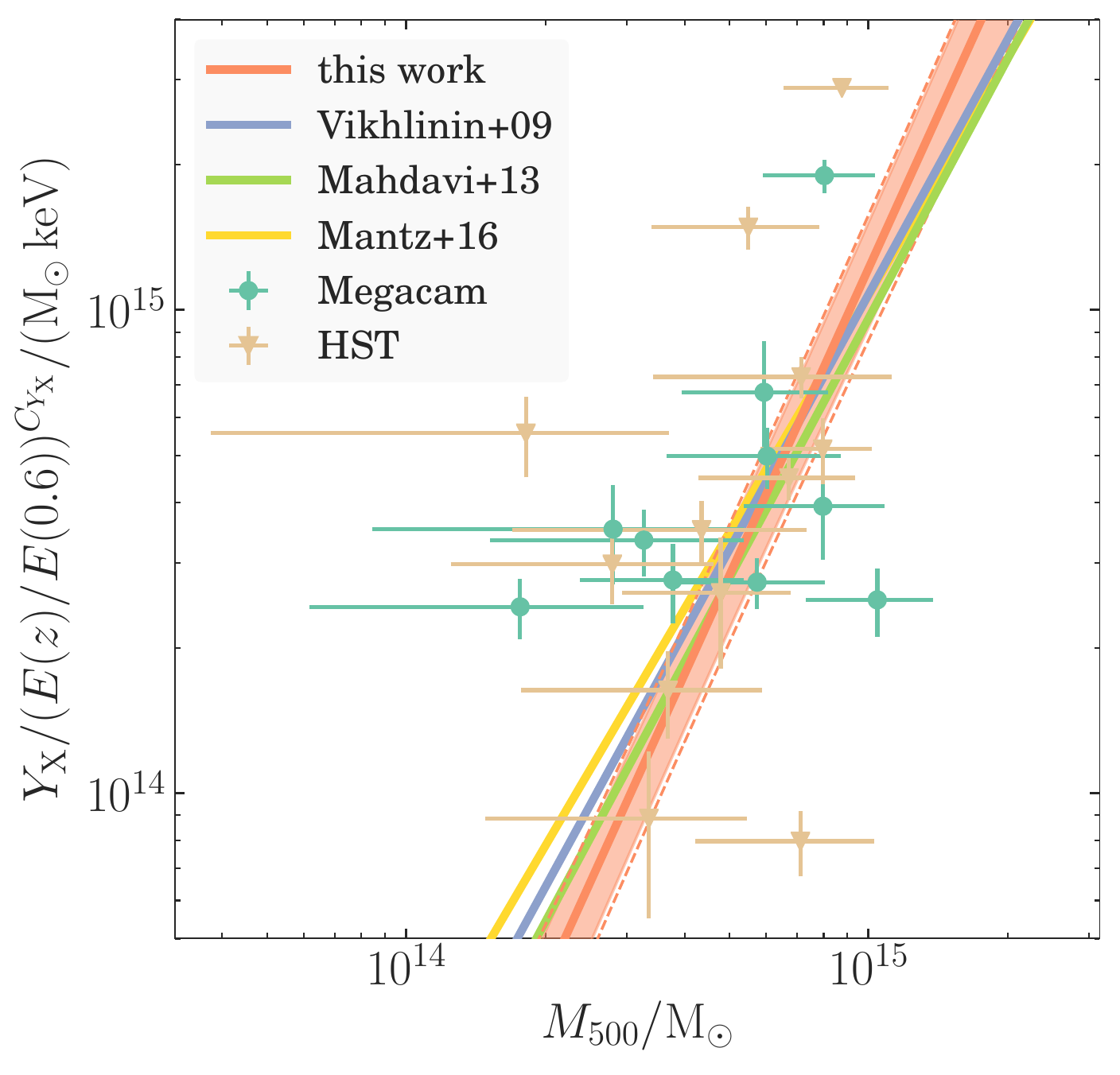}
  \caption{\yx-weak lensing mass scaling relation. Our result is shown in red,
    the \yx-mass relations of \citet{vikhlinin09b}, \citet{mahdavi13}, and
    \citet{mantz16} are shown in blue, green, and yellow, respectively, for
    comparison. The \yx\ values are evaluated at the reference cosmology and
    best fit scaling relations parameters. We extract values for
    $M_\mathrm{WL}$ and \yx\ as described in the text. The masses on the
    $x$-axis are debiased according to eq.~(\ref{eq:19}). We note that the
    values displayed here are only used for illustration purposes; the
    analysis pipeline does not use these values but follows the forward
    modelling approach described in Sect.~\ref{sec:scal-rel-pipeline}. The
    horizontal errorbars take only the shape noise component into account. The
    vertical error bars are also only the observational error. The 19 clusters
    observed with Megacam are shown as green circles, while the 13 clusters
    observed with HST \citep{schrabback18} are shown as light brown
    triangles. Shaded regions indicate the uncertainty on our best fit
    parameters. The dashed lines indicate the best fit intrinsic scatter
    added in quadrature to the parameter uncertainties. We show the intrinsic
    scatter only for our result; its contribution to the other scaling
    relations is almost exactly the same.}
  \label{fig:mwl-yx-scal-rel}
\end{figure}

\begin{figure}
  \includegraphics[width=\columnwidth]{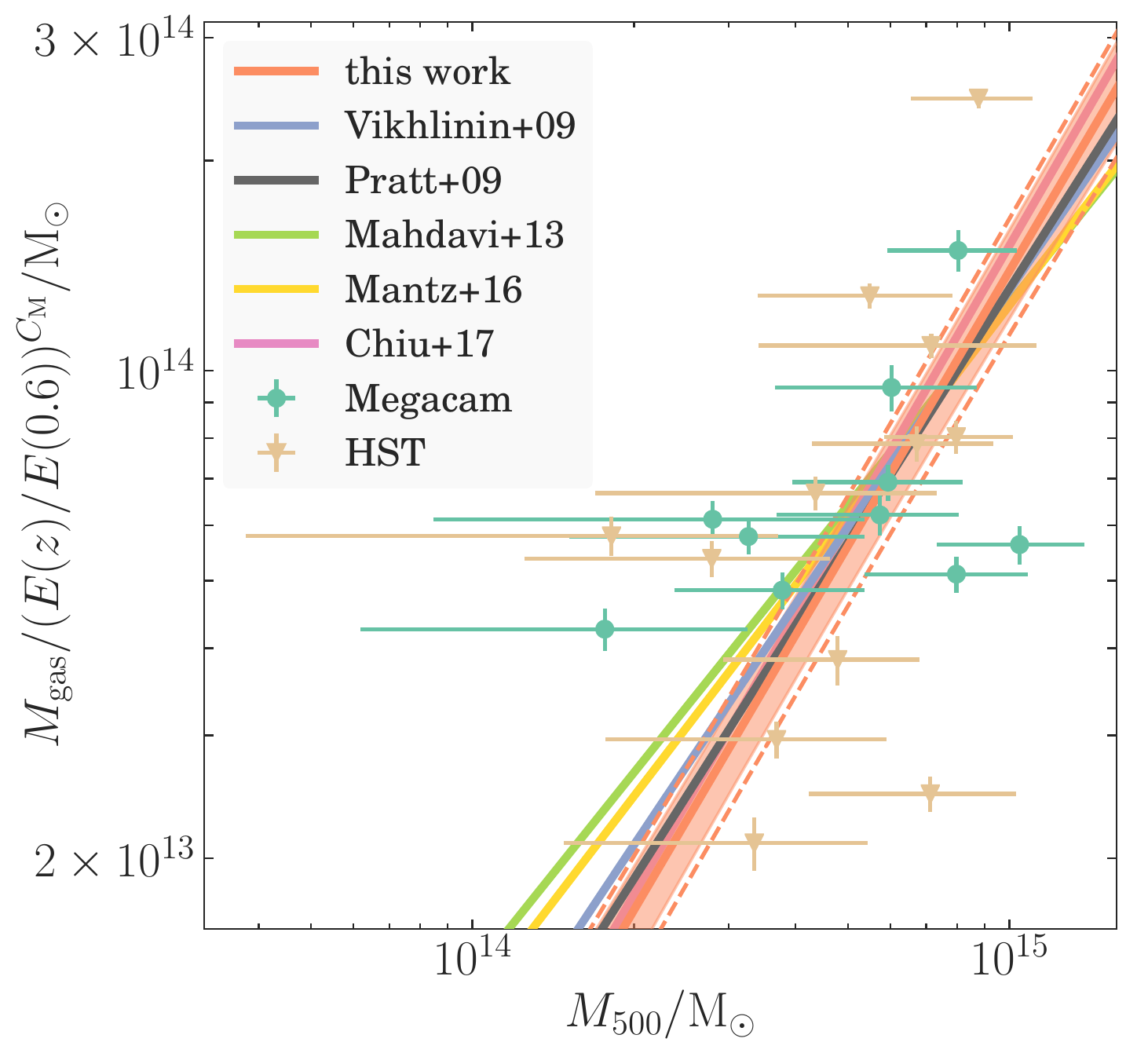}
  \caption{Same as Fig.~\ref{fig:mwl-yx-scal-rel} but for the \mgas-mass
    scaling relation. We compare our best fit relation, where the slope was
    set by a galaxy cluster number counts analysis \citep{dehaan16}, with that
    of \citet{vikhlinin09b}, \citet{pratt09}, \citet{mahdavi13}, the WtG team
    \citep{mantz16}, and \citet{chiu17}.}
  \label{fig:mwl-mgas-scal-rel}
\end{figure}

The mass-slope and redshift evolution parameters follow their prior
probability distribution in the observable on which an informative prior was
imposed. The \by\ constraint is determined purely by the \bsz-\by
degeneracy. Likewise the \csz\ constraint is governed by the \csz-\cy (\cm)
degeneracy. The \csz\ values derived in this way differ by $0.3\sigma$ between
the \yx\ (\csz = $0.96^{+0.41}_{-0.43}$) and \mgas\
($\csz = 0.79\pm0.43$) chains. Both values are higher than the \csz\
prior in \citet{dehaan16}, but even the higher $\csz(\yx)$ value deviates by
only $0.5\sigma$. All \csz\ posteriors of \citet{dehaan16} agree with our
values at better than $1\sigma$.

Our modelling of the weak-lensing bias and scatter (Sect.~\ref{sec:sze-x-ray})
introduces numerous nuisance parameters that we are not able to constrain with
the data. They all follow the priors. Similarly we are not able to distinguish
between various sources of scatter in our data. As the degeneracy between
\dsz\ and \dy\ (\dm) shows, we are only able to put limits on the sum of their
squares, i.e. the total scatter of the scaling relations.

It is expected from numerical simulations that the intrinsic scatter of the
weak-lensing and SZE measurements are correlated
\citep[e.g.,][]{shirasaki16}. For the current data, however, we cannot
constrain any of the three correlation coefficients. Furthermore, for all
parameters the constraints obtained by leaving the correlation coefficients
free are indistinguishable from those where we set all correlation
coefficients to zero.

\begin{figure}
  \includegraphics[width=\columnwidth]{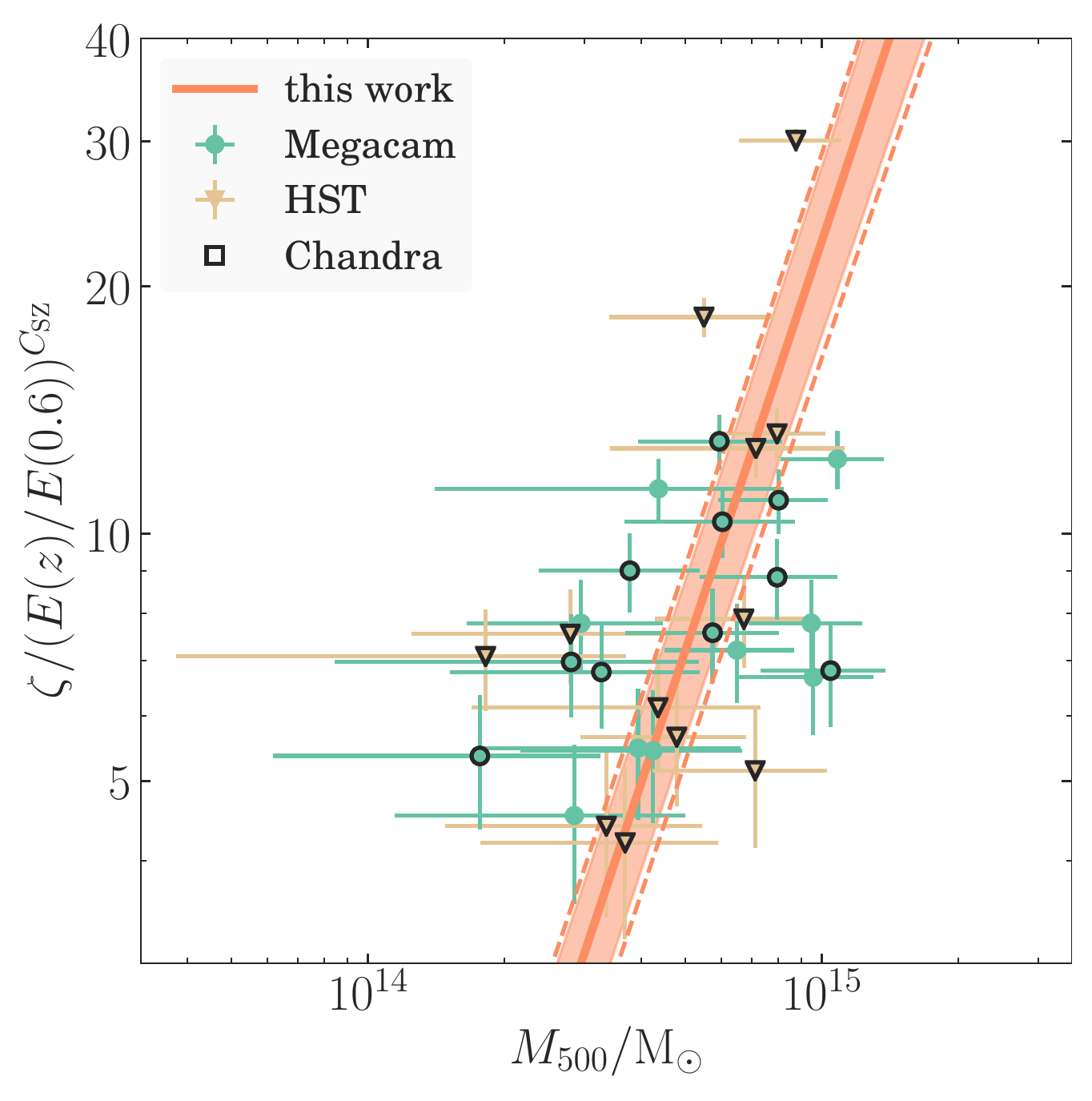}
  \caption{The $\zeta$--mass scaling relation and estimates $\hat{\zeta}$ and
    $\mfive$ for the 32 clusters with weak lensing data. Points marked in
    black are clusters with \textit{Chandra} X-ray data used in the scaling
    relation analysis, i.e. all clusters shown in
    Figs.~\ref{fig:mwl-yx-scal-rel} and~\ref{fig:mwl-mgas-scal-rel}.}
  \label{fig:mwl-zeta-scal-rel}
\end{figure}

In Figs.~\ref{fig:mwl-yx-scal-rel}--\ref{fig:mwl-zeta-scal-rel} we show the
scaling relations (\ref{eq:23})--(\ref{eq:25}) with marginalised uncertainties
in comparison to the data points. In these plots green circles indicate the 19
clusters followed-up with Megacam while light brown triangles are the 13 HST
observations from \citetalias{schrabback18}. In all of these figures, the
distributions of the Megacam and HST data points appear to be consistent with
each other. This visual impression is confirmed by finding consistent
normalisations of the scaling relations when the sample is split in redshift
at $z=0.6$ (Bocquet et al., in prep.).

\begin{figure}
  \includegraphics[width=\columnwidth]{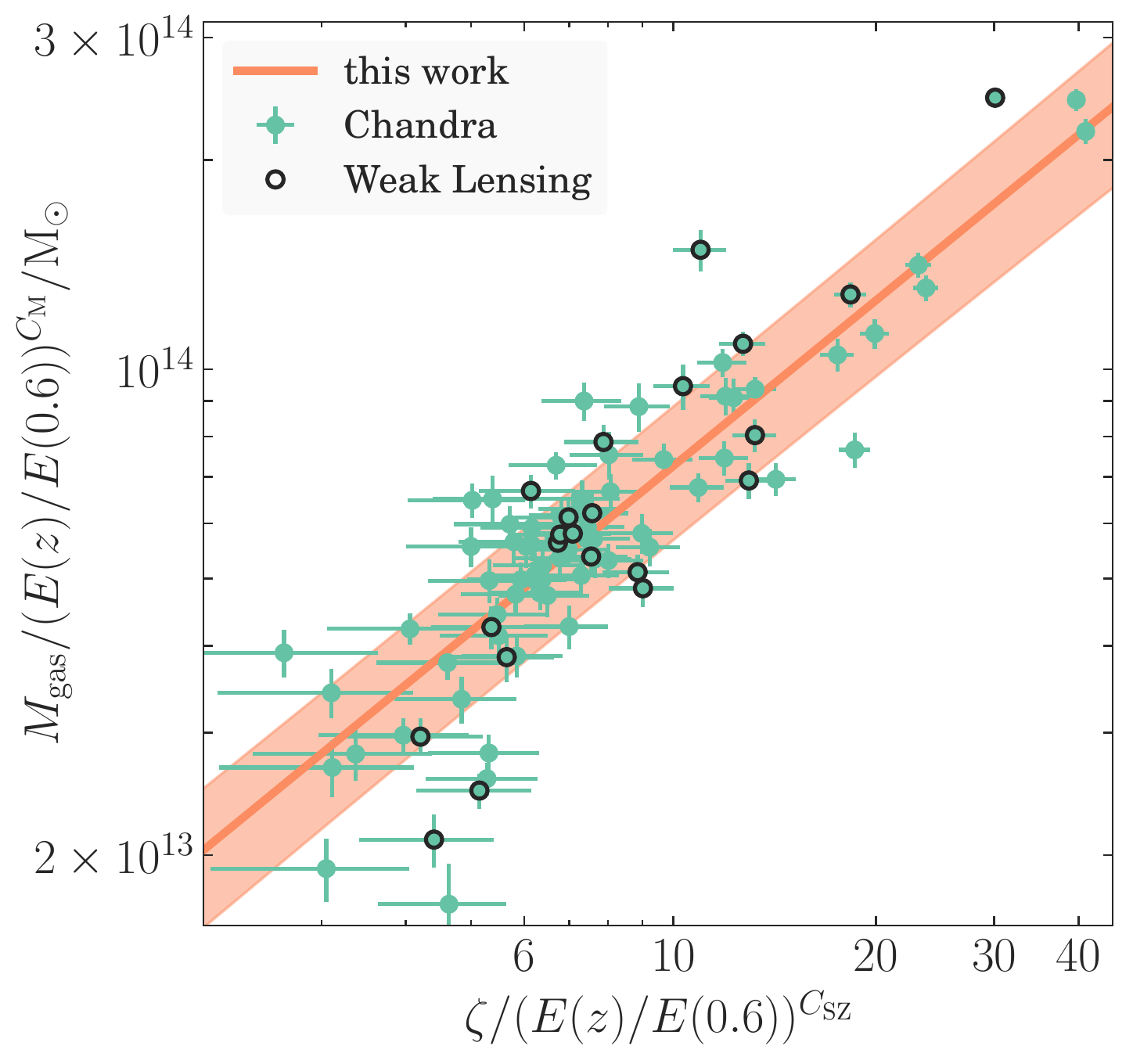}
  \caption{The derived $\zeta\text{--}\mgas$ scaling relation
    (eq.~(\ref{eq:33})). Cluster data points show the redshift evolution
    corrected estimate $\hat{\zeta}$ (eq.~\ref{eq:32}). We show only the
  parameter uncertainty and not the intrinsic scatter for this relation.}
  \label{fig:zeta-mgas-scal-rel}
\end{figure}

Note that we do not directly observe the weak lensing mass $M_\mathrm{wl}$,
and that the X-ray observables are radial profiles and not scalar
quantities. In the following, we briefly describe how we extracted the
quantities displayed in the figures. The X-ray measurements consist of a
temperature measurement of the hot ICM and a radial gas mass profile
$\mgas(r)$. Both quantities can be combined to give the radial $\yx(r)$
profile. In principle the temperature also varies radially but this is slow
enough to be accurately approximated by a global average temperature. In the
case of \mgas, the scaling relation~(\ref{eq:25}) relates the gas mass to the
cluster total mass \mfive, from which the radius $r_{500}$ can be uniquely
determined. Assuming the best-fit scaling relation parameters for the
\mgas--mass relation, we can now solve for \mgas\ by solving the implicit
equation $\mgas^\text{data}(r) = \mgas^\text{MOR}(r)$. We can then obtain the
mean and standard deviation of the recovered distribution in Mgas for
Figs.~\ref{fig:mwl-mgas-scal-rel} and~\ref{fig:zeta-mgas-scal-rel}. The same
procedure is used for \yx\ and eq.~(\ref{eq:24}) and
Fig.~\ref{fig:mwl-yx-scal-rel}.

Our likelihood framework also never computes a weak lensing mass that best
fits the observed radial shear profile $\gt(r)$. Instead it computes how
probable it is to find the observed shear profile given the mass predicted
from the scaling relations. To nevertheless be able to plot weak lensing
masses we perform maximum likelihood fits to the contamination corrected shear
profiles and use their location and uncertainty when plotting weak lensing
masses.

Furthermore, when we plot cluster data points in $\zeta$--mass scaling
relations we also need an estimate of $\zeta$ for each cluster. We obtain this
from the observable SNR $\xi$ via
\begin{equation}
  \label{eq:32}
  \hat{\zeta} = \sqrt{\xi^2 - 3} / f_\mathrm{field} \;,
\end{equation}
where $f_\mathrm{field}$ is a scaling factor to correct for the different
depths of fields in the 2500\,sq.\,deg. SPT-SZ survey.

Figures~\ref{fig:mwl-yx-scal-rel}--\ref{fig:zeta-mgas-scal-rel} show the
predicted scaling relations for the underlying cluster population and are not
corrected for our SZE selection. This is most obvious in
Fig.~\ref{fig:zeta-mgas-scal-rel} where the two low-scatter mass proxies
$\hat{\zeta}$ and $\mgas$ are plotted against each other for a cluster
population selected in $\xi$. The Eddington bias is clearly visible in the
lower left corner of this plot from the points falling below the best fit
line, i.e. they are preferentially scattered towards higher $\hat{\zeta}$. We
remind the reader that the scaling relation analysis takes this bias into
account through the shape of the mass function and the SPT cluster selection
function. The scaling relation plotted in Fig.~\ref{fig:zeta-mgas-scal-rel} is
obtained by combining eqs.~(\ref{eq:23}) and~(\ref{eq:25}) into
\begin{equation}
\label{eq:33}
  \frac{\mgas}{5 \times 10^{14}\,\msun} = \am  \left(\frac{6}{7}\right)^\bm 
  \left(\frac{\zeta}{\asz}\right)^{\bm / \bsz}\;,
\end{equation}
and omitting the redshift evolution terms, because they are taken care of when
the plotted data are rescaled to a common redshift.

Our estimates for the normalisations of the X-ray scaling relations show good
agreement with previous studies \citep{vikhlinin09b, pratt09, mahdavi13,
  mantz16}. For the mass--\yx\ relation this holds over the entire mass range
under investigation here. For the mass--\mgas\ relation the sometimes
significantly different slopes lead to good agreement only in the vicinity of
our pivot point $M_\mathrm{p} = 5\times 10^{14}\,\hm\,\msun$ and marginal
discrepancies at the extreme ends of the mass range under investigation
here. This is particularly obvious for the relations of \citet{mahdavi13}, who
find a slope slightly smaller than but consistent with self-similarity, and
\citet{mantz16}, whose slope is very nearly exactly self-similar. However, at
our pivot $\mfive^\mathrm{piv} = 5 \times 10^{14}\,\msun$ we agree with all
cited studies within our mutual uncertainties.

We note again that we are not able to constrain the slope \bm\ with our present
data set. Rather our value for the slope is determined by the prior we put on
\bsz\ -- based on the cosmology analysis of \citet{dehaan16} -- and the
degeneracy between \bsz\ and \bm. Future weak lensing analyses of SPT selected
clusters covering a wider $\xi$ and thus mass range will enable us to
constrain the slope directly from weak lensing observations instead of only
through self-calibration in a cosmological framework, as in \citet{dehaan16}
and \citet{mantz16}. 

In Fig.~\ref{fig:mwl-zeta-scal-rel} we show the scaling relation between
cluster mass and debiased SPT detection significance $\zeta$. In this plot we
also highlight those clusters with \textit{Chandra} data used in the scaling
relation analysis. We find no indication that the 10 clusters from the Megacam
sample without X-ray follow-up come from a different population.

Finally, we compare our mass estimate for the stack of all 19 Megacam clusters
to that of a previous study using gravitational magnification instead of shear
\citep{chiu16b}, who found a mass estimate of
$\mfive = (5.37 \pm 1.56) \times 10^{14}\,\msun$. This is in very good
agreement with our weighted mean mass
$\mfive = (5.96 \pm 0.61) \times 10^{14}\,\msun$ for these clusters.

\section{Conclusion}
\label{sec:conclusion}
In this paper we describe the observations and weak lensing analysis of 19
clusters from the 2500~sq.~deg. SPT-SZ survey. We pay particular attention to
controlling systematic uncertainties in the weak lensing analysis and provide
stringent upper limits for a large number of systematic uncertainties and
avoided confirmation bias by carrying out a blind analysis. The upper limit of
our total systematic error budget is $5.4\%$ (68\% confidence) and is dominated
by uncertainties stemming from the modelling of haloes as NFW profiles.

We used $N$-body simulations to calibrate our mass modelling method. The
sources of systematic errors in this approach are the uncertainty in this
calibration, the mass--concentration relation, and the miscentering
distribution. Future analyses could mitigate these either by employing a
larger suite of simulations and an improved understanding of the sources of
discrepancies of published $M\text{--}c$ relations, or by using other mass
estimators that avoid these complications. \citet{hoekstra15} for example used
the aperture mass \citep{fahlman94,schneider96} to mitigate these
problems. This, however, is done at the cost of increased statistical
uncertainties, so that future studies will have to carefully weigh the cost
and benefits of using either the aperture mass or an NFW modelling
approach. We emphasise that in our present work we are still dominated by
statistical and not by systematic errors. The total uncertainty (systematic
and statistical) on the mass scale is $\sigma_{\mtwo} = \asz^{1/\bsz} = 8.9\%$.

We combined the weak-lensing data of our 19 clusters with those of 13 clusters
from the SPT-SZ survey at high redshift observed with HST
\citepalias{schrabback18} and \textit{Chandra} X-ray data to calibrate
mass--observable scaling relations. We described an extension of the scaling
relations framework of \citet{bocquet15a} to include weak lensing
information. An important feature of our method is its ability to correct for
Eddington bias while at the same time not using cluster number counts to
self-calibrate mass--observable relations.

The normalisation of the mass--SZE relation is in good agreement with the prior used in
the latest SPT cosmology analysis \citep{dehaan16}, which are based on an external
calibration of this mass--observable relation. Future SPT cosmology analyses (Bocquet et
al., in prep.) will now be able to use an internal calibration of the absolute mass
scale, i.e. a calibration that is performed on the same clusters used for obtaining
cosmological constraints. Also, our values for the normalisations of the mass--X-ray
scaling relations all agree within $1\sigma$ with those found by other authors
\citep{vikhlinin09b, pratt09, mahdavi13, mantz16, chiu17}. For example, at
$\yx = 5\times10^{14}\,\msun\,\mathrm{keV}$ our \mfive\ normalisation is $2.4\%$ higher
than that of \citet{vikhlinin09b} and $6.3\%$ lower than that of \citet{dehaan16}. At
$\mgas = 6\times10^{13}\,\msun$ we obtain \mfive\ values $4.6\%$ higher than
\citet{vikhlinin09b}.

At the same time our choice to avoid self-calibration of the mass--observable
scaling relation from cluster number counts limits our ability to constrain
the slopes and evolution parameters of these relations with a cluster sample
of the present size. We therefore chose to impose informative priors on these
quantities based on the self-calibration results of the SPT-SZ cluster
cosmology analysis.

We already have secured more follow-up data, including HST data, so that we
can expect to overcome this limitation in the near future. Particularly, the
planned combination of SPT-SZ data with the shear catalogues of the DES survey
\citep{zuntz17} combined with an expanded SZE cluster sample from the SPTpol
experiment \citep{austermann12} should allow us to extract meaningful
constraints on the slope of the mass--SZE scaling relation and lead to a more
stringent estimation of the mass--observable scaling relation normalisations.

\section*{Acknowledgements}
\label{sec:acknowledgments}
We acknowledge the support by the DFG Cluster of Excellence ``Origin and
Structure of the Universe'' and the Transregio program TR33 ``The Dark
Universe''. The emcee runs have been carried out on the computing facilities
of the Computational Center for Particle and Astrophysics (C2PAP), located at
the Leibniz Supercomputer Center (LRZ) in Garching. The South Pole Telescope
is supported by the National Science Foundation through grant
PLR-1248097. Partial support is also provided by the NSF Physics Frontier
Center grant PHY-1125897 to the Kavli Institute of Cosmological Physics at the
University of Chicago, the Kavli Foundation and the Gordon and Betty Moore
Foundation grant GBMF 947. DA and TS acknowledge support from the German
Federal Ministry of Economics and Technology (BMWi) provided through DLR under
projects 50 OR 1210, 50 OR 1308, 50 OR 1407, and 50 OR 1610. Work at Argonne
National Laboratory was supported under U.S. Department of Energy contract
DE-AC02-06CH11357. DR is supported by a NASA Postdoctoral Program Senior
Fellowship at NASA's Ames Research Center, administered by the Universities
Space Research Association under contract with NASA. CR and SR acknowledge
support from the Australian Research Council’s Discovery Projects funding
scheme (DP150103208). A.~Saro is supported by the ERC-StG ‘ClustersXCosmo’,
grant agreement 71676. We thank Peter Capak for providing the
COSMOS30+UltraVISTA photo-$z$ catalogue ahead of publication. Based on data
products from observations made with ESO Telescopes at the La Silla Paranal
Observatory under ESO programme ID 179.A-2005 and on data products produced by
TERAPIX and the Cambridge Astronomy Survey Unit on behalf of the UltraVISTA
consortium. This research made use of \textsc{Astropy}, a community-developed
core Python package for Astronomy \citep{astropy2013} and \textsc{topcat}
\citep{taylor05}. Figures~\ref{fig:contours_Yx}, \ref{fig:contours_Mgas}, and
\ref{fig:normalisation_comp} were created with \textsc{pyGTC}
\citep{bocquet16b}.

\bibliography{../../BIBTEX/oir.bib}

\appendix
\section{PSF residual and shear systematic figures}
\label{sec:psf-residual-shear}
\begin{figure*}
  \centering
  \includegraphics[width=\textwidth]{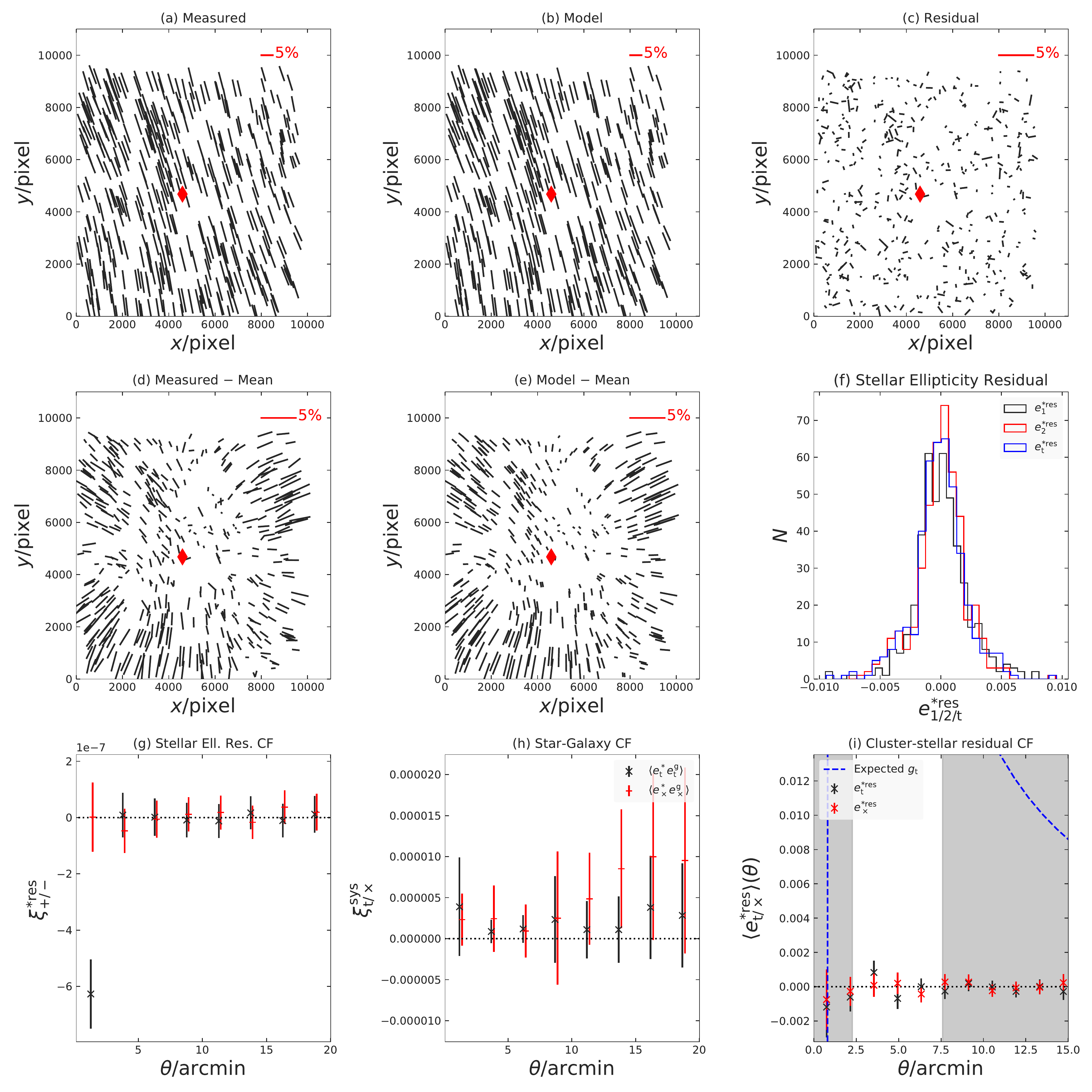}
  \caption{Same as Fig.~\ref{fig:psf-diagnostics-2031_2} for a
    typical, randomly chosen exposures. In this case exposure 2 of
    SPT-CL\,J0234$-$5831.}
  \label{fig:psf-diagnostics-0234_2}
\end{figure*}
\clearpage

\newpage
\section{Mass reconstructions and shear profiles}
\label{sec:mass-reconstr-prof}

\begin{figure*}
   \subfigure[Surface mass density of SPT-CL\,J0234$-$5831.]{
   \includegraphics[width=\columnwidth]{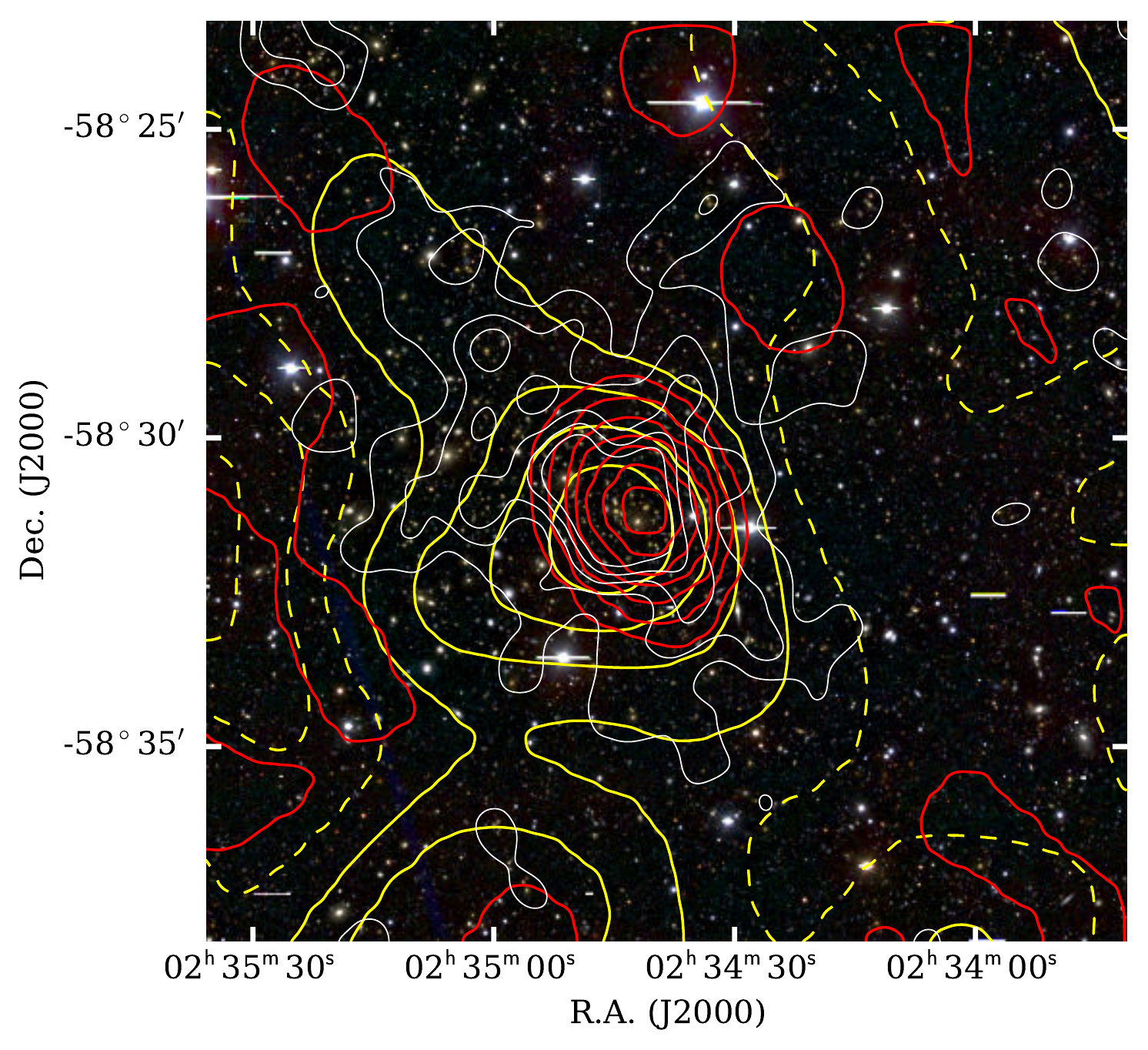}
   \label{fig:1}}
   \hfill
   \subfigure[Tangential shear profile of SPT-CL\,J0234$-$5831.]{
   \includegraphics[width=\columnwidth]{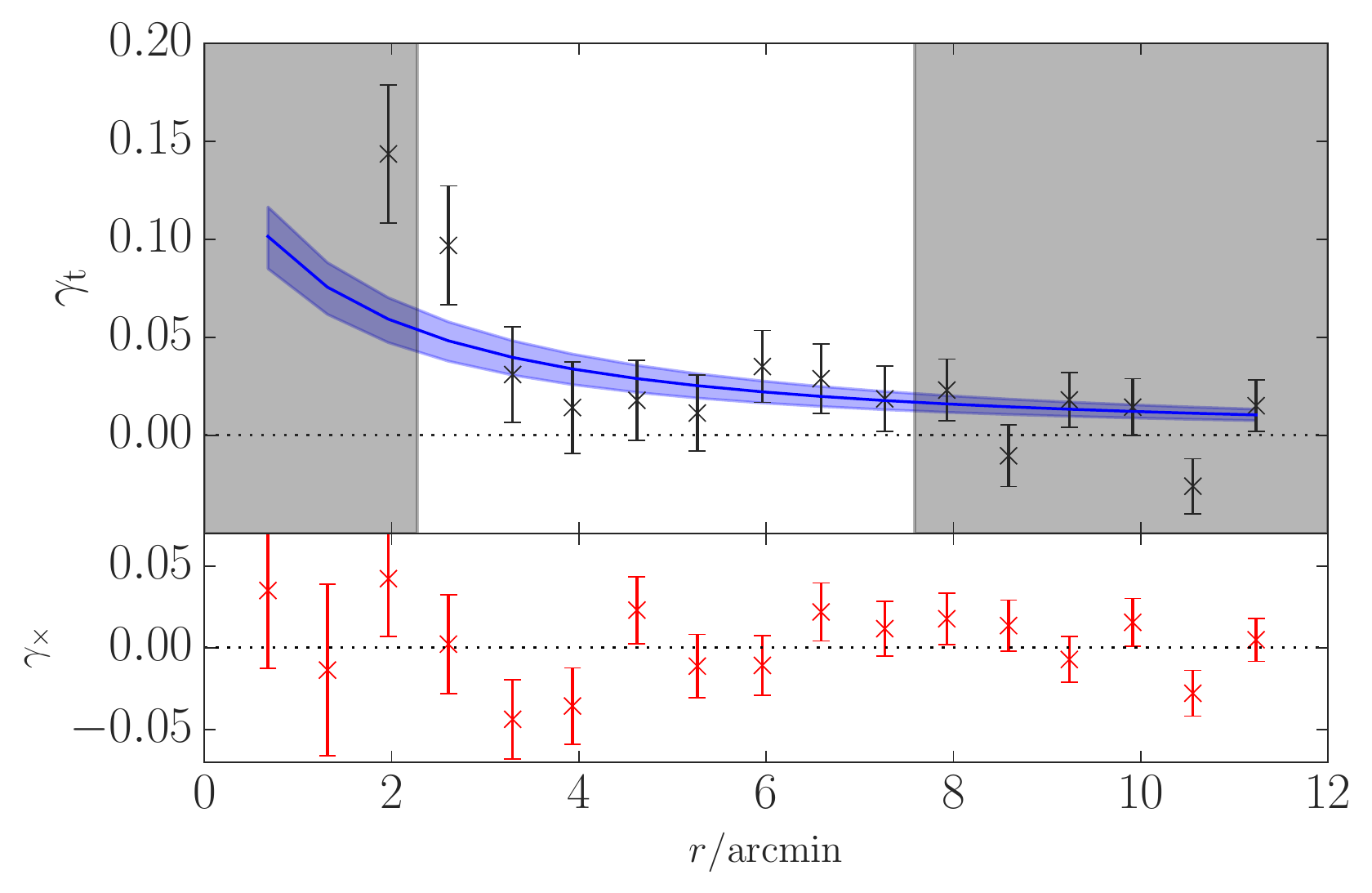}
   \label{fig:gammat0234}}
   \input{first_shear_kappa_caption}
   \label{fig:kappa-shear-0234}
 \end{figure*}

\begin{figure*}
   \subfigure[Surface mass density of SPT-CL\,J0240$-$5946.]{
   \includegraphics[width=\columnwidth]{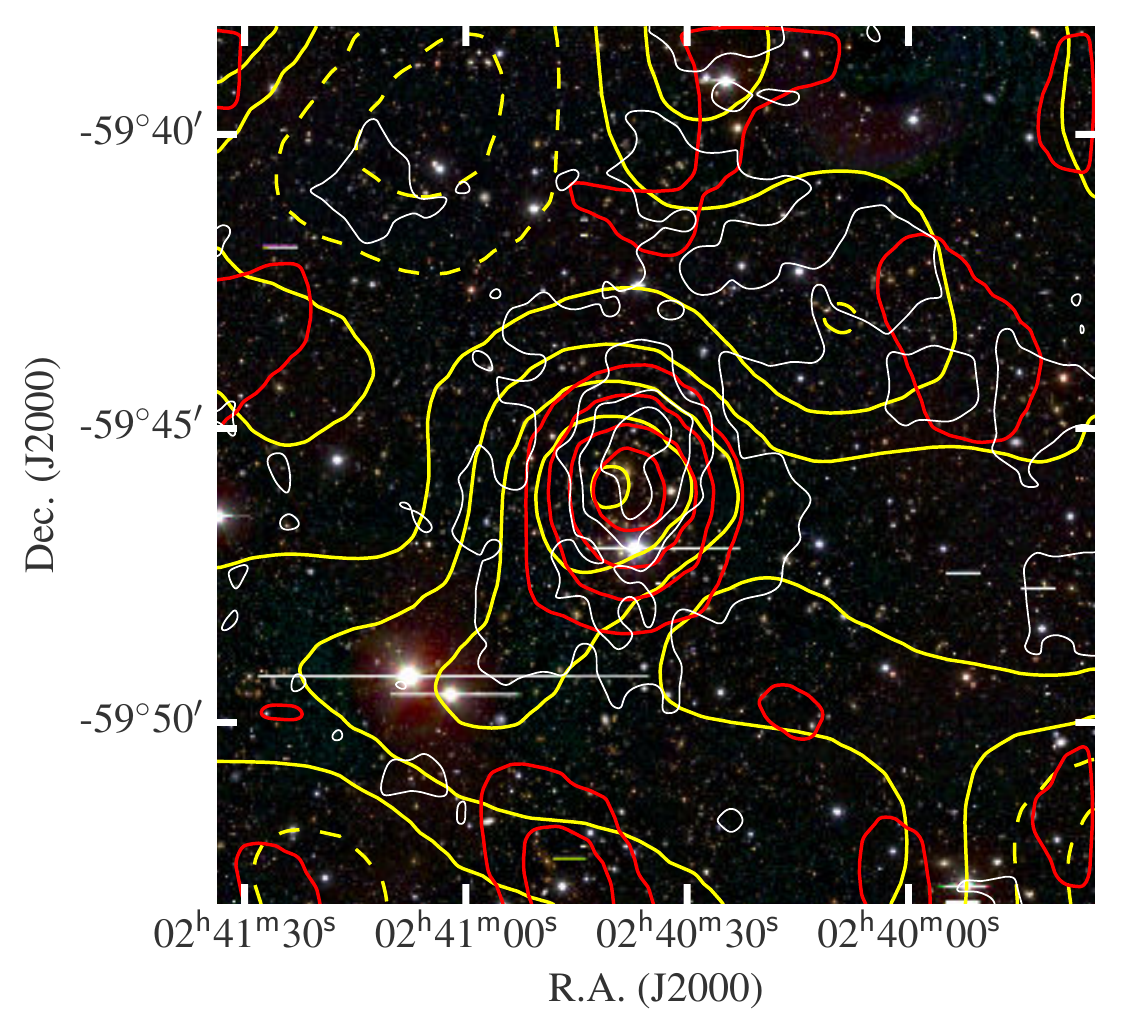}
   \label{fig:2}}
   \hfill
   \subfigure[Tangential shear profile of SPT-CL\,J0240$-$5946.]{
   \includegraphics[width=\columnwidth]{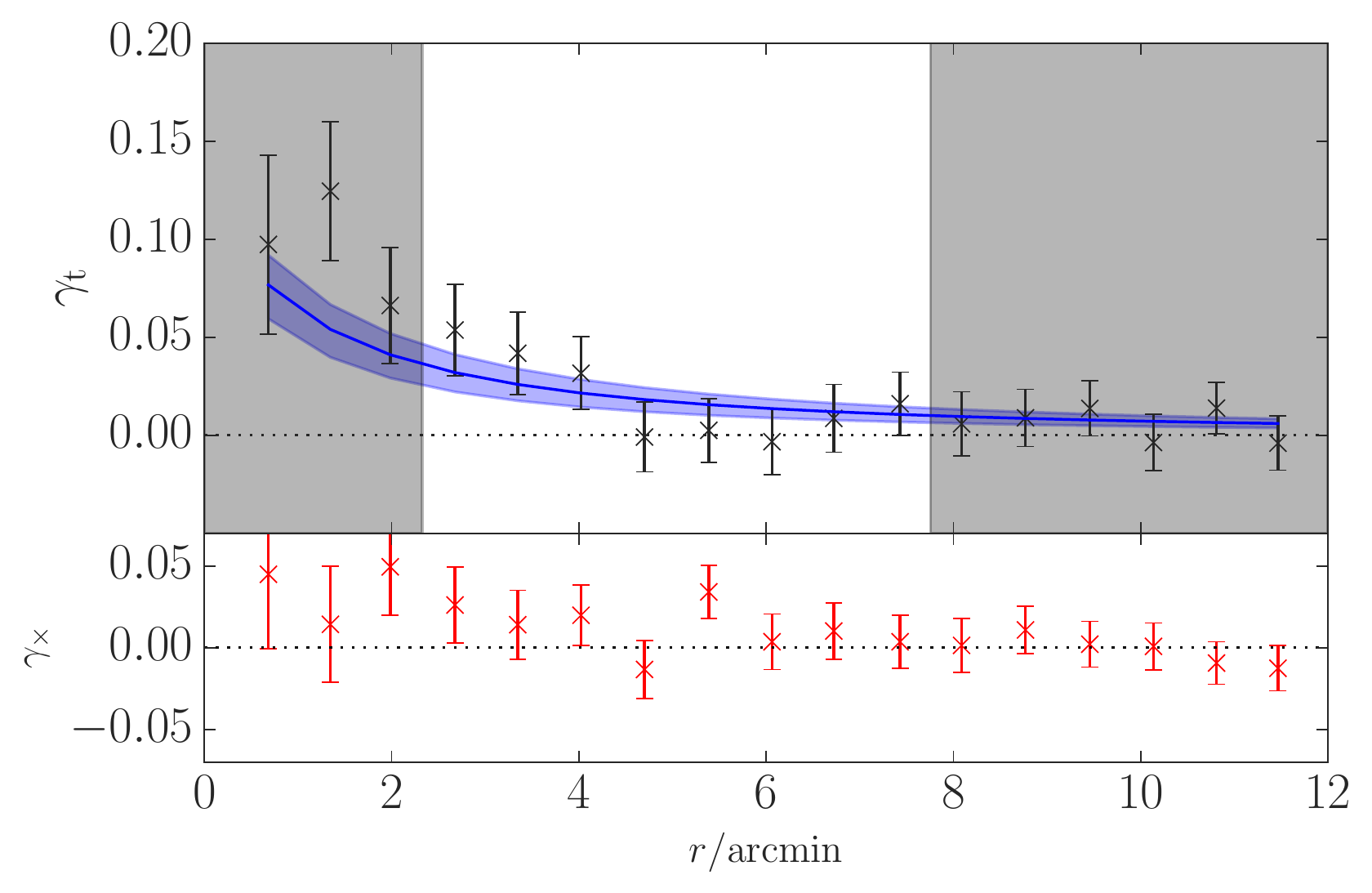}
   \label{fig:gammat0240}}
   \caption{Same as Figure~\ref{fig:kappa-shear-0234} for SPT-CL\,J0240$-$5946.}
   \label{fig:kappa-shear-0240}
 \end{figure*}

\clearpage
\begin{figure*}
   \subfigure[Surface mass density of SPT-CL\,J0254$-$5857.]{
   \includegraphics[width=\columnwidth]{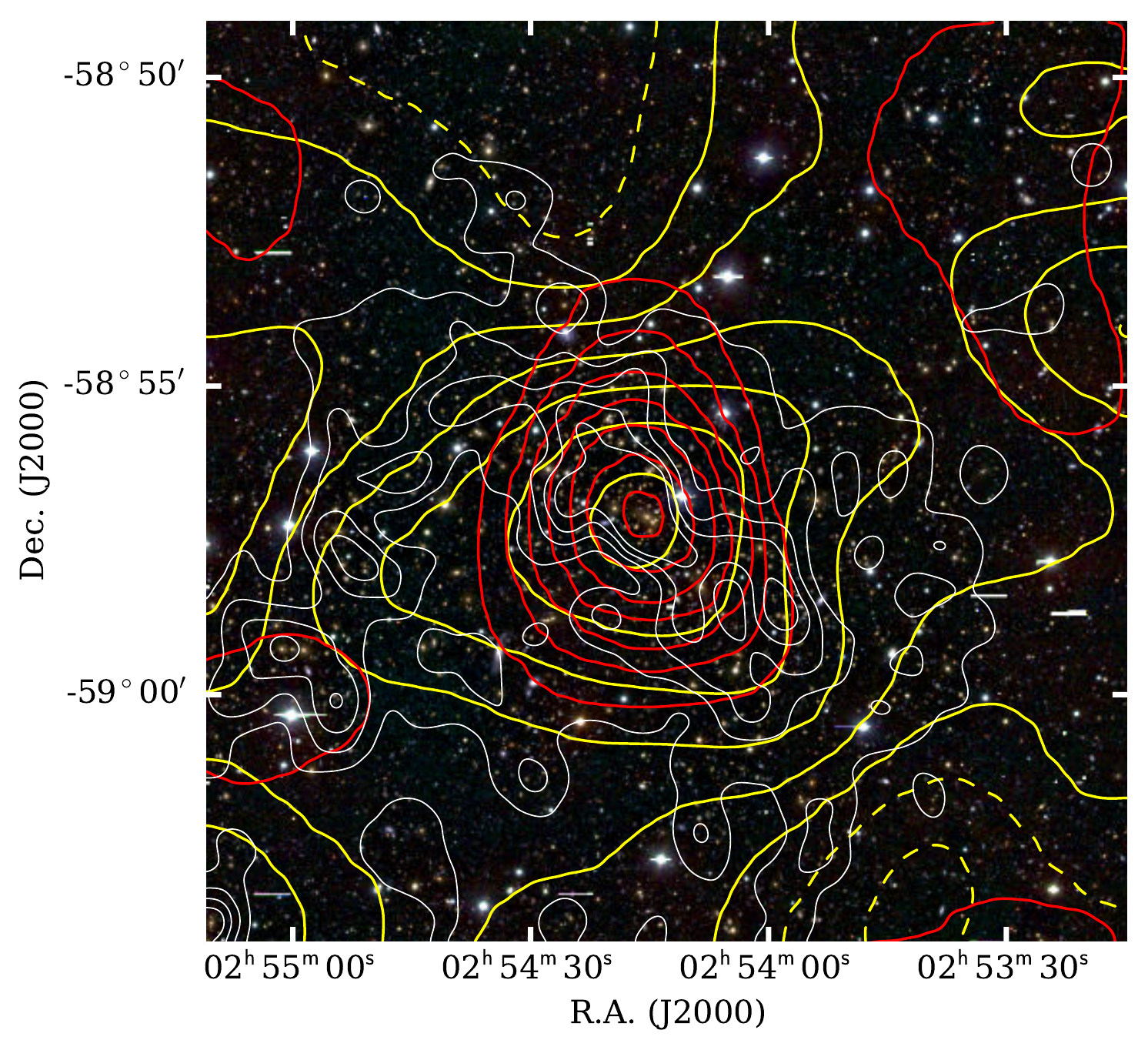}
   \label{fig:3}}
   \hfill
   \subfigure[Tangential shear profile of SPT-CL\,J0254$-$5857.]{
   \includegraphics[width=\columnwidth]{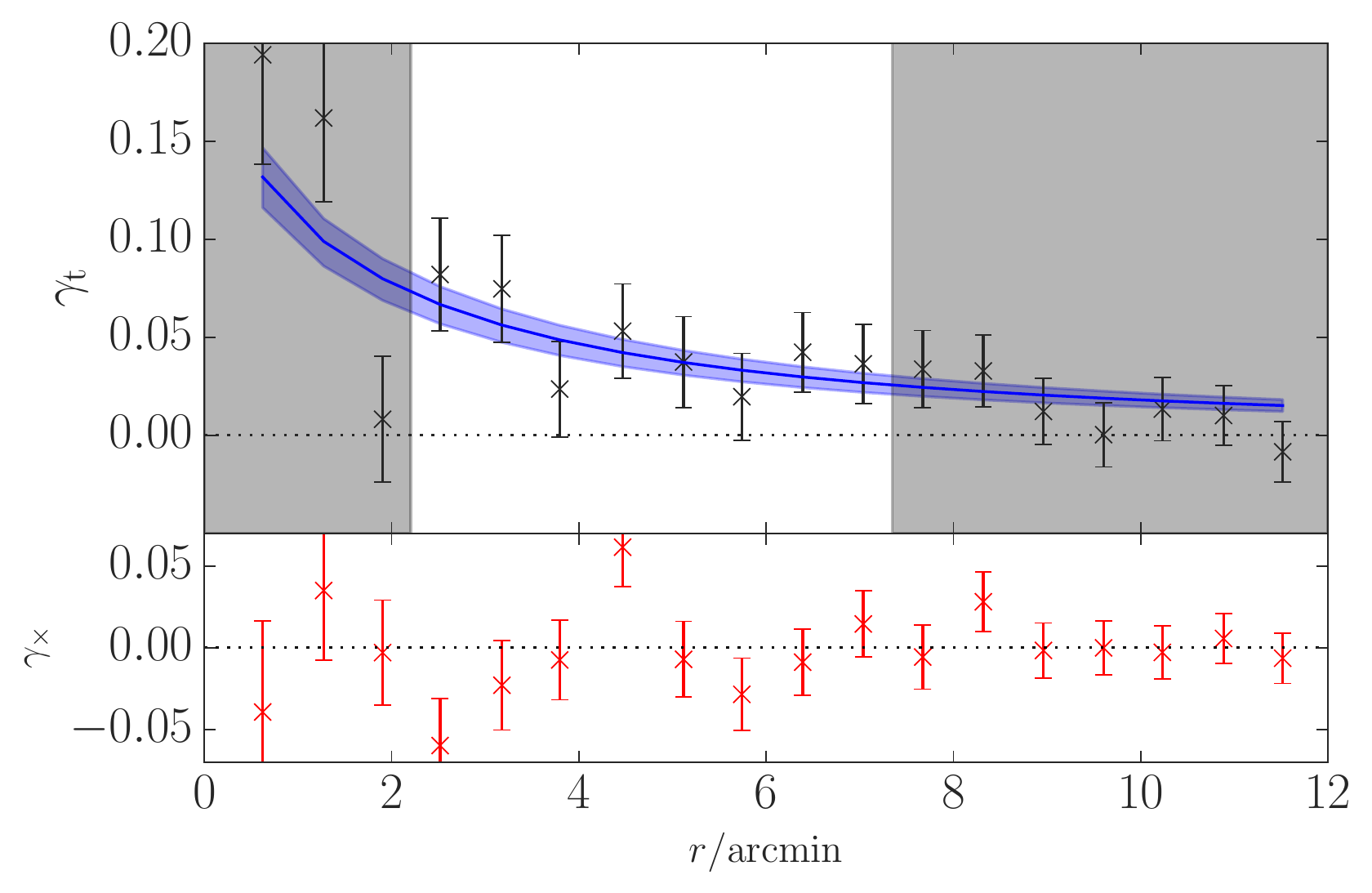}
   \label{fig:gammat0254}}
   \caption{Same as Figure~\ref{fig:kappa-shear-0234} for SPT-CL\,J0254$-$5857.}
   \label{fig:kappa-shear-0254}
 \end{figure*}

\begin{figure*}
   \subfigure[Surface mass density of SPT-CL\,J0307$-$6225.]{
   \includegraphics[width=\columnwidth]{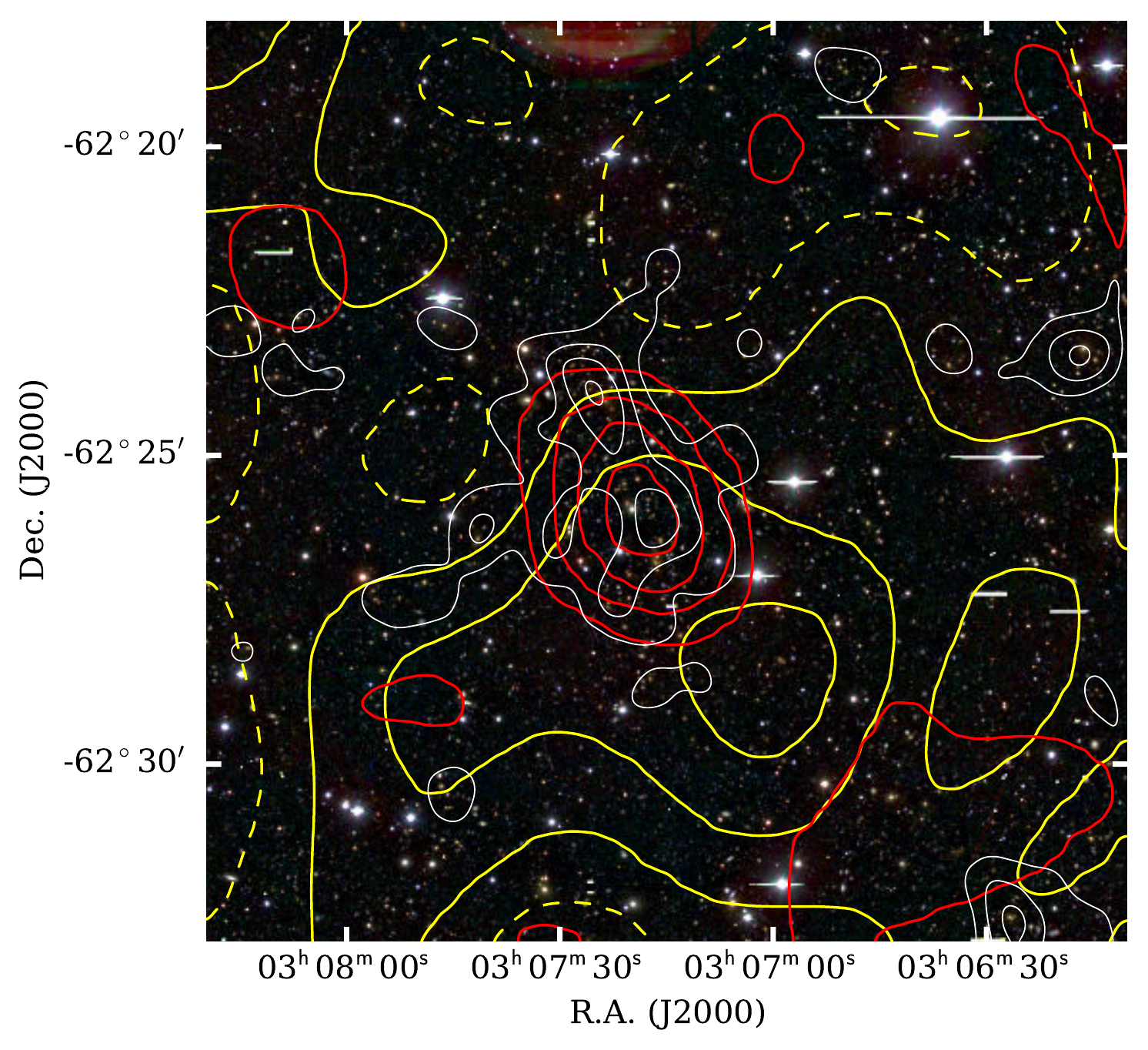}
   \label{fig:4}}
   \hfill
   \subfigure[Tangential shear profile of SPT-CL\,J0307$-$6225.]{
   \includegraphics[width=\columnwidth]{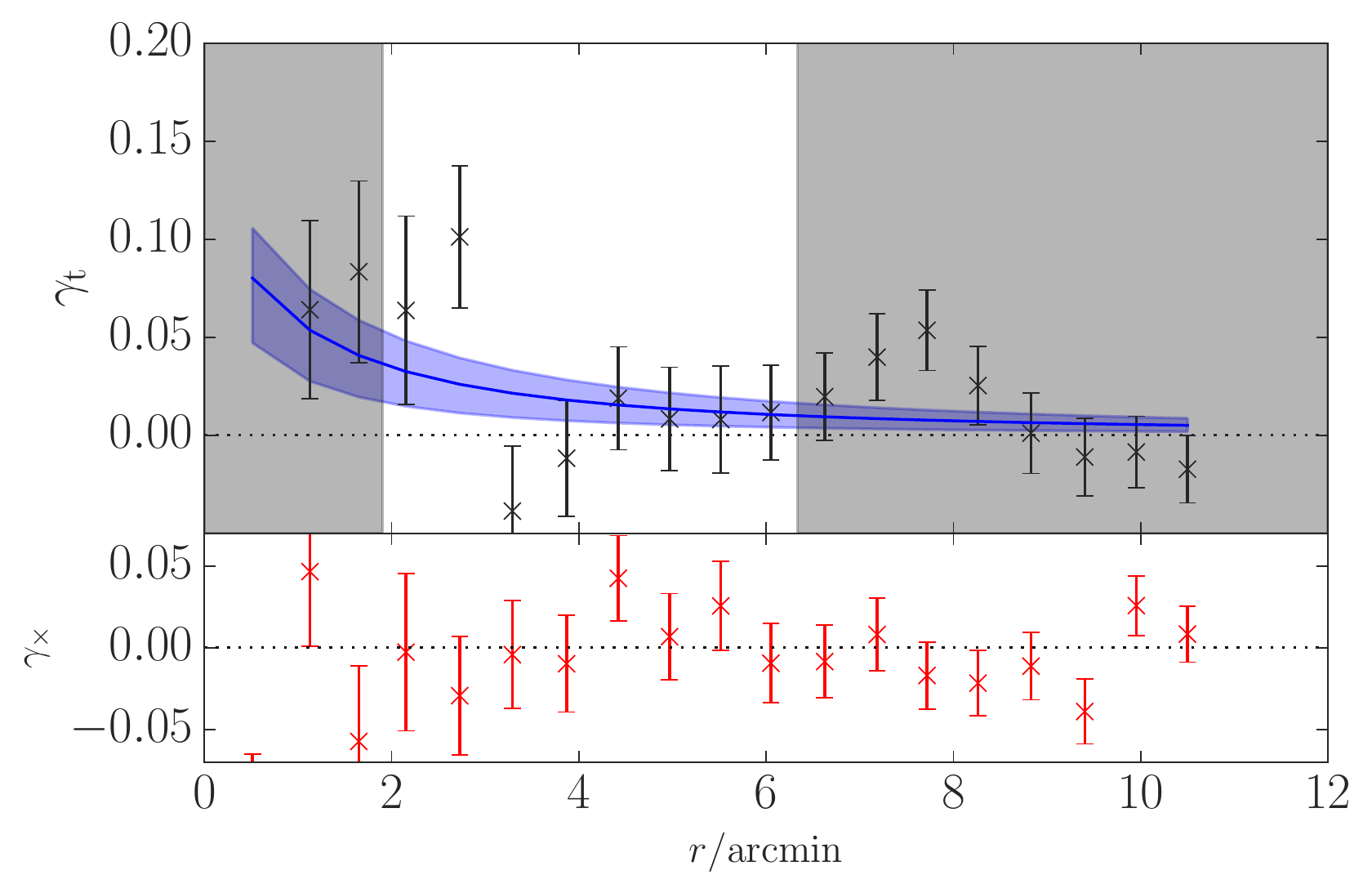}
   \label{fig:gammat0307}}
   \caption{Same as Figure~\ref{fig:kappa-shear-0234} for SPT-CL\,J0307$-$6225.}
   \label{fig:kappa-shear-0307}
 \end{figure*}

\clearpage
\begin{figure*}
   \subfigure[Surface mass density of SPT-CL\,J0317$-$5935.]{
   \includegraphics[width=\columnwidth]{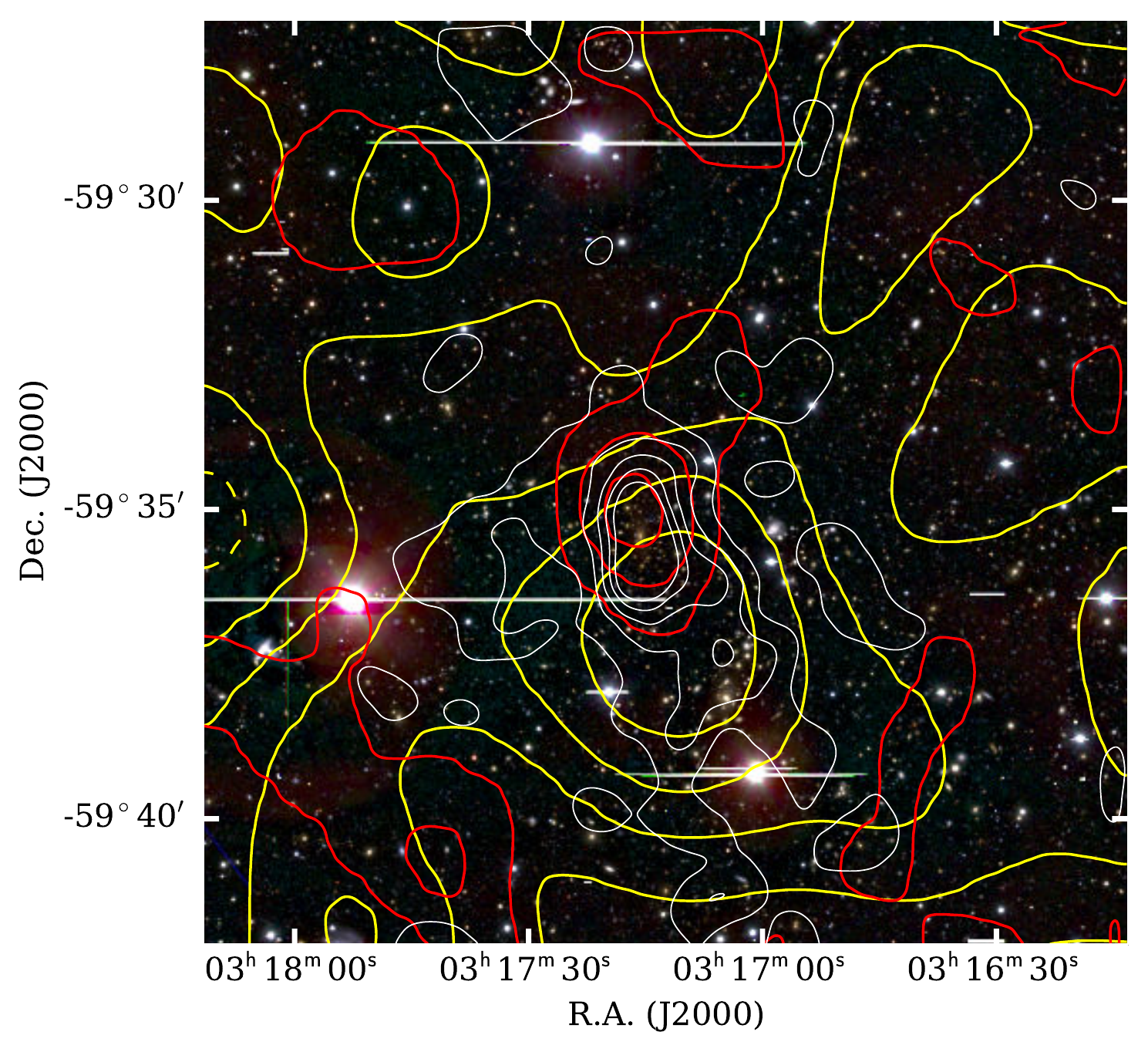}
   \label{fig:5}}
   \hfill
   \subfigure[Tangential shear profile of SPT-CL\,J0317$-$5935.]{
   \includegraphics[width=\columnwidth]{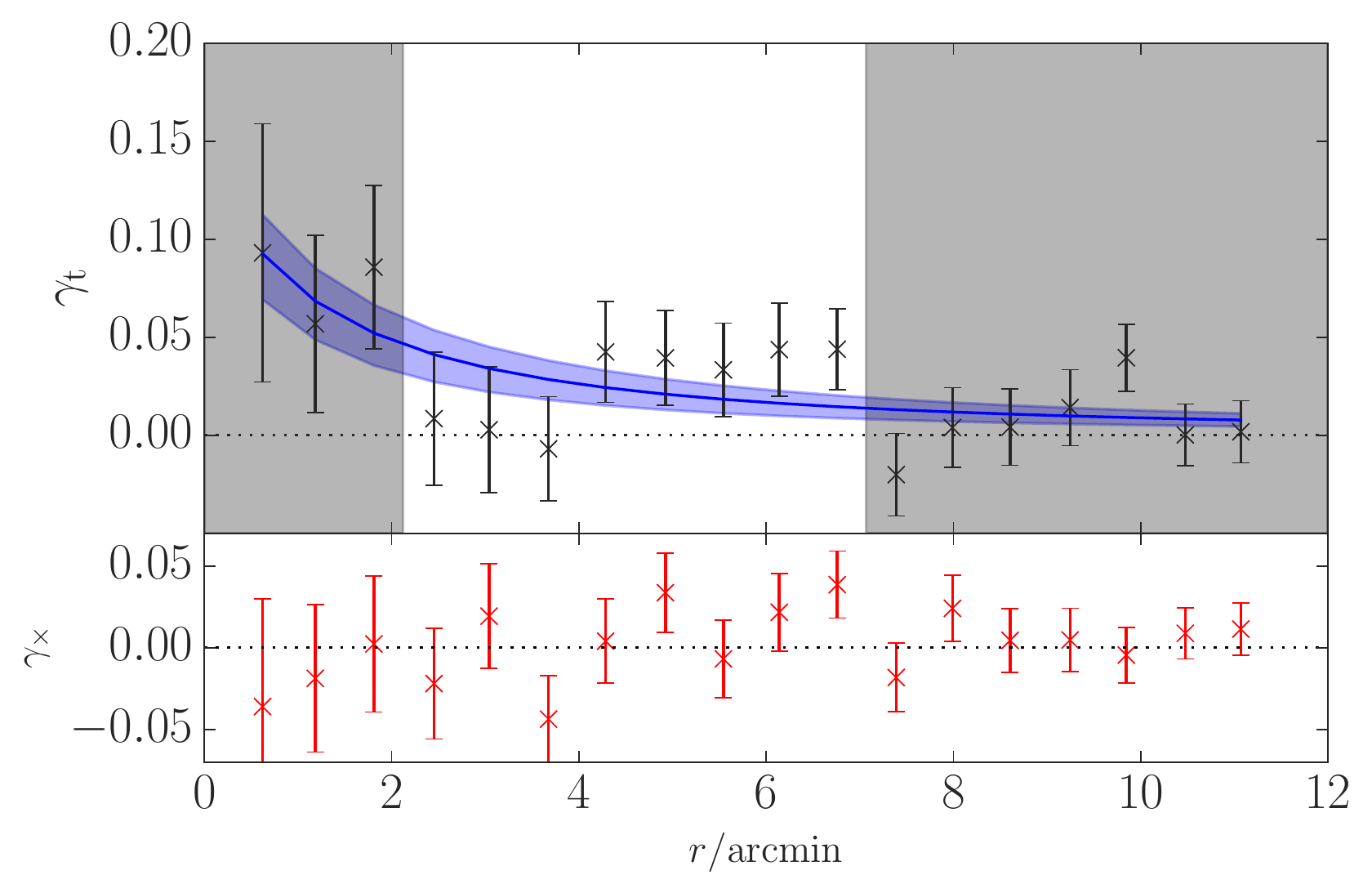}
   \label{fig:gammat0317}}
   \caption{Same as Figure~\ref{fig:kappa-shear-0234} for SPT-CL\,J0317$-$5935.}
   \label{fig:kappa-shear-0317}
 \end{figure*}

\begin{figure*}
   \subfigure[Surface mass density of SPT-CL\,J0346$-$5439.]{
   \includegraphics[width=\columnwidth]{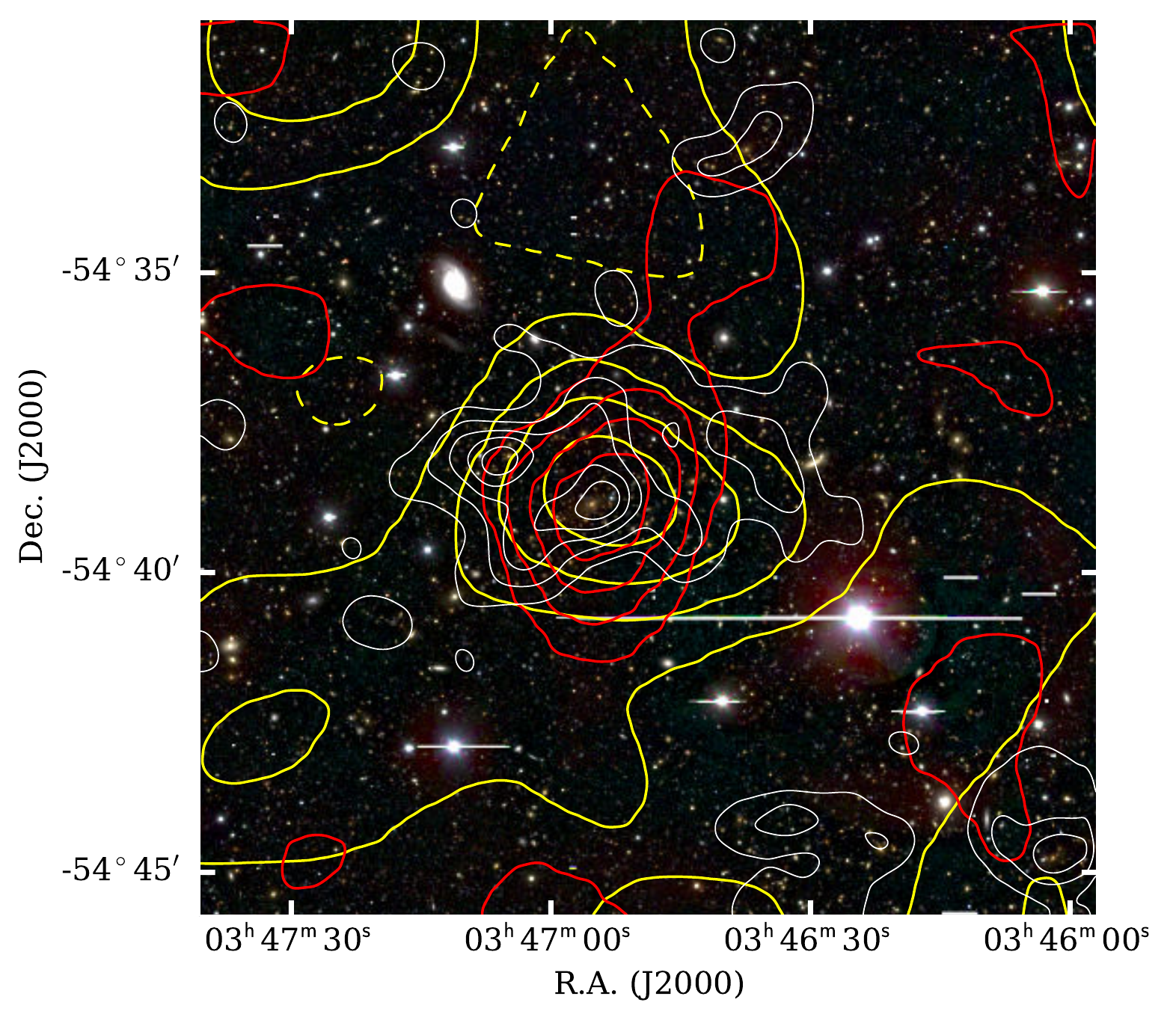}
   \label{fig:6}}
   \hfill
   \subfigure[Tangential shear profile of SPT-CL\,J0346$-$5439.]{
   \includegraphics[width=\columnwidth]{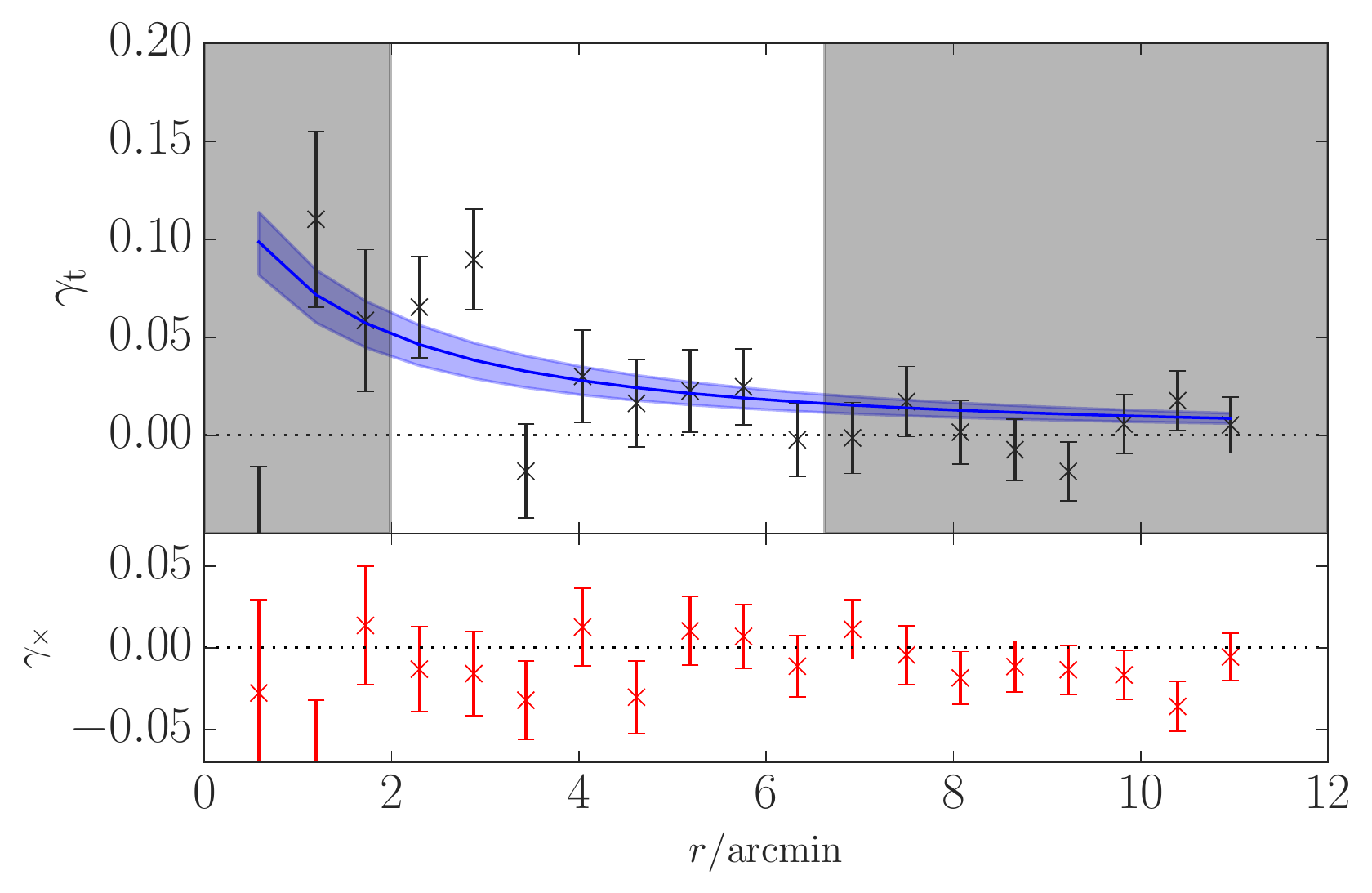}
   \label{fig:gammat0346}}
   \caption{Same as Figure~\ref{fig:kappa-shear-0234} for SPT-CL\,J0346$-$5439.}
   \label{fig:kappa-shear-0346}
 \end{figure*}

\clearpage
\begin{figure*}
   \subfigure[Surface mass density of SPT-CL\,J0348$-$4515.]{
   \includegraphics[width=\columnwidth]{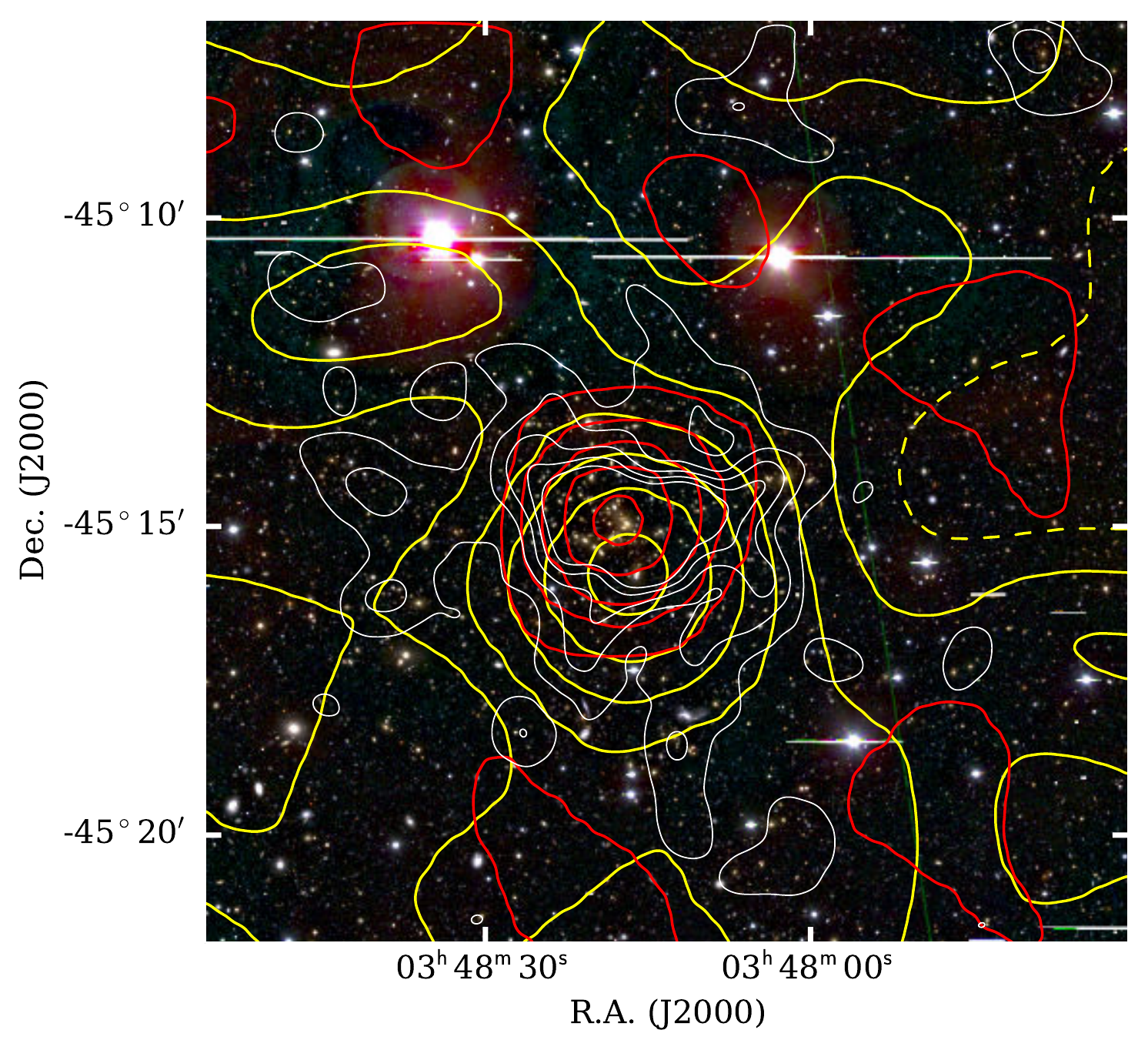}
   \label{fig:7}}
   \hfill
   \subfigure[Tangential shear profile of SPT-CL\,J0348$-$4515.]{
   \includegraphics[width=\columnwidth]{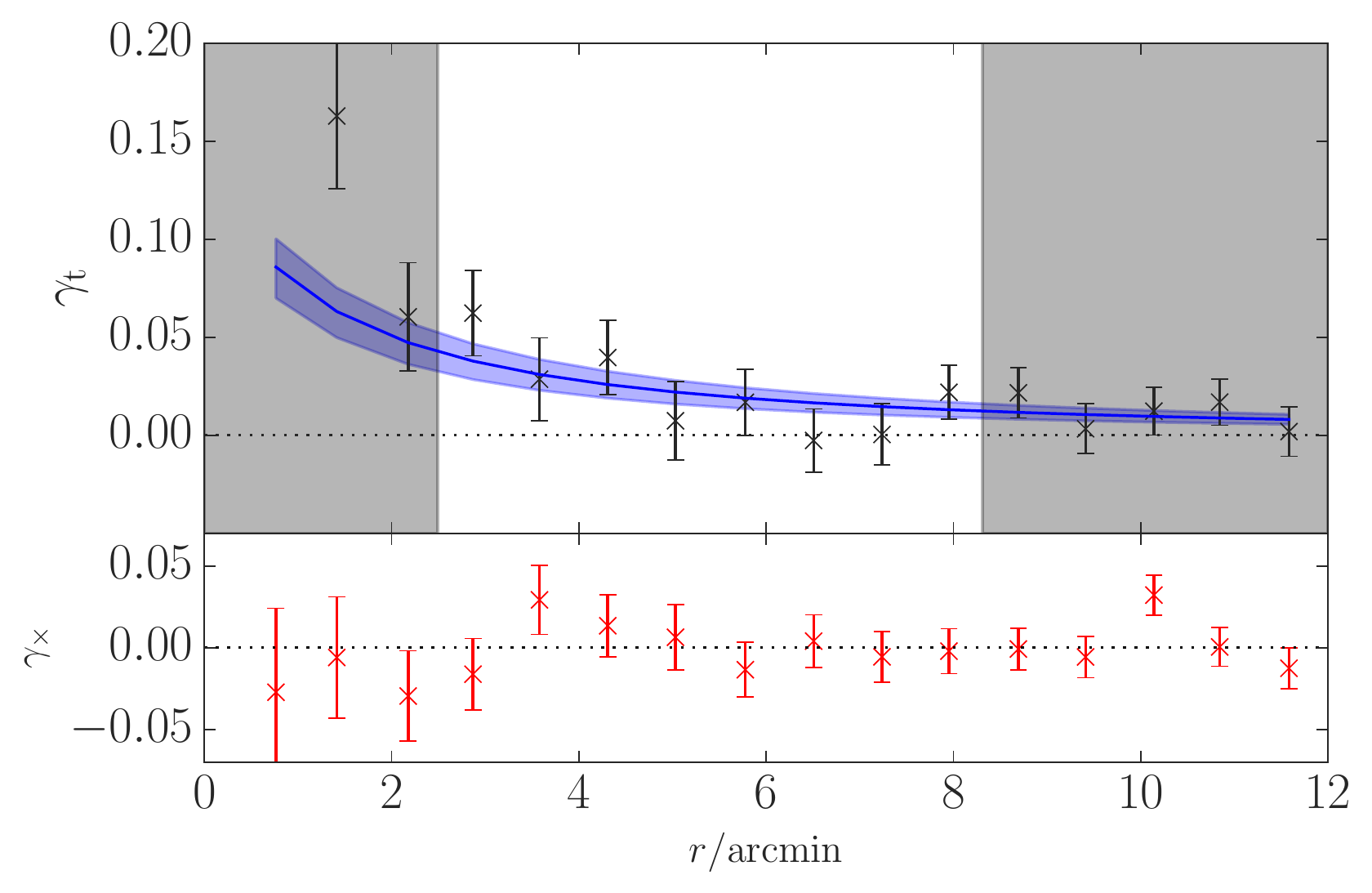}
   \label{fig:gammat0348}}
   \caption{Same as Figure~\ref{fig:kappa-shear-0234} for SPT-CL\,J0348$-$4515.}
   \label{fig:kappa-shear-0348}
 \end{figure*}

\begin{figure*}
   \subfigure[Surface mass density of SPT-CL\,J0426$-$5455.]{
   \includegraphics[width=\columnwidth]{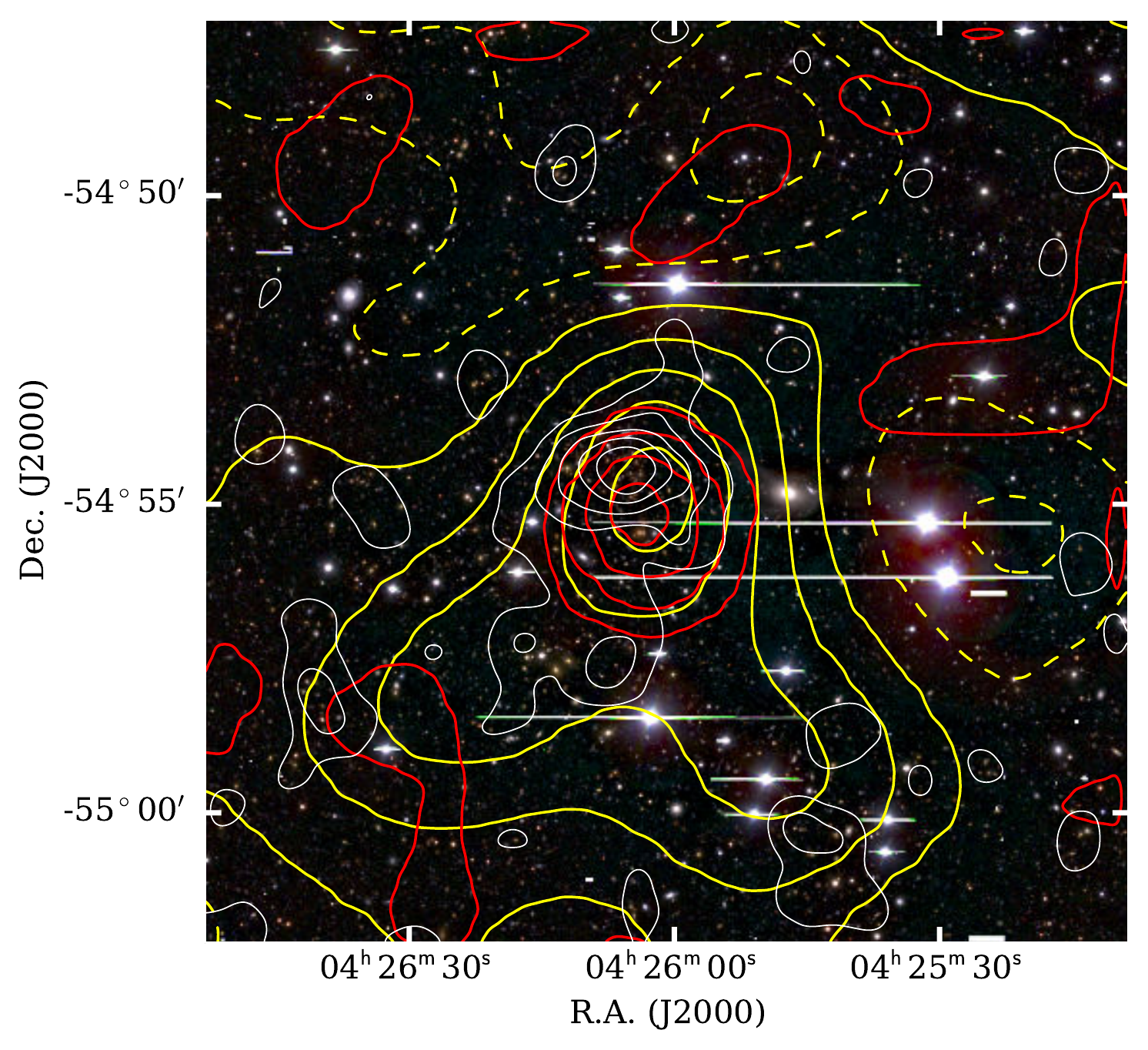}
   \label{fig:8}}
   \hfill
   \subfigure[Tangential shear profile of SPT-CL\,J0426$-$5455.]{
   \includegraphics[width=\columnwidth]{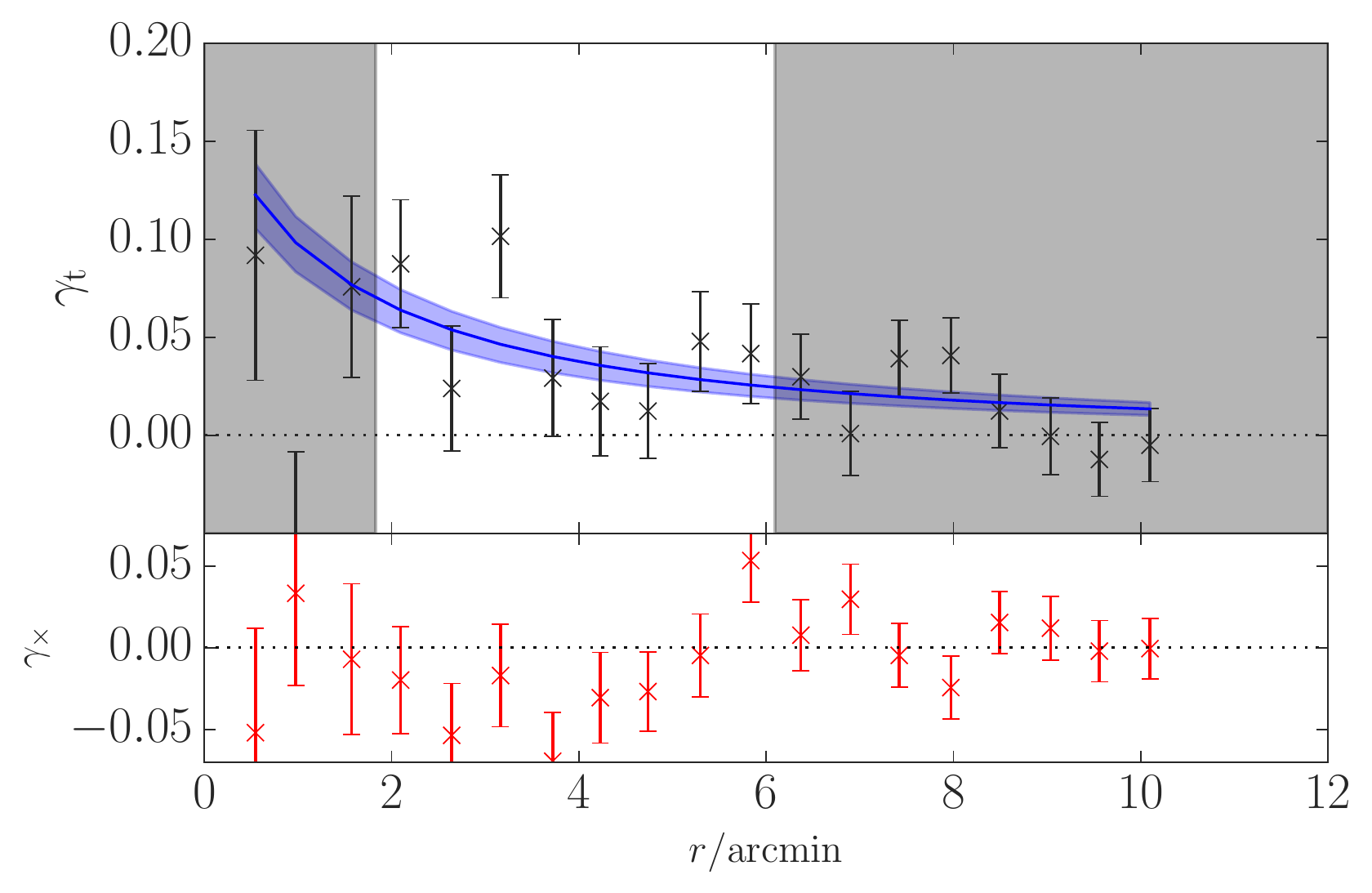}
   \label{fig:gammat0426}}
   \caption{Same as Figure~\ref{fig:kappa-shear-0234} for SPT-CL\,J0426$-$5455.}
   \label{fig:kappa-shear-0426}
 \end{figure*}

\clearpage
\begin{figure*}
   \subfigure[Surface mass density of SPT-CL\,J0509$-$5342.]{
   \includegraphics[width=\columnwidth]{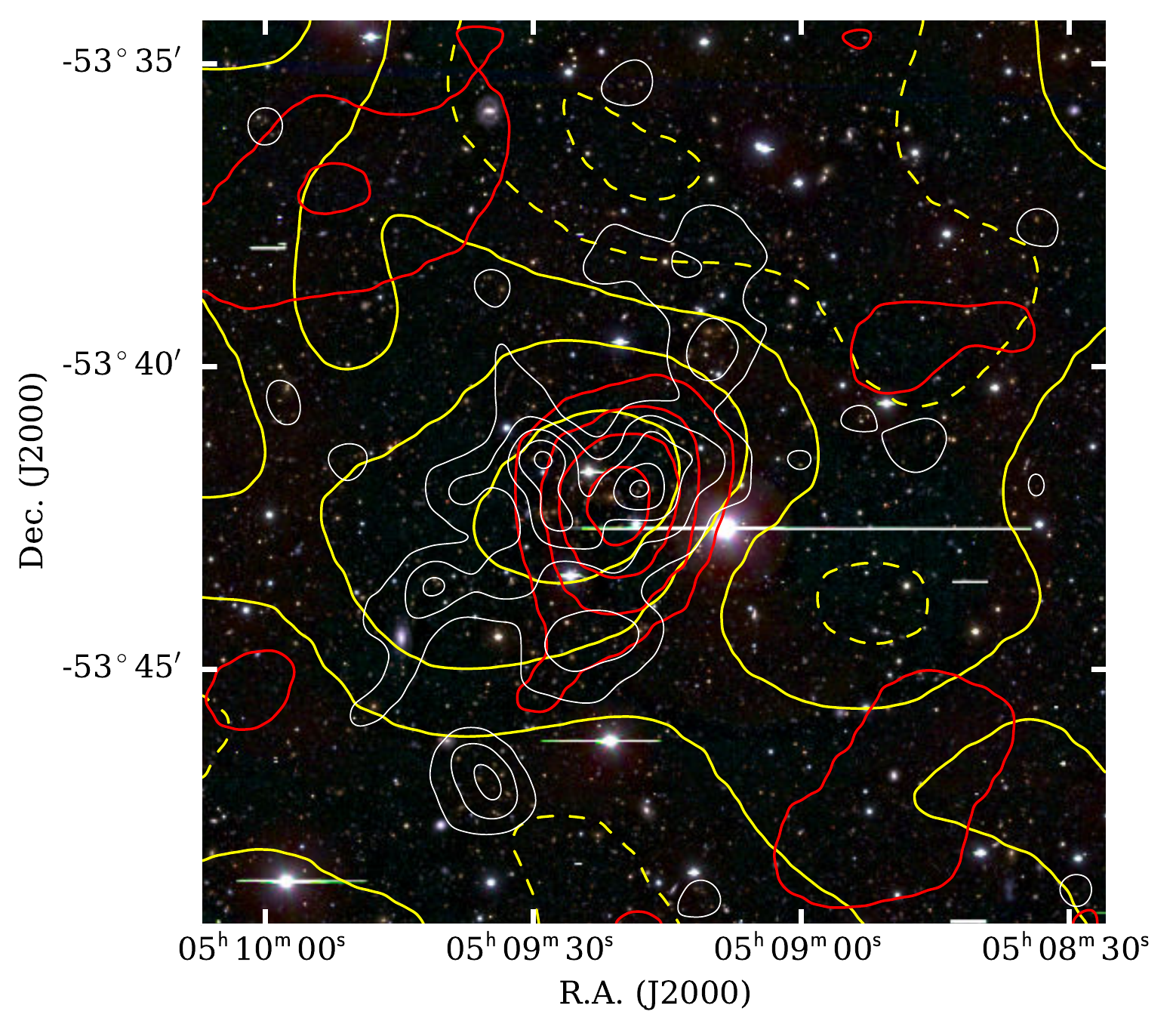}
   \label{fig:9}}
   \hfill
   \subfigure[Tangential shear profile of SPT-CL\,J0509$-$5342.]{
   \includegraphics[width=\columnwidth]{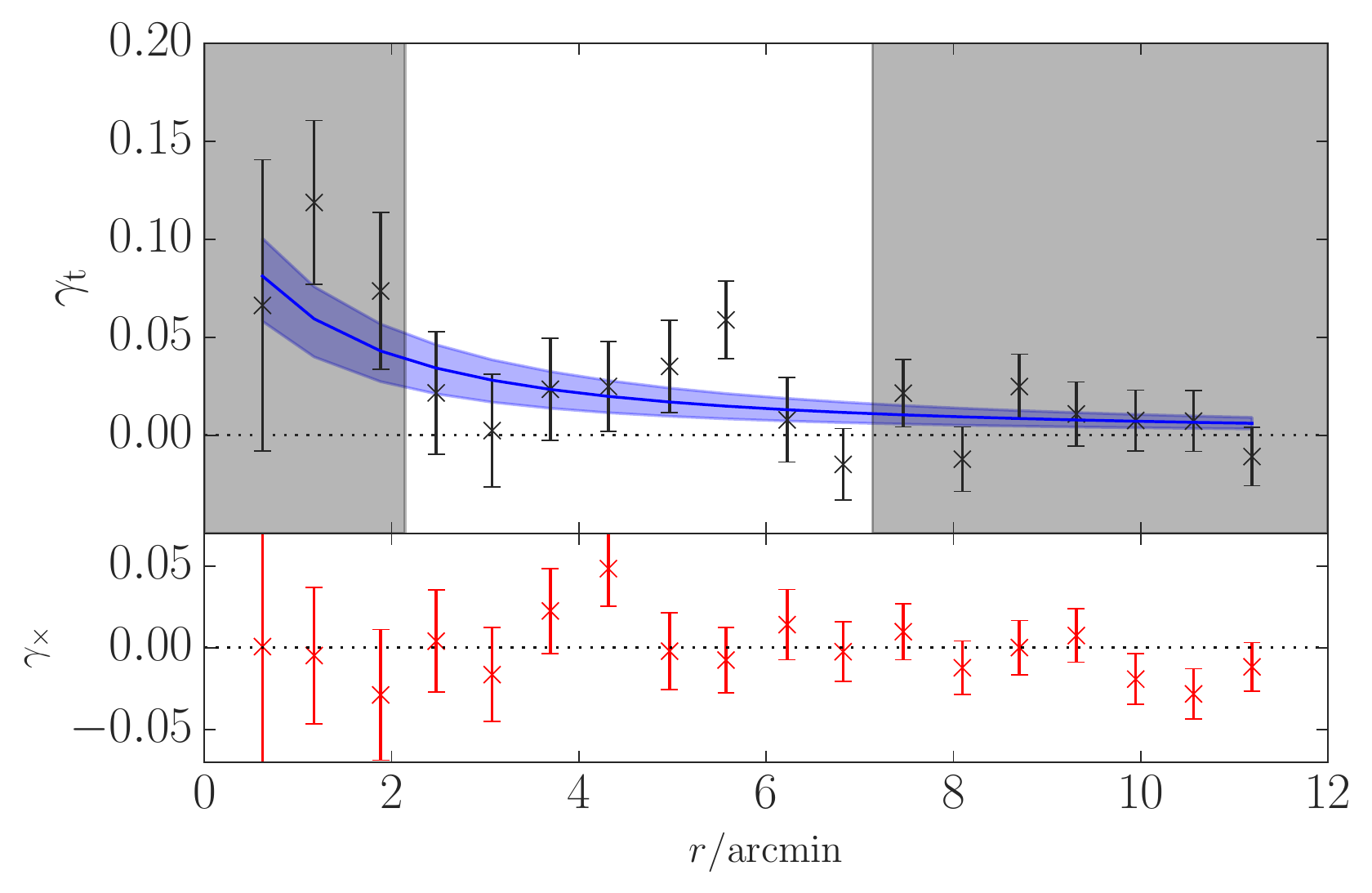}
   \label{fig:gammat0509}}
   \caption{Same as Figure~\ref{fig:kappa-shear-0234} for SPT-CL\,J0509$-$5342.}
   \label{fig:kappa-shear-0509}
 \end{figure*}

\begin{figure*}
   \subfigure[Surface mass density of SPT-CL\,J0516$-$5430.]{
   \includegraphics[width=\columnwidth]{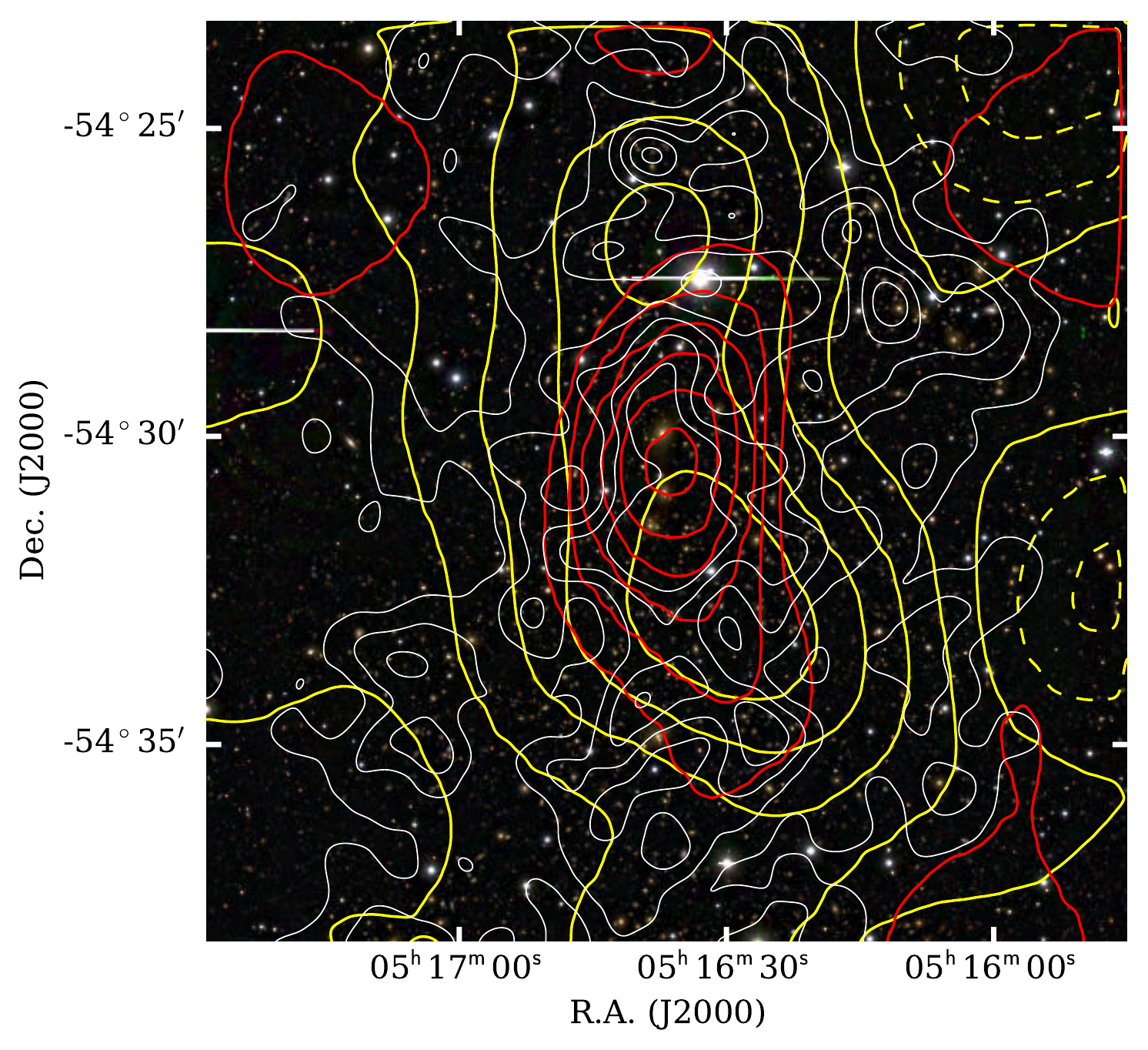}
   \label{fig:10}}
   \hfill
   \subfigure[Tangential shear profile of SPT-CL\,J0516$-$5430.]{
   \includegraphics[width=\columnwidth]{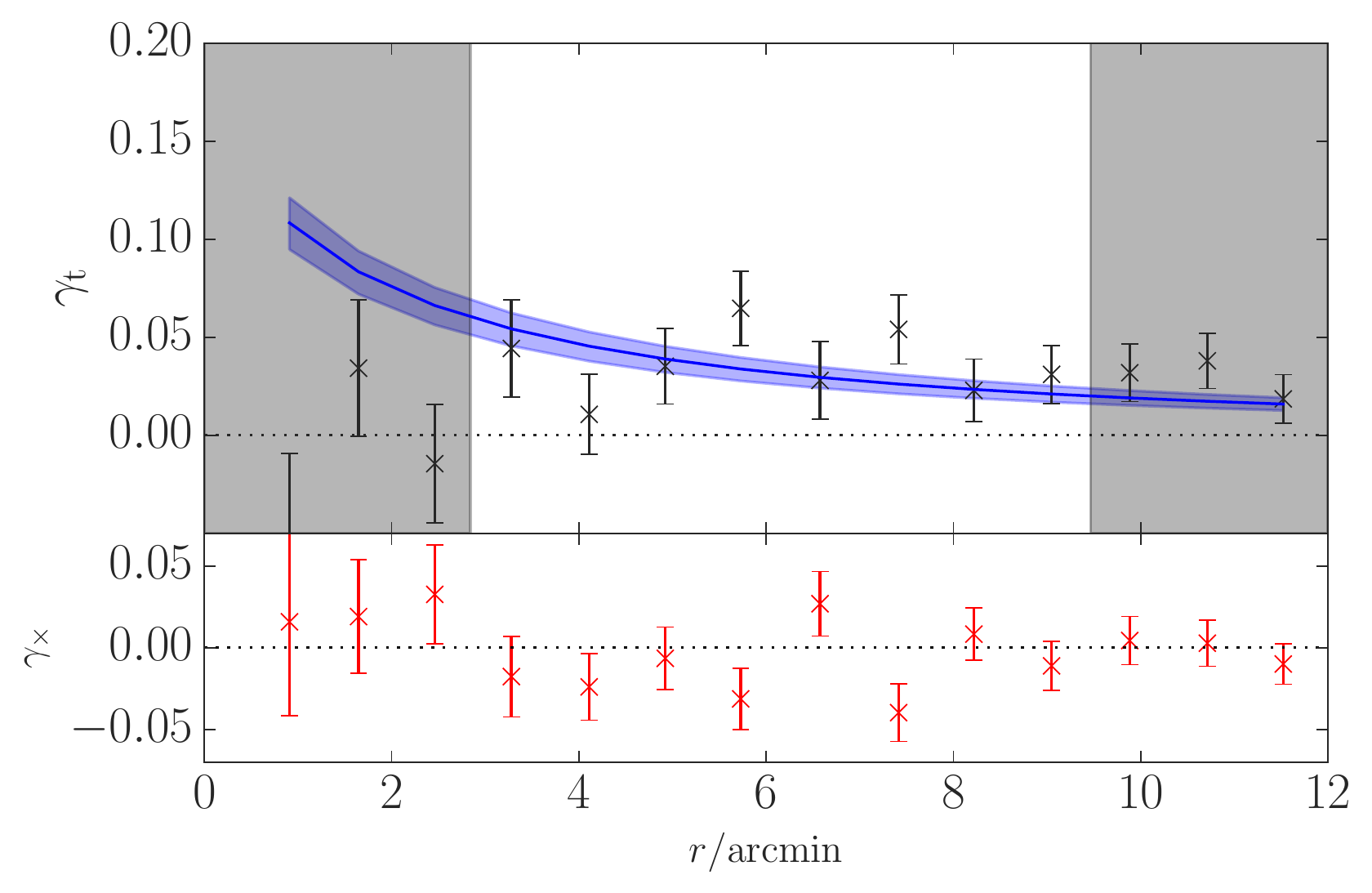}
   \label{fig:gammat0516}}
   \caption{Same as Figure~\ref{fig:kappa-shear-0234} for SPT-CL\,J0516$-$5430.}
   \label{fig:kappa-shear-0516}
 \end{figure*}

\clearpage
\begin{figure*}
   \subfigure[Surface mass density of SPT-CL\,J0551$-$5709.]{
   \includegraphics[width=\columnwidth]{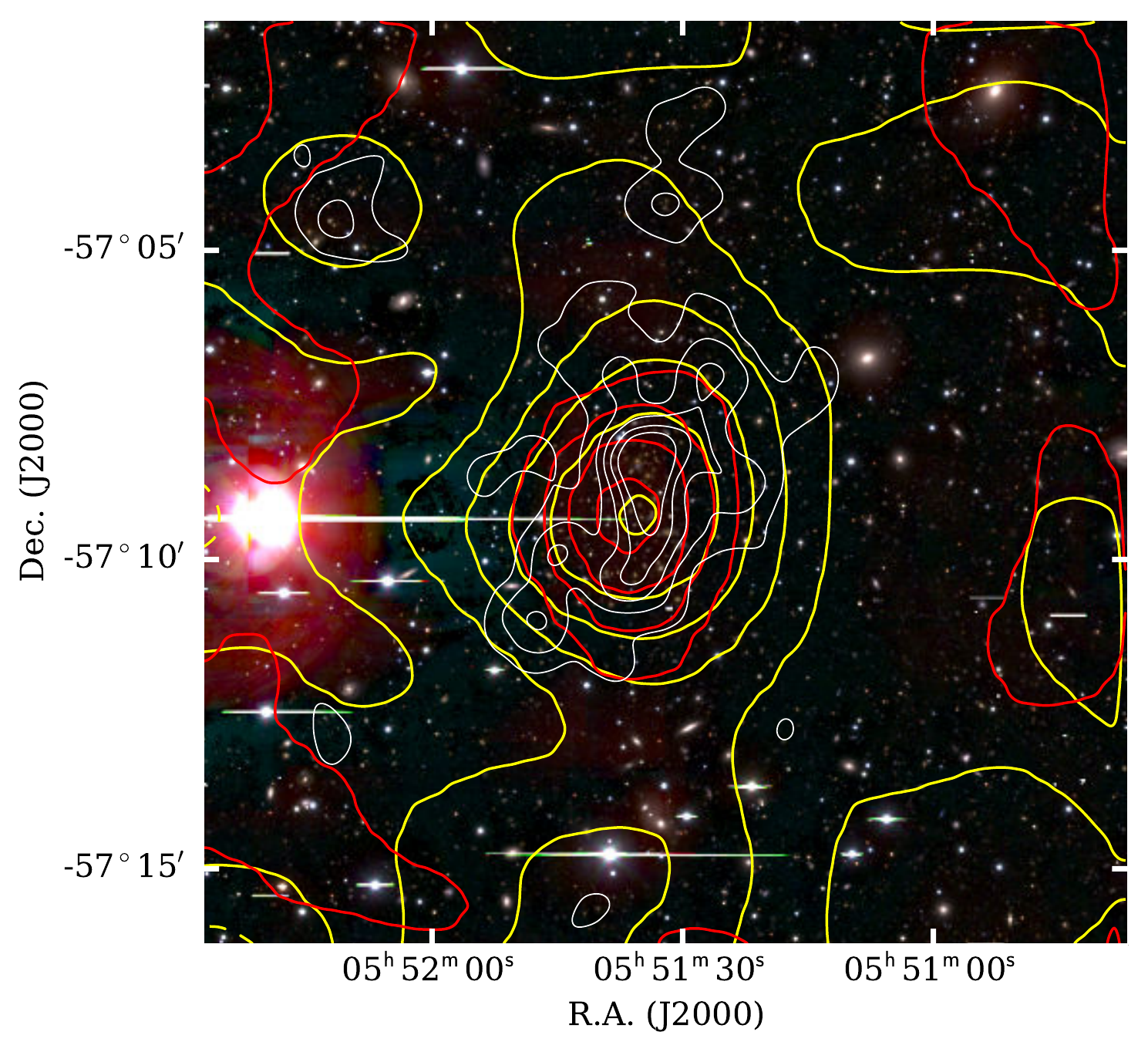}
   \label{fig:11}}
   \hfill
   \subfigure[Tangential shear profile of SPT-CL\,J0551$-$5709.]{
   \includegraphics[width=\columnwidth]{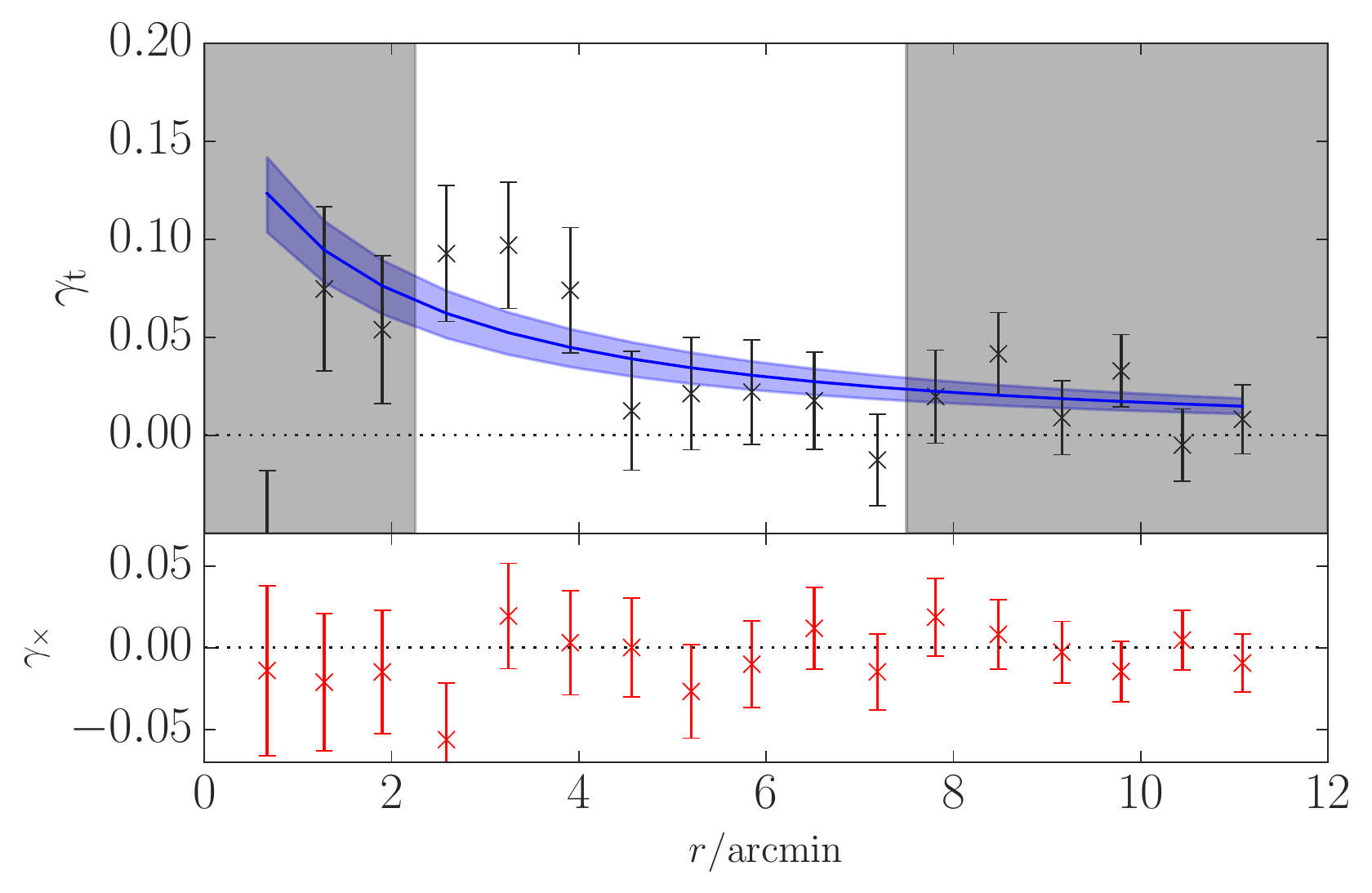}
   \label{fig:gammat0551}}
   \caption{Same as Figure~\ref{fig:kappa-shear-0234} for SPT-CL\,J0551$-$5709.}
   \label{fig:kappa-shear-0551}
 \end{figure*}

\begin{figure*}
   \subfigure[Surface mass density of SPT-CL\,J2022$-$6323.]{
   \includegraphics[width=\columnwidth]{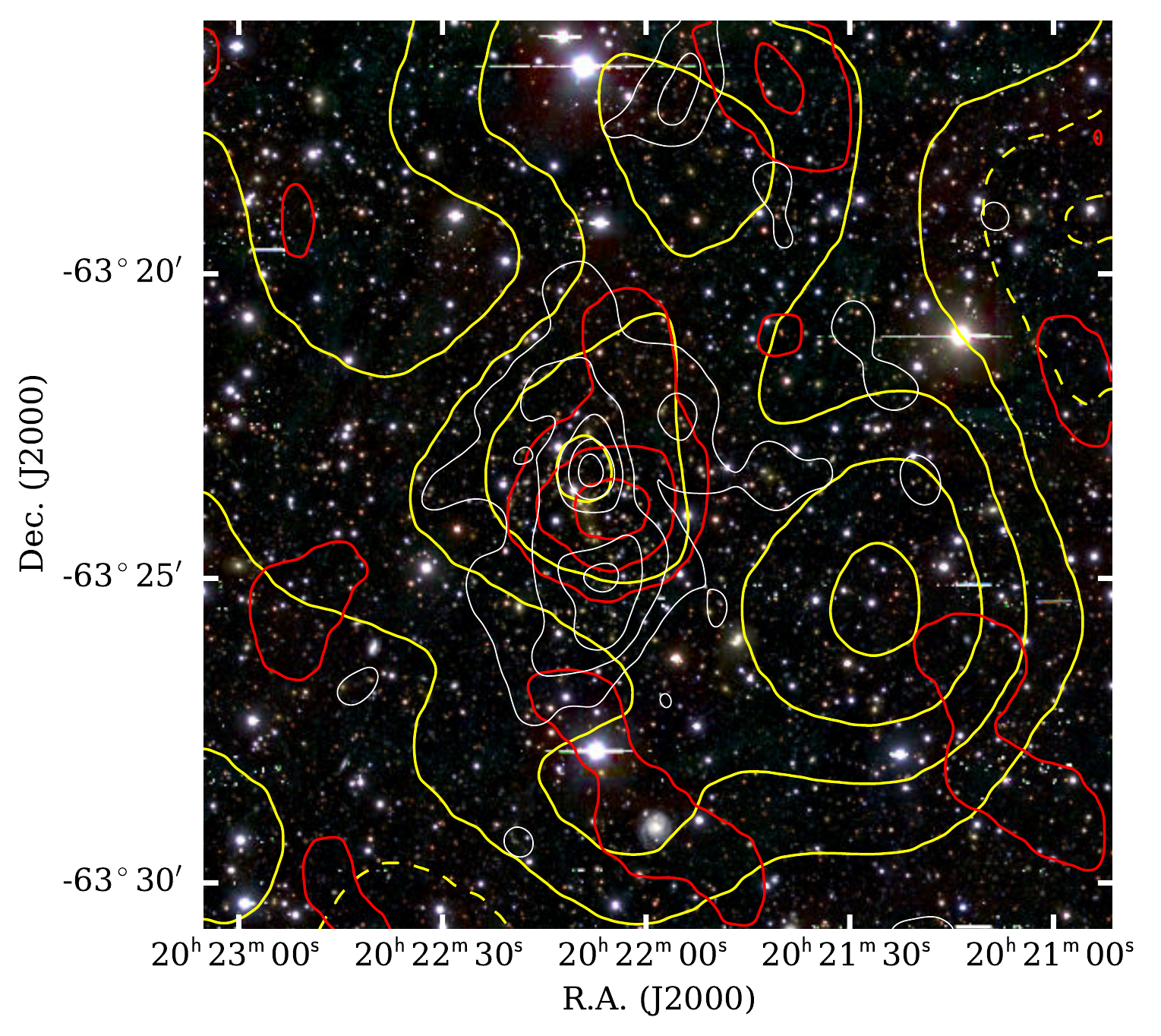}
   \label{fig:12}}
   \hfill
   \subfigure[Tangential shear profile of SPT-CL\,J2022$-$6323.]{
   \includegraphics[width=\columnwidth]{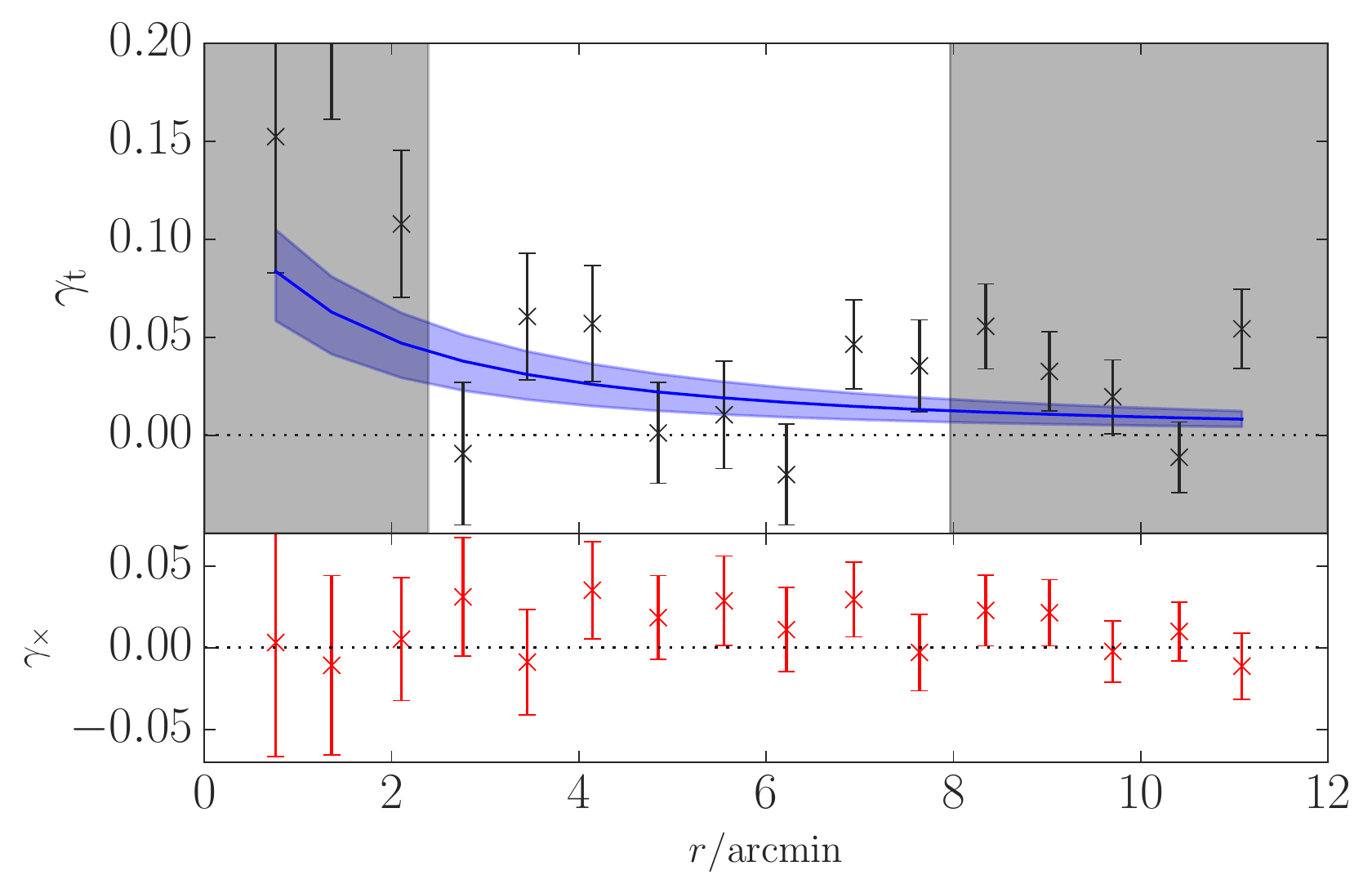}
   \label{fig:gammat2022}}
   \caption{Same as Figure~\ref{fig:kappa-shear-0234} for SPT-CL\,J2022$-$6323.}
   \label{fig:kappa-shear-2022}
 \end{figure*}

\clearpage
\begin{figure*}
   \subfigure[Surface mass density of SPT-CL\,J2030$-$5638.]{
   \includegraphics[width=\columnwidth]{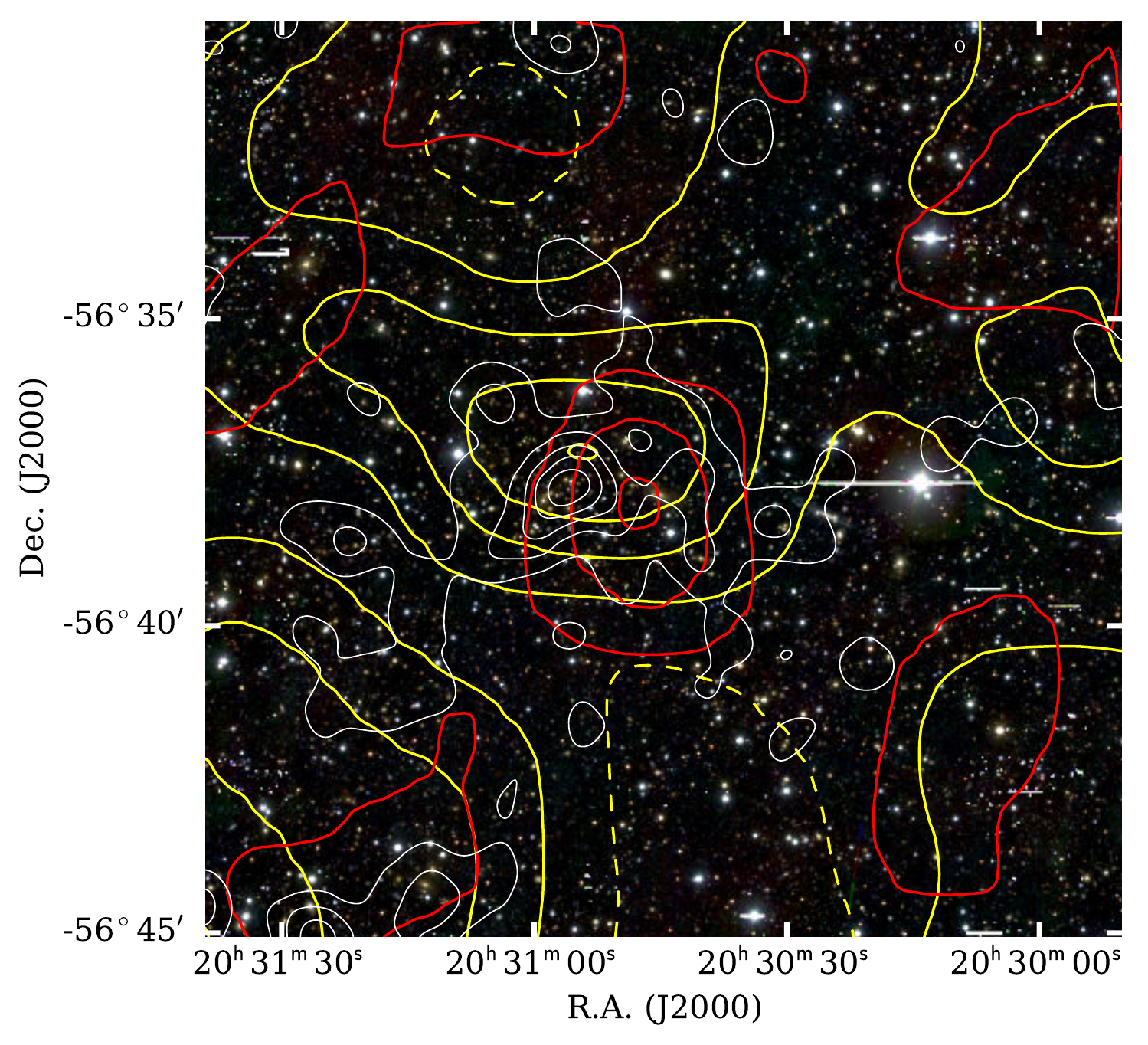}
   \label{fig:13}}
   \hfill
   \subfigure[Tangential shear profile of SPT-CL\,J2030$-$5638.]{
   \includegraphics[width=\columnwidth]{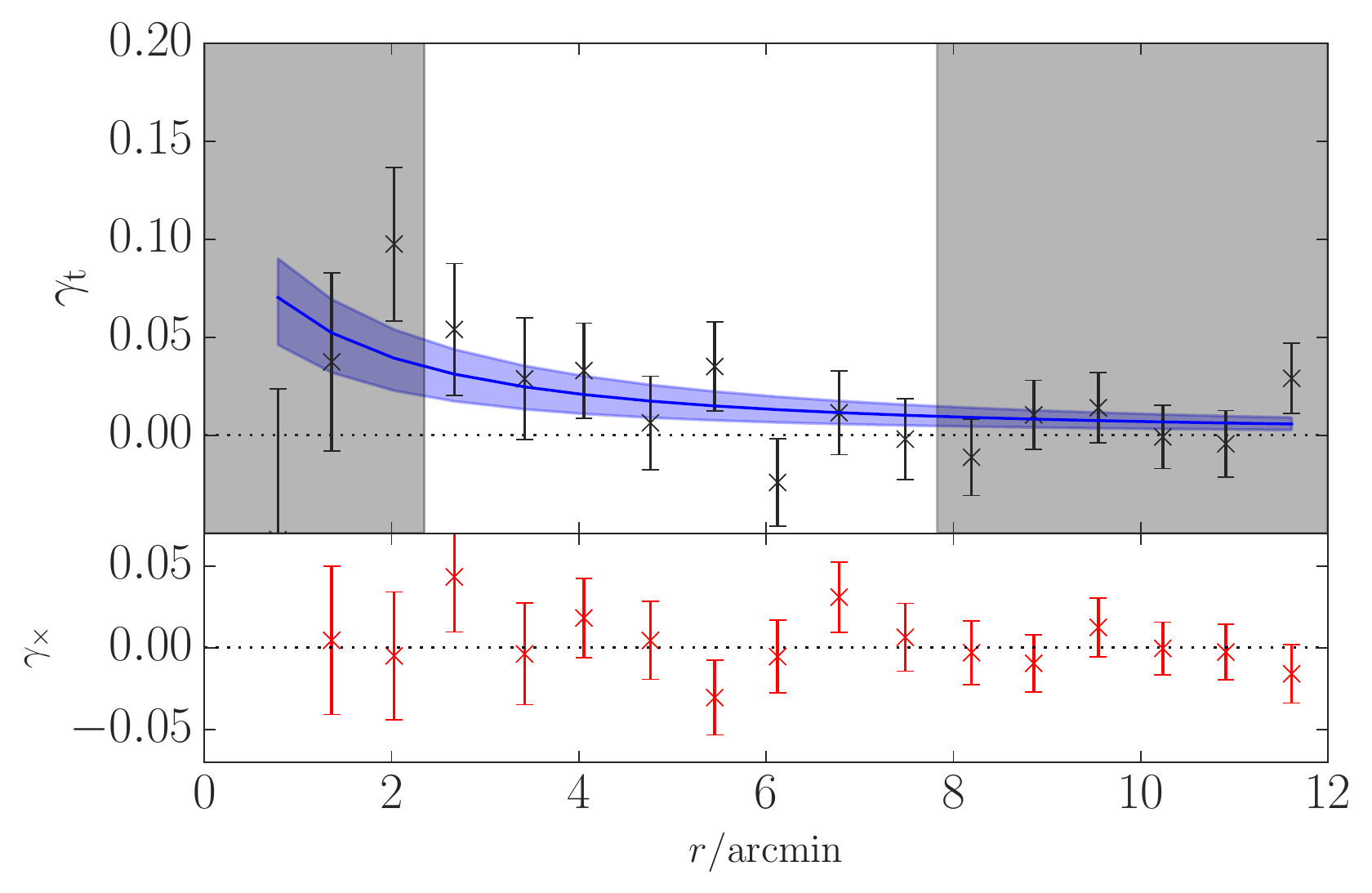}
   \label{fig:gammat2030}}
   \caption{Same as Figure~\ref{fig:kappa-shear-0234} for SPT-CL\,J2030$-$5638.}
   \label{fig:kappa-shear-2030}
 \end{figure*}

\begin{figure*}
   \subfigure[Surface mass density of SPT-CL\,J2032$-$5627.]{
   \includegraphics[width=\columnwidth]{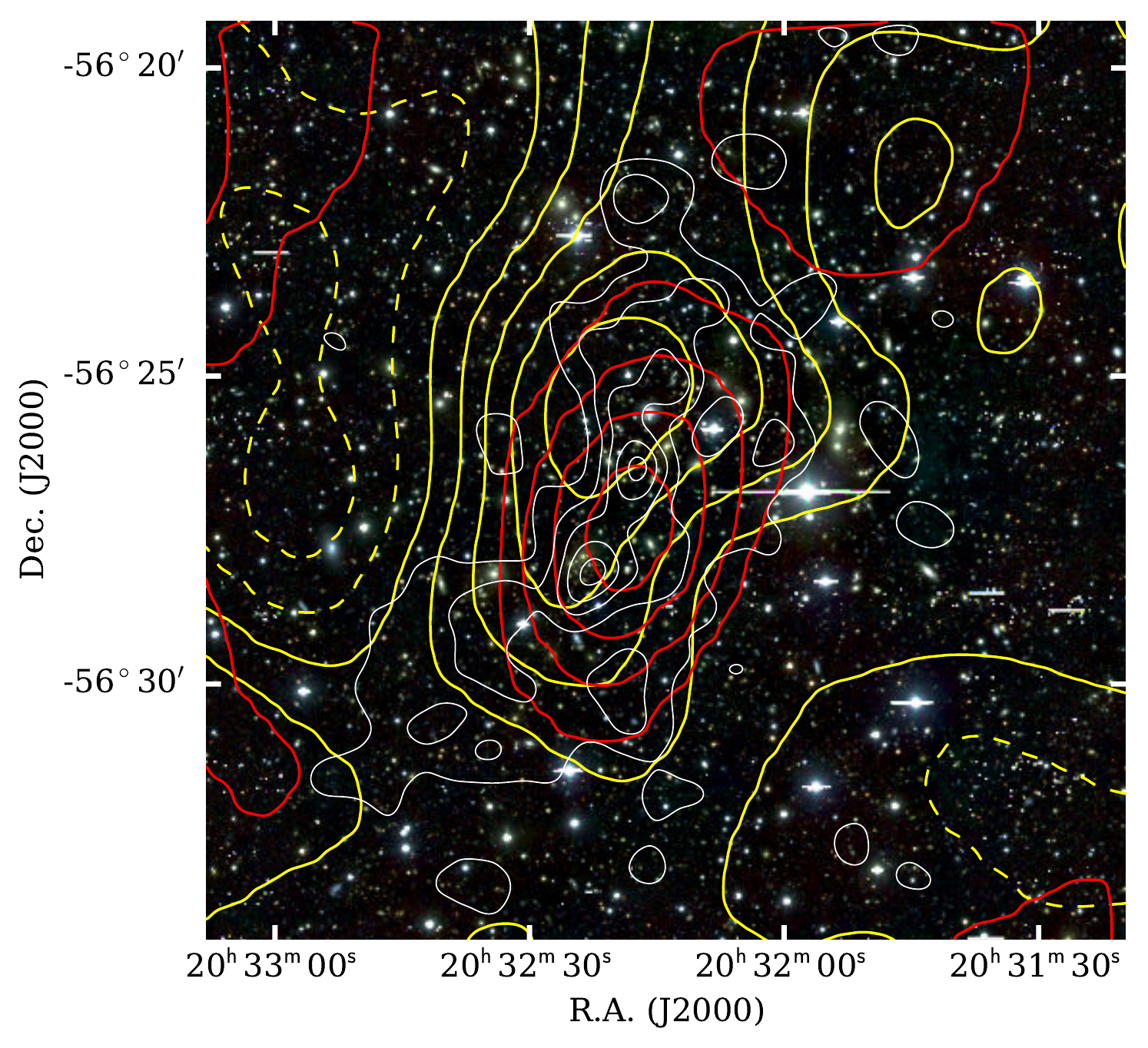}
   \label{fig:14}}
   \hfill
   \subfigure[Tangential shear profile of SPT-CL\,J2032$-$5627.]{
   \includegraphics[width=\columnwidth]{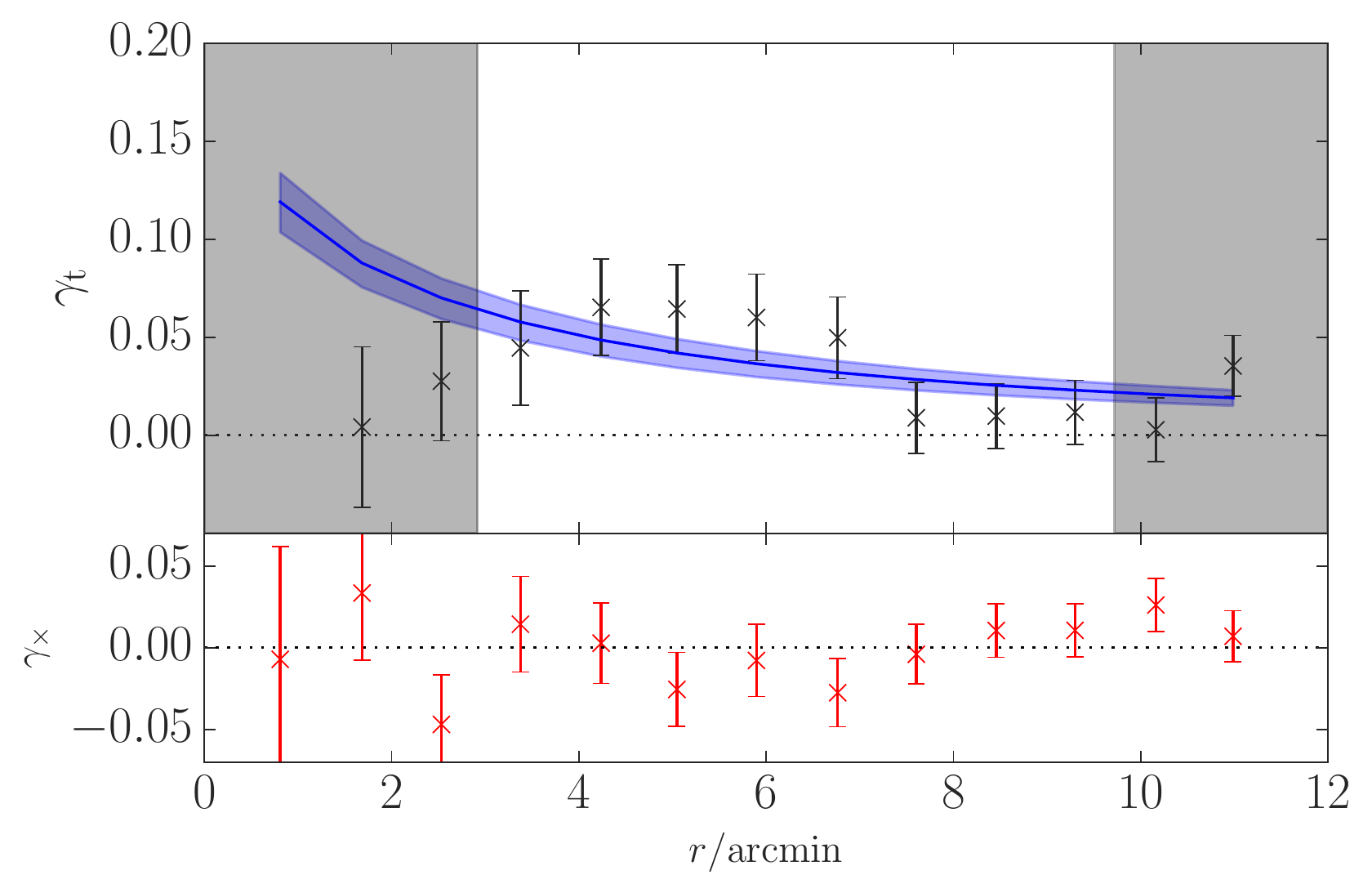}
   \label{fig:gammat2032}}
   \caption{Same as Figure~\ref{fig:kappa-shear-0234} for SPT-CL\,J2032$-$5627.}
   \label{fig:kappa-shear-2032}
 \end{figure*}

\clearpage
\begin{figure*}
   \subfigure[Surface mass density of SPT-CL\,J2135$-$5726.]{
   \includegraphics[width=\columnwidth]{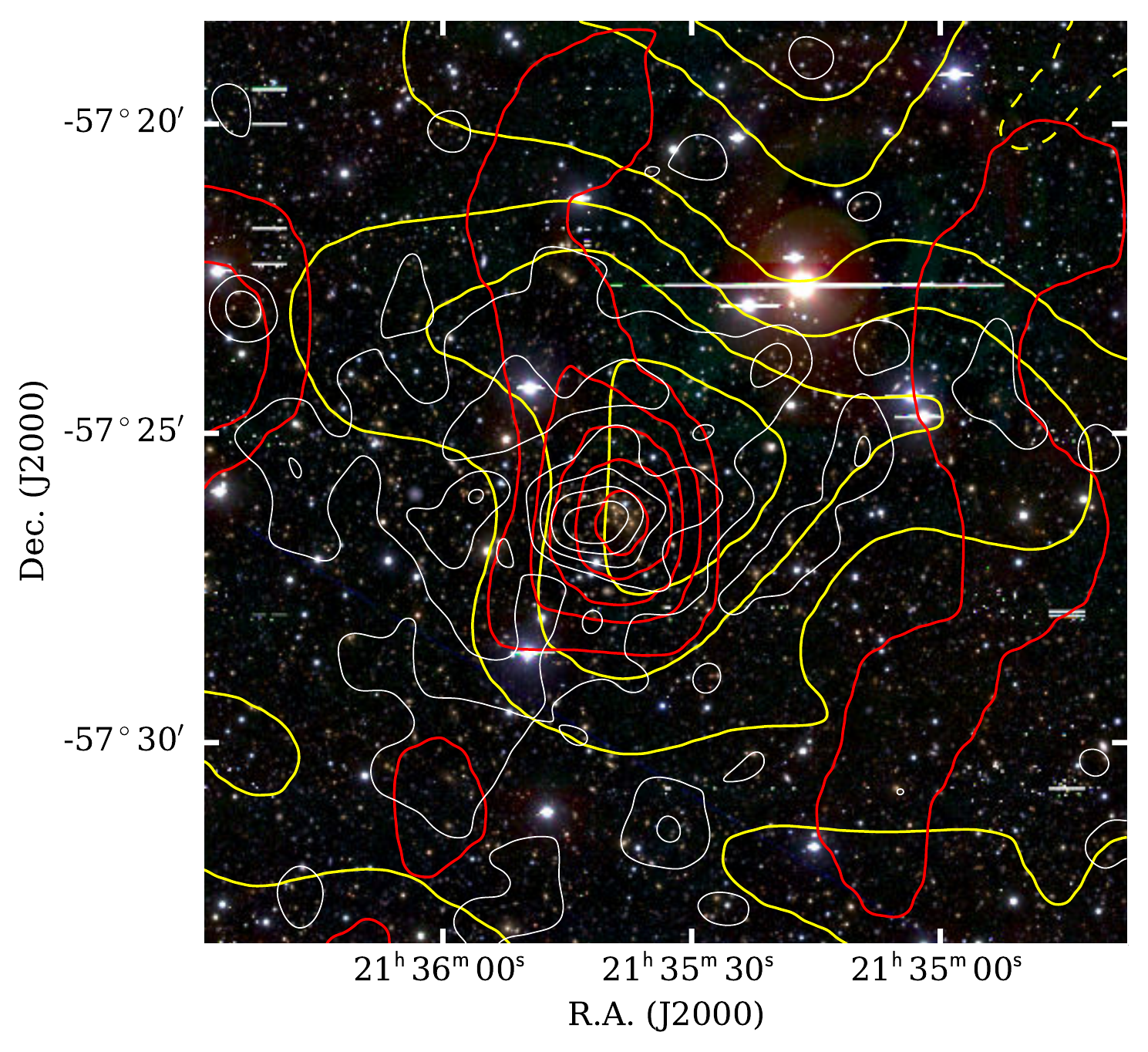}
   \label{fig:15}}
   \hfill
   \subfigure[Tangential shear profile of SPT-CL\,J2135$-$5726.]{
   \includegraphics[width=\columnwidth]{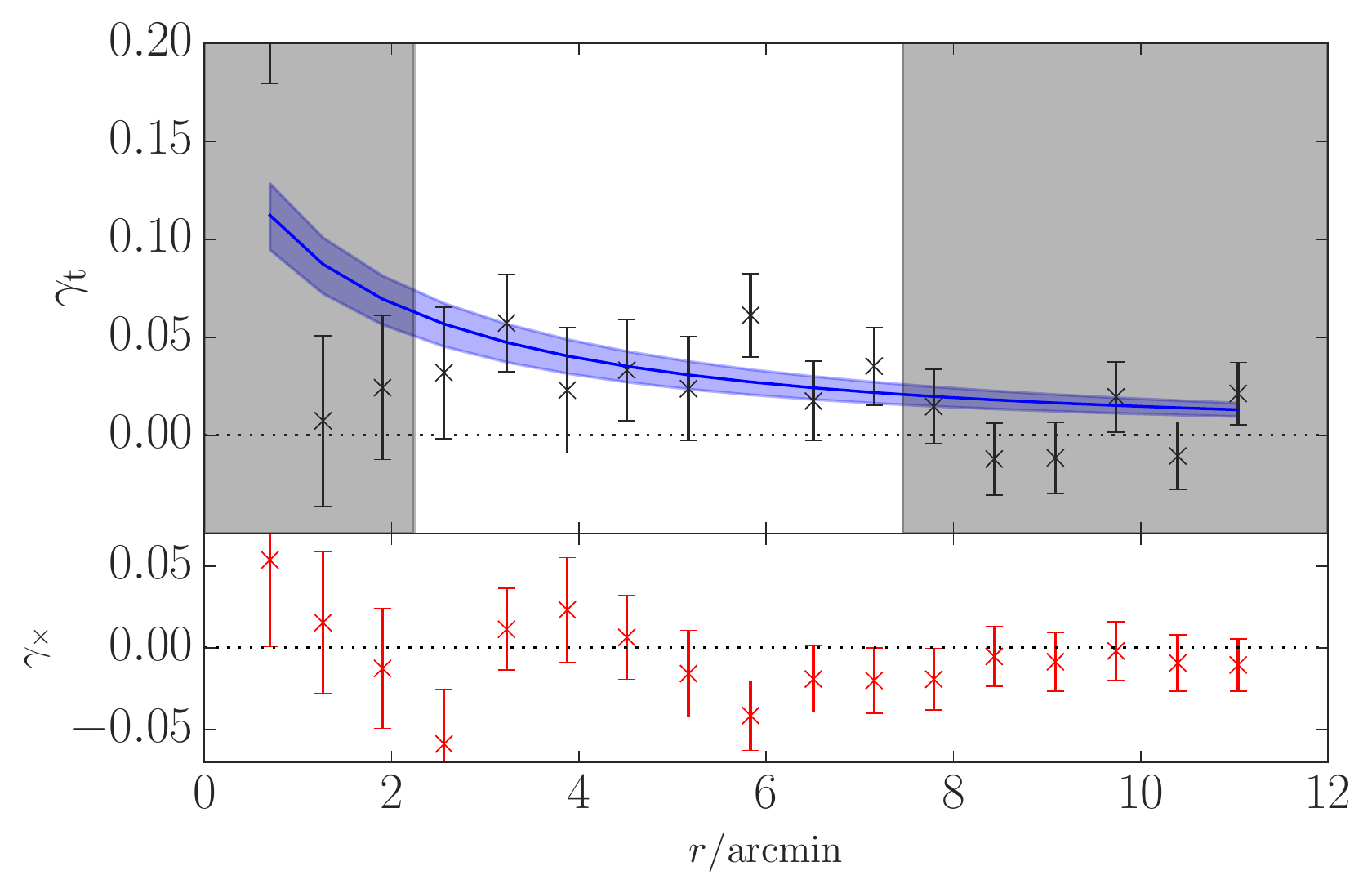}
   \label{fig:gammat2135}}
   \caption{Same as Figure~\ref{fig:kappa-shear-0234} for SPT-CL\,J2135$-$5726.}
   \label{fig:kappa-shear-2135}
 \end{figure*}

\begin{figure*}
   \subfigure[Surface mass density of SPT-CL\,J2138$-$6008.]{
   \includegraphics[width=\columnwidth]{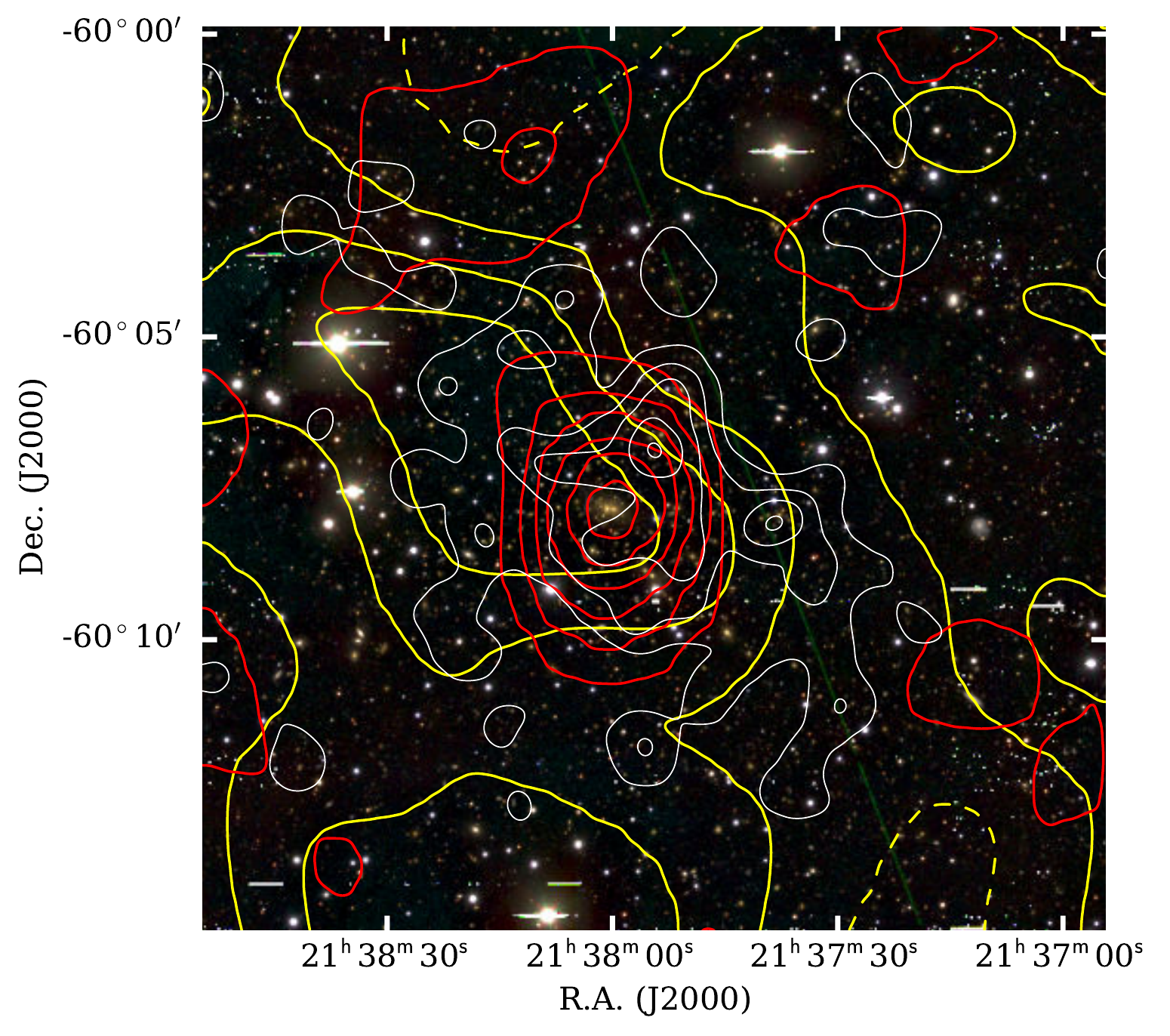}
   \label{fig:16}}
   \hfill
   \subfigure[Tangential shear profile of SPT-CL\,J2138$-$6008.]{
   \includegraphics[width=\columnwidth]{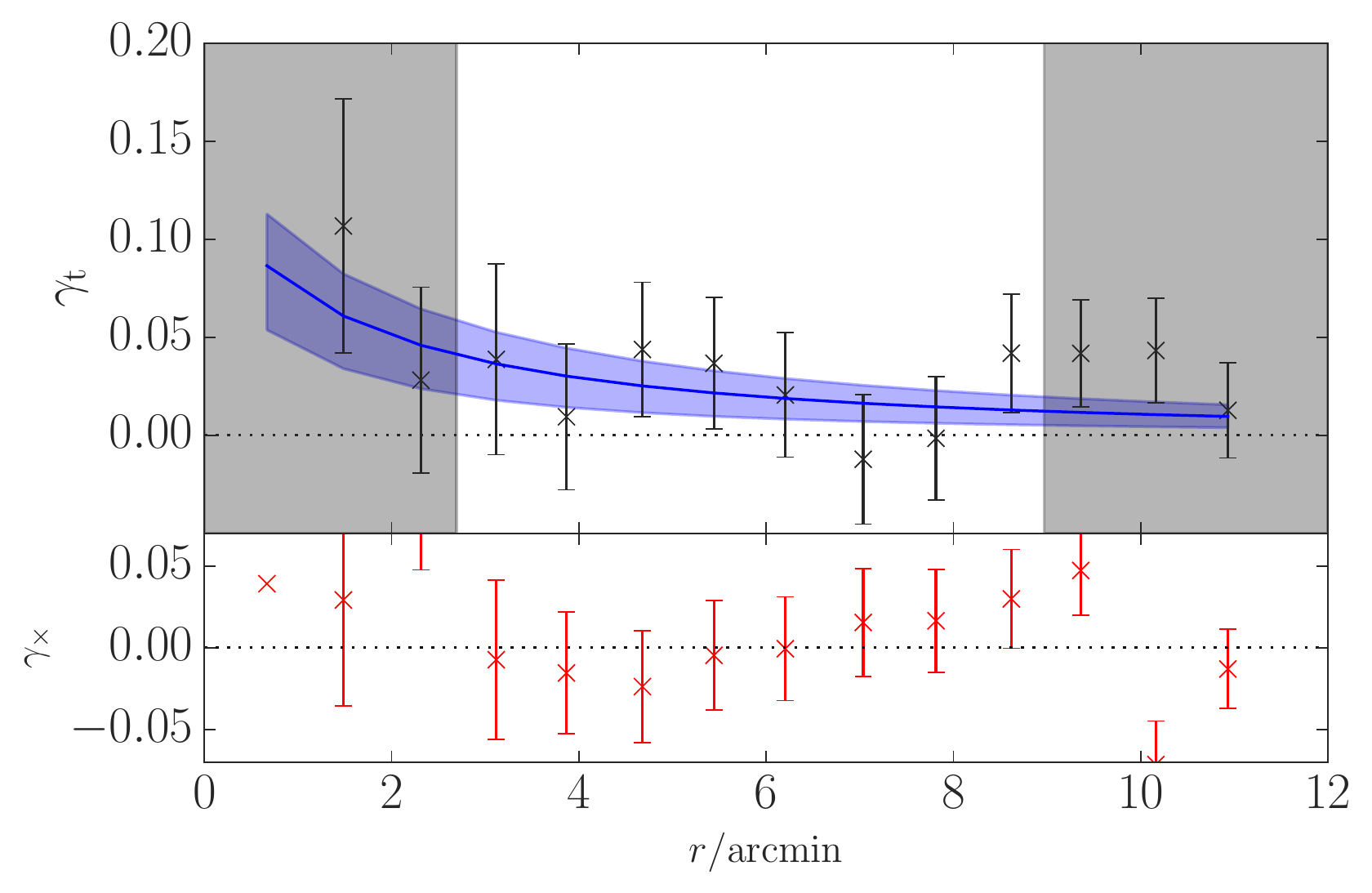}
   \label{fig:gammat2138}}
   \caption{Same as Figure~\ref{fig:kappa-shear-0234} for SPT-CL\,J2138$-$6008.}
   \label{fig:kappa-shear-2138}
 \end{figure*}

\clearpage
\begin{figure*}
   \subfigure[Surface mass density of SPT-CL\,J2145$-$5644.]{
   \includegraphics[width=\columnwidth]{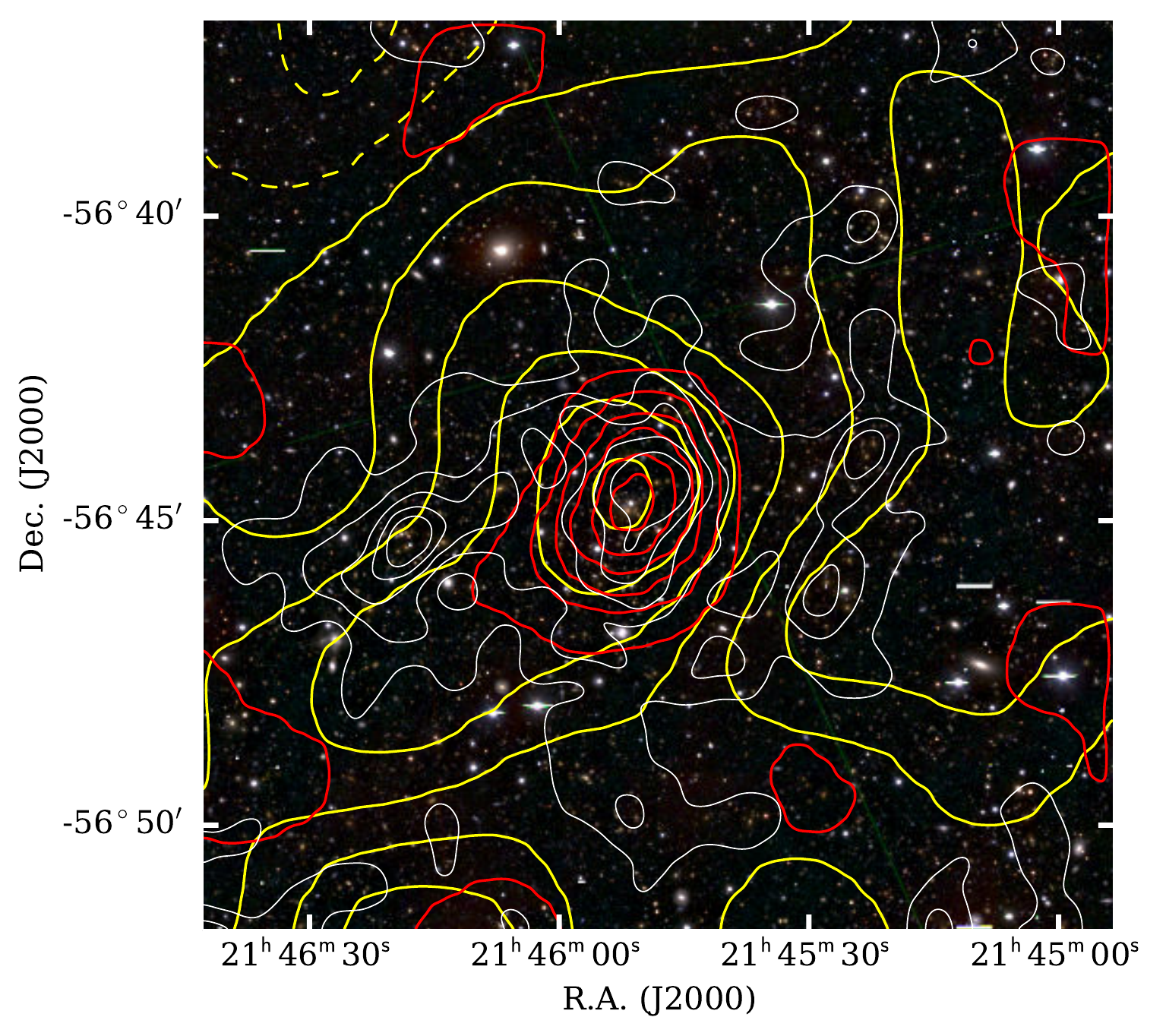}
   \label{fig:17}}
   \hfill
   \subfigure[Tangential shear profile of SPT-CL\,J2145$-$5644.]{
   \includegraphics[width=\columnwidth]{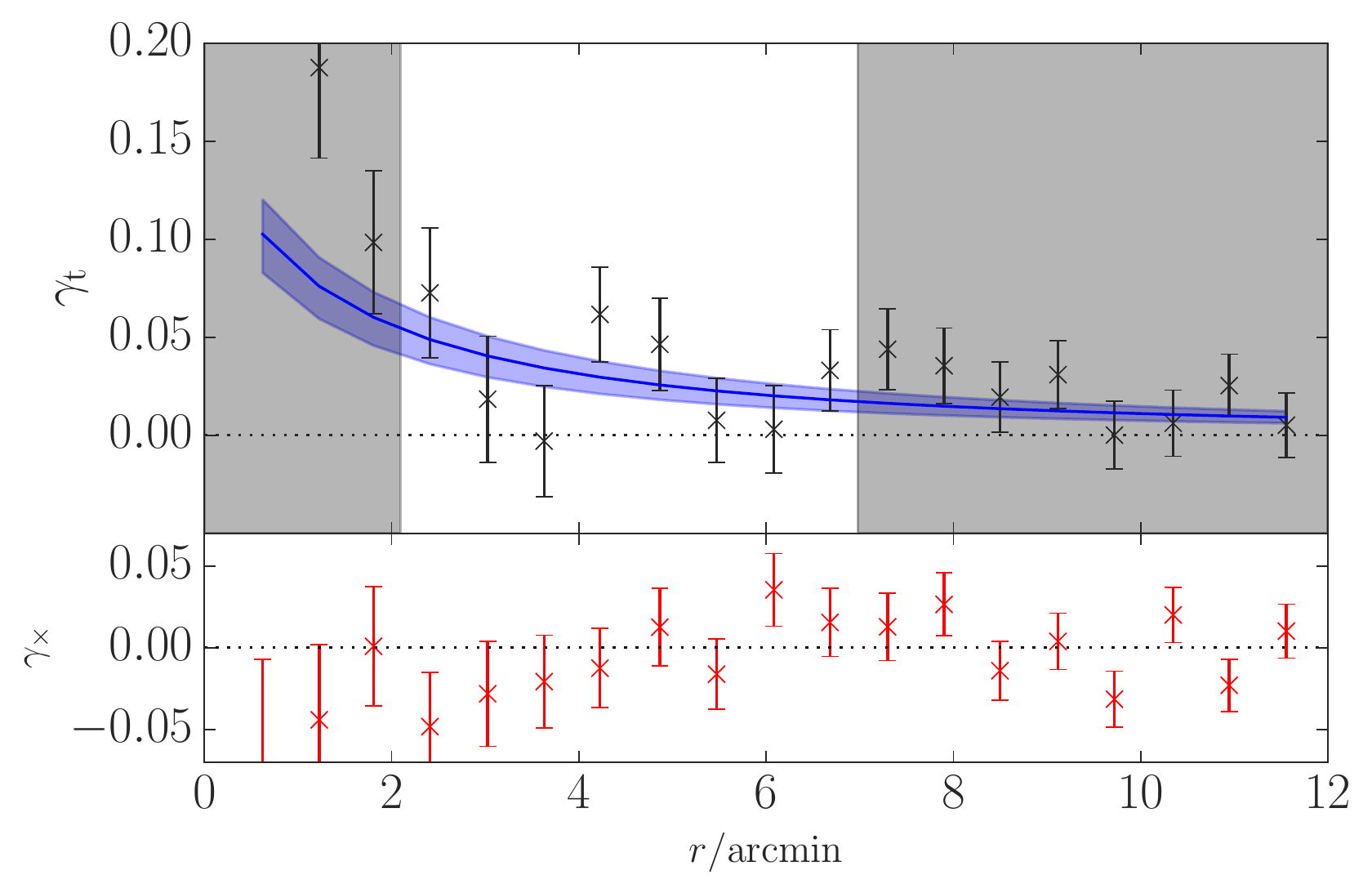}
   \label{fig:gammat2145}}
   \caption{Same as Figure~\ref{fig:kappa-shear-0234} for SPT-CL\,J2145$-$5644.}
   \label{fig:kappa-shear-2145}
 \end{figure*}

\begin{figure*}
   \subfigure[Surface mass density of SPT-CL\,J2332$-$5358.]{
   \includegraphics[width=\columnwidth]{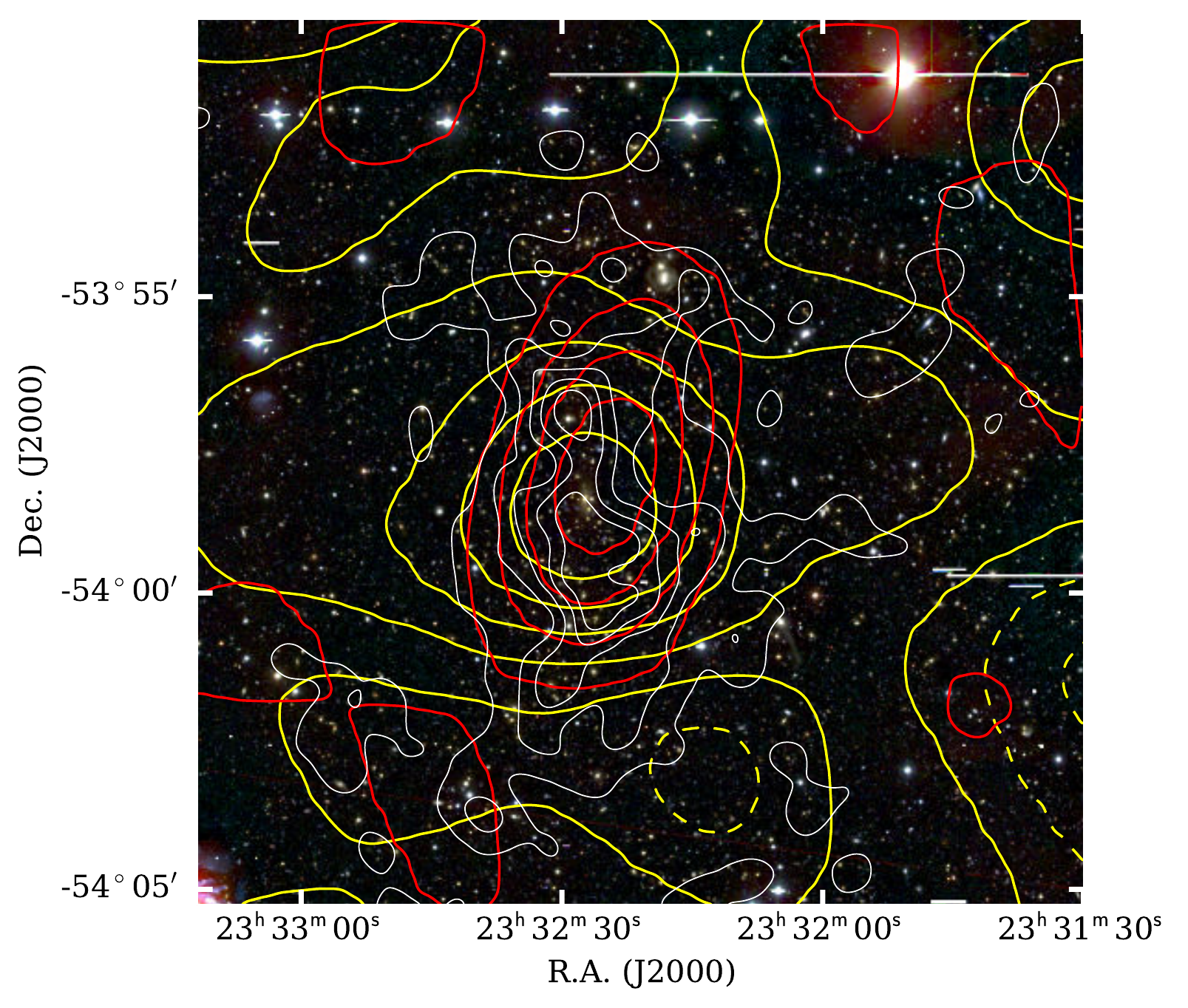}
   \label{fig:18}}
   \hfill
   \subfigure[Tangential shear profile of SPT-CL\,J2332$-$5358.]{
   \includegraphics[width=\columnwidth]{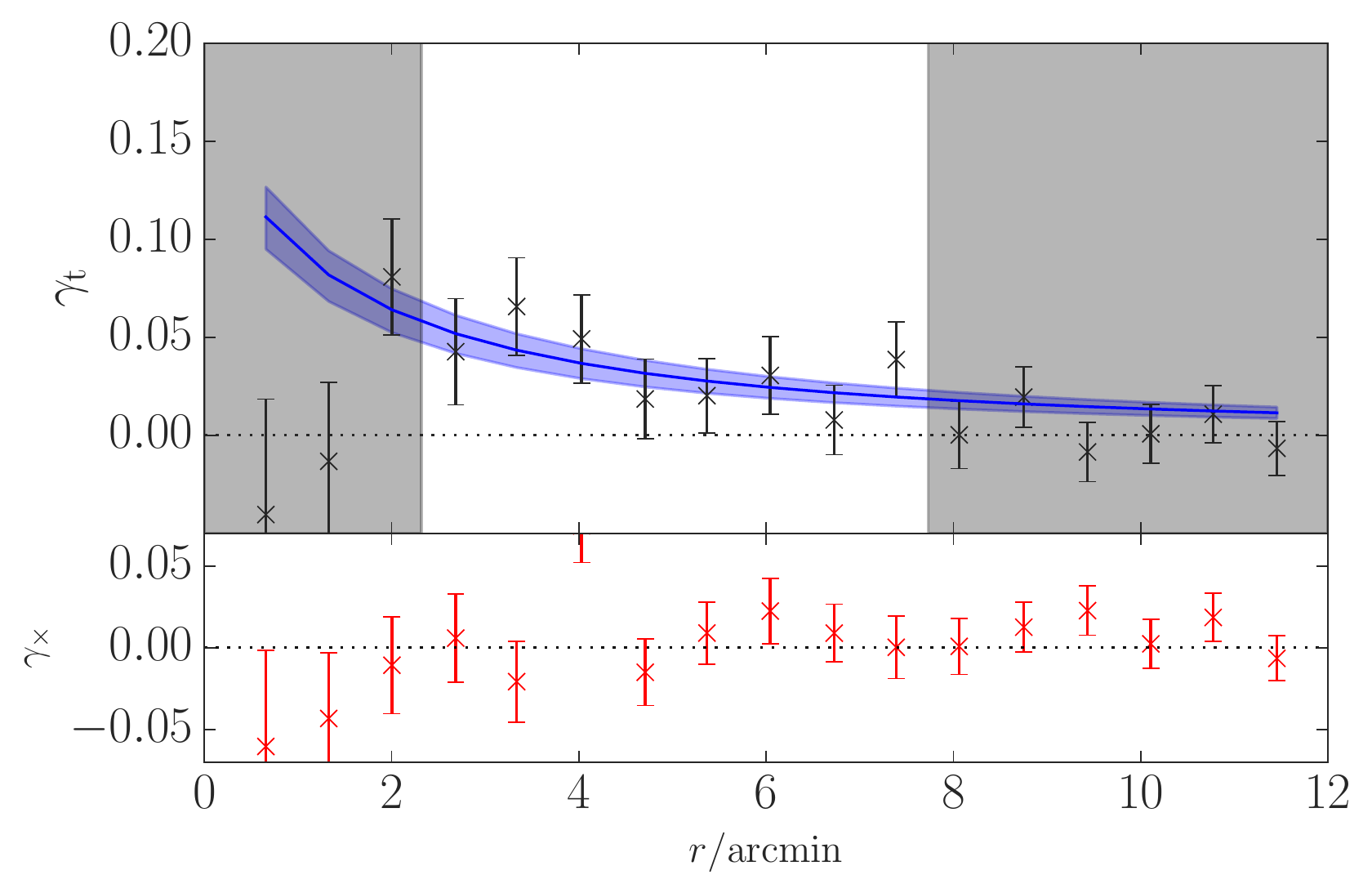}
   \label{fig:gammat2332}}
   \caption{Same as Figure~\ref{fig:kappa-shear-0234} for SPT-CL\,J2332$-$5358.}
   \label{fig:kappa-shear-2332}
 \end{figure*}

\clearpage
\begin{figure*}
   \subfigure[Surface mass density of SPT-CL\,J2355$-$5055.]{
   \includegraphics[width=\columnwidth]{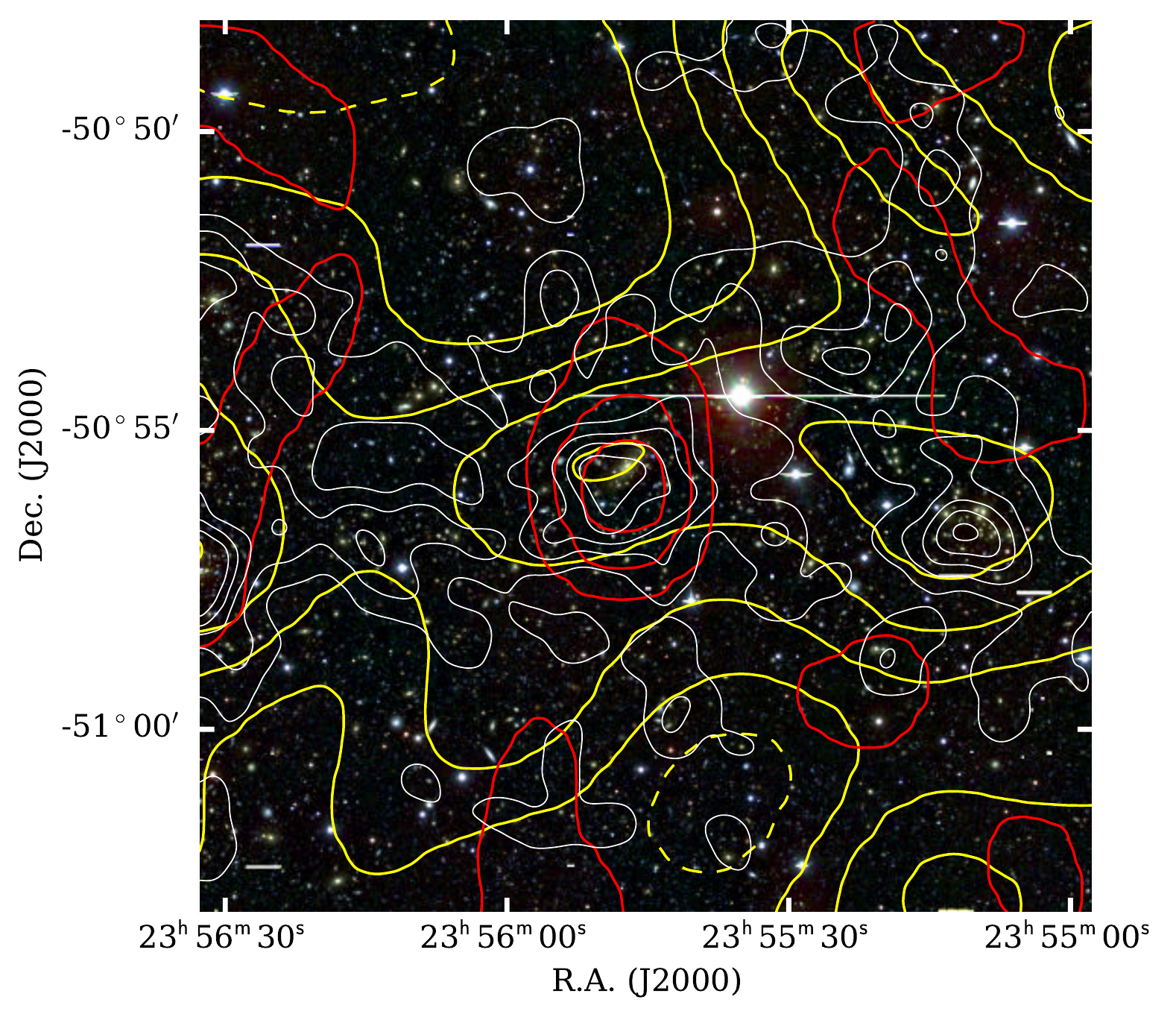}
   \label{fig:19}}
   \hfill
   \subfigure[Tangential shear profile of SPT-CL\,J2355$-$5055.]{
   \includegraphics[width=\columnwidth]{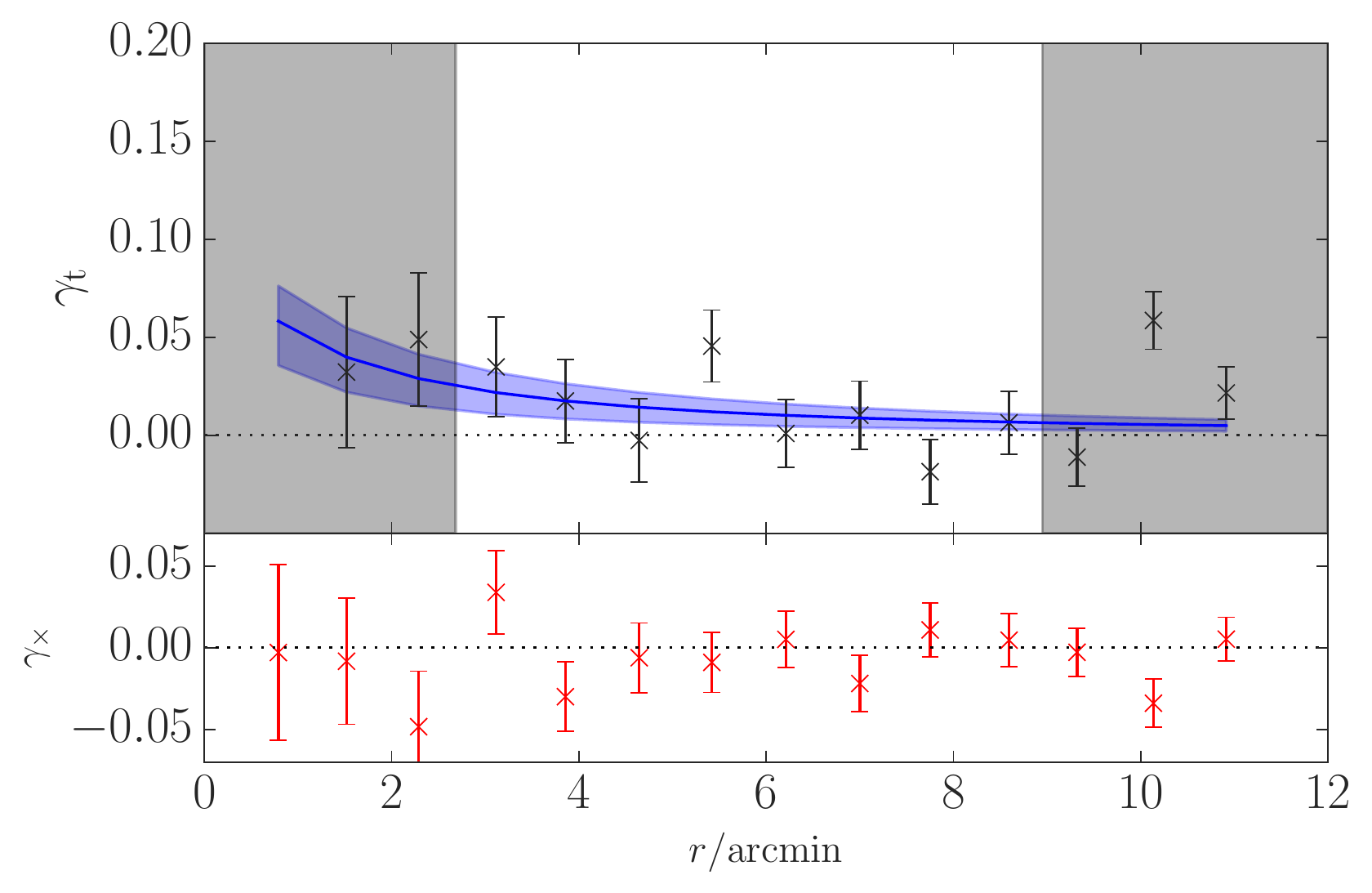}
   \label{fig:gammat2355}}
   \caption{Same as Figure~\ref{fig:kappa-shear-0234} for SPT-CL\,J2355$-$5055.}
   \label{fig:kappa-shear-2355}
 \end{figure*}

\clearpage

\section*{Affiliations}
\altaffilmark{\LMU}Faculty of Physics, Ludwig-Maximilians-Universit\"at,
Scheinerstr.\
1, 81679 Munich, Germany \\
\altaffilmark{\ECUniverse}Excellence Cluster Universe, Boltzmannstr.\ 2, 85748
Garching, Germany \\
\altaffilmark{\ANL}Argonne National Laboratory, High-Energy Physics Division,
9700 S. Cass
Avenue, Argonne, IL, USA 60439 \\
\altaffilmark{\KICPChicago}Kavli Institute for Cosmological Physics,
University of
Chicago, 5640 South Ellis Avenue, Chicago, IL, USA 60637 \\
\altaffilmark{\AIfA}Argelander-Institut f{\"u}r Astronomie, Auf dem H{\"u}gel
71, 53121
Bonn, Germany \\
\altaffilmark{\Leiden}Leiden Observatory, Leiden University, Niels Bohrweg 2,
2300 CA Leiden, The Netherlands\\
\altaffilmark{\MPE}Max-Planck-Institut f\"{u}r extraterrestrische Physik,
Giessenbachstr.\ 85748 Garching, Germany\\
\altaffilmark{\Stanford}Department of Physics, Stanford University, 382 Via
Pueblo Mall,
Stanford, CA 94305\\
\altaffilmark{\SLAC}SLAC National Accelerator Laboratory, 2575 Sand Hill Road,
Menlo Park, CA 94025\\
\altaffilmark{\KIPAC}Kavli Institute for Particle Astrophysics and Cosmology,
Stanford University, 452 Lomita Mall, Stanford, CA 94305\\
\altaffilmark{\FNAL}Fermi National Accelerator Laboratory, Batavia, IL 60510-0500\\
\altaffilmark{\AAUChicago}Department of Astronomy and Astrophysics, University
of Chicago, 5640 South Ellis Avenue, Chicago, IL 60637\\
\altaffilmark{\UMKC}Department of Physics and Astronomy, University of
Missouri, 5110 Rockhill Road, Kansas City, MO 64110\\
\altaffilmark{\MIT}Kavli Institute for Astrophysics and Space Research,
Massachusetts Institute of Technology, 77 Massachusetts Avenue,
Cambridge, MA 02139\\
\altaffilmark{\UFlorida}Department of Astronomy, University of Florida,
Gainesville, FL 32611\\
\altaffilmark{\Berkeley}Department of Physics, University of California,
Berkeley, CA 94720\\
\altaffilmark{\StonyBrook}{Department of Physics and Astronomy, Stony Brook
  University, Stony Brook, NY 11794, USA}\\
\altaffilmark{\Arizona}{Steward Observatory, University of Arizona, 933 North
  Cherry Avenue, Tucson, AZ 85721}\\
\altaffilmark{\Melbourne}School of Physics, University of Melbourne,
Parkville, VIC 3010, Australia\\
\altaffilmark{\CASA}Center for Astrophysics and Space Astronomy, Department of
Astrophysical and Planetary Science, University of Colorado, Boulder, C0
80309, USA\\ 
\altaffilmark{\Ames}NASA Ames Research Center, Moffett Field, CA 94035, USA\\
\altaffilmark{\INAFTrieste}INAF-Osservatorio Astronomico di Trieste, via
G. B. Tiepolo 11, 34143 Trieste, Italy\\
\altaffilmark{\LSST}{LSST, 950 North Cherry Avenue, Tucson, AZ 85719}
\altaffilmark{\CfA}Harvard-Smithsonian Center for Astrophysics, 60 Garden
Street, Cambridge, MA 02138\\
\altaffilmark{\Harvard}{Department of Physics, Harvard University, 17 Oxford
  Street, Cambridge, MA 02138} 
\end{document}